\documentclass[aps,pra,twocolumn,amsmath,amssymb,superscriptaddress,tightenlines]{revtex4}
\usepackage{graphicx}% Include figure files
\usepackage{epstopdf}
\usepackage{dcolumn}% Align table columns on decimal point
\usepackage{bm}% bold math
\usepackage{amssymb}
\usepackage{bm}
\usepackage{amsmath, bm}
\usepackage{upgreek}
\usepackage[colorlinks,citecolor=blue,linkcolor=blue,hyperindex]{hyperref}

\begin{document}

\title{The sixteenfold way and the quantum Hall effect at half-integer filling factors}
\author{Ken K. W. Ma %, P. T. Zucker, 
and D. E. Feldman}
\affiliation{Department of Physics, Brown University, Providence, Rhode Island 02912, USA}
\date{\today}


%----------------------- Abstract ------------------------------

\begin{abstract}
Fractional quantum Hall states at half-integer filling factors have been observed in many systems beyond the $5/2$ and $7/2$ plateaus in GaAs quantum wells. This includes bilayer states in GaAs, several half-integer plateaus in ZnO-based heterostructures, and quantum Hall liquids in graphene. In all cases, Cooper pairing of composite fermions is believed to explain the plateaus. The nature of Cooper pairing and the topological order on those plateaus are hotly debated. Different orders are believed to be present in different systems. This makes it important to understand experimental signatures of all proposed orders. We review the expected experimental signatures for all possible composite-fermion states at half-integer filling. We address Mach-Zehnder interferometry, thermal transport, tunneling experiments, and Fabry-P\'{e}rot interferometry. For this end, we introduce a uniform description of the topological orders of Kitaev's sixteenfold way in terms of their wave functions, effective Hamiltonians, and edge theories.
\end{abstract}

\maketitle

%-------------------------- Introduction -----------------------------

\section{Introduction} \label{sec:intro}

Experimental discovery of the fractional quantum Hall effect (FQHE)~\cite{Tsui1982} has started a new chapter in condensed matter physics and opened the field of topological matter. One major development was the observation of a quantized Hall plateau at the filling factor $\nu=5/2$~\cite{Willet1987, Pan1999}. This has led to many innovative theoretical ideas. One of them is non-Abelian statistics of elementary excitations ~\cite{MR1991, Wen1991, Greiter1992, Nayak1996} which may open a path to topological quantum computing \cite{Kitaev2003, Das-Sarma}. Another related idea is topological superconductivity \cite{Hasan-Kane, Qi-Zhang, Sato-Ando}.

A powerful approach to quantum Hall physics involves composite fermions~\cite{Jain_book}. In particular, odd-denominator FQHE can be understood as the integer quantum Hall 
effect of composite fermions. Such an explanation fails for half-integer states, where composite fermions are subject to zero effective magnetic field. Instead, a gapped half-integer liquid can be seen as a superconductor built from Cooper pairs of composite fermions~\cite{Read-Green}.

Different pairing channels give rise to different topological orders. The correspondence is not one to one. Indeed, pairing can occur in an infinite number of angular momentum channels. At the same time, topological order depends only on the types of anyons in the system and their mutual statistics. Kitaev's classification \cite{Kitaev} reveals 16 possibilities. Understanding which ones are relevant for available materials has proved a major challenge.

The bilayer state at $\nu=1/2$ in GaAs \cite{331-exp-1,331-exp-2} is believed to be the Abelian 331 liquid~\cite{Halperin331,  331-num-1,331-num-2,331-num-3}. The nature of the single-layer 5/2-liquid in GaAs remains controversial. The existing experimental results are consistent with the non-Abelian PH-Pfaffian 
liquid~\cite{Son2015, Son2016, Zucker2016, APf_Lee2007, x18a-Ashwin, x18b-Bonderson} 
at charge densities $n\sim 2~-~3\times 10^{11}~{\rm cm}^{-2}$. Yet, numerics ~\cite{x1-Morf, new-num, Zaletel2015, Rezayi-PRL, Luo2017, Yang-APf} supports the non-Abelian Pfaffian \cite{MR1991} and anti-Pfaffian 
orders \cite{APf_Lee2007, APf_Levin2007}. 
Besides, some experiments \cite{Lin2012, Baer2014, Fu2016} were interpreted as compatible with the Abelian 113 and 331 states~\cite{Halperin331, Guang2013, Guang2014}. To make matters even more puzzling, experimental evidence \cite{Pan2014, Samkharadze} exists for a different topological order at low electron densities $n<0.5\times 10^{11}~{\rm cm}^{-2}$. Little is known about the fragile $7/2$ state in 
GaAs~\cite{7/2-2002, Liu-7/2} and several recently discovered half-integer states in ZnO~\cite{Falson2015, Falson2018}. It was argued that the SU(2)$_2$ topological order is present in graphene~\cite{221_graphene, 221_graphene_exp}. A recent thermal conductance experiment \cite{Kasahara} also supports a topological order from Kitaev's classification in the spin-liquid material $\alpha-{\rm RuCl}_3$. Thus, several of the 16 theoretical possibilities are currently seen as viable candidates for real materials. Not enough evidence exists to dismiss the remaining orders of Kitaev's sixteenfold way. This makes it crucial to understand possible experimental signatures of all 16 orders and provides the main motivation for this paper.

Not all available experimental probes are equally useful to distinguish the 16 topological orders. For example, the quasiparticle charge of $e/4$ was reported by several groups on the $5/2$ plateau in GaAs~\cite{Dolev2008, Willett2009, Willett2010, Dolev2010, Radu2008, Lin2012, Baer2014, Venkatachalam2011, Fu2016}. This does not shed light on the topological order since the same quasiparticle charge is predicted in all 16 states. Similarly, the preponderance of the experimental evidence~\cite{NMR1, NMR2, optics, polarized-CF, Stern-optical} points at a spin-polarized FQHE liquid in GaAs at $n\sim 2~-~3\times 10^{11}~{\rm cm}^{-2}$. All 16 topological orders are compatible with a fully polarized liquid. One can get more information from quasiparticle tunneling \cite{Radu2008, Lin2012, Baer2014, Fu2016, Guang2013, Wen_book, Overbosch2008}, thermal conductance \cite{Read-Green, Banerjee2018, Kane_thermal, Cappelli_thermal, Ken2019}, upstream noise experiments \cite{Feifei2008, Bid2010, Dolev2011,  WF2011, Gross2012, WF2013}, and 
interferometry~\cite{Das-Sarma, Willett2010, Stern2006, Bonderson2006, Feldman2006, KT_noise, Ponomarenko2007, Ponomarenko2010, Chenjie2010, Guang2015, Chamon1997, Kane2003-MZ, Chung2006, Bonderson2007, Ilan2008, Bishara2009, Ilan2009, 
Bonderson_PRB2010, Stern_PRB2010, KT2006, Kang-FPI, Simon-noise, Deviatov2013, KT2008, Potter-2012}.  
These are the types of experiments we consider below. Interferometry in the Mach-Zehnder geometry \cite{MZ-2003} exhibits particularly interesting behavior.

Kitaev's classification addresses a neutral system, such as a spin liquid~\cite{Kitaev}. Its extension to a charged FQHE system involves subtleties which we handle below.
A simultaneous discussion of 16 topological orders requires their uniform description. Simple wave functions are known for some of the orders, such as Pfaffian~\cite{MR1991} and PH-Pfaffian~\cite{Zucker2016}. We use the known wave functions for the Pfaffian and 113 states to generate similar wave functions for all other topological orders. This is possible due to mother-daughter relations among all non-Abelian states and among all Abelian states. The same mother-daughter relations give a simple way to construct edge theories for all 16 orders and to iteratively generate effective Hamiltonians in a coupled-stripe construction. Coupled-wire constructions have already been used for half-integer states~\cite{Teo-Kane, Kane-Stern-Halperin, Fuji-Furusaki}. Our iterative approach is different.

The paper is organized as follows. In Sec.~\ref{sec:5/2}, we begin by introducing our construction of topological orders for half-integer FQH liquids. In particular, we formulate  mother-daughter relations among the orders and show a way to systematically generate wave functions for multiple orders. Then, in Sec.~\ref{sec:sixteen_fold}, we show explicitly that the resulting topological orders satisfy the sixteenfold way. Based on this result, we address multiple experimental  signatures of all orders in  Secs.~\ref{sec:EXP}, ~\ref{sec:MZ_experiment}, and~\ref{sec:summary}. The physics of Mach-Zehnder interferometry is especially rich and subtle. Its discussion occupies Sec.~\ref{sec:MZ_experiment}, with the more technical points being addressed in the Appendix. We summarize experimental signatures in Table \ref{tab:summary} in Sec.~\ref{sec:summary}.  In Sec.~\ref{sec:CW_construction}, we relate different topological orders iteratively and construct their effective Hamiltonians from a system of coupled quantum Hall stripes in the Pfaffian state. Any other topological order of the sixteenfold way could also be used as a starting point instead of the Pfaffian order. We conclude our work in Sec.~\ref{sec:conclusion}. 

%Appendix A contains a discussion of finite-temperature Mach-Zehnder interferometry.  The coupled-stripe construction for Jain states is given in Appendix B.

% ------------------------------- Generation of orders --------------------------------

\section{Topological orders of the sixteenfold way}  \label{sec:5/2}

We focus on an FQHE system with a filling factor $\nu=n+1/2$, where $n$ is an integer. In the simplest picture, $n$ filled spin-resolved Landau levels do not affect topological properties; FQHE physics is due to one half-filled spin-resolved Landau level. It is unclear, if such picture captures the relevant microscopic physics. For example, Coulomb interaction of electrons in different Landau levels is strong in GaAs at $\nu=5/2$. This results in Landau level mixing (LLM) effects~\cite{new-num, Zaletel2015, Rezayi-PRL, Luo2017, Bishara_LLM, Peterson_LLM, Simon_LLM, Sodeman_LLM}. In a uniform system without LLM, Pfaffian and anti-Pfaffian FQHE liquids have exactly the same energy~\cite{APf_Lee2007, APf_Levin2007}. Arbitrarily weak LLM breaks this degeneracy. Moreover, it was argued that strong LLM can help stabilize the PH-Pfaffian order~\cite{Zucker2016, Milovanovic_LLM}. The above picture also assumes a single-component (in particular, spin-polarized) wave function. The existing evidence does support spin polarized half-integer states in ZnO~\cite{Falson2015} and in GaAs~\cite{NMR1, NMR2, optics, polarized-CF} at the electron densities $n\sim 2~-~3\times 10^{11}{\rm cm}^{-2}$. At the same time, the accepted description of the $1/2$ state in bilayers assumes a two-component wave function with equal populations of the two layers~\cite{Halperin331}.

The above points reflect great difficulty of writing a realistic wave function for an experimentally relevant system. This difficulty becomes even more daunting if one attempts to incorporate disorder effects. Such effects are present in all samples and may be crucial for the nature of topological order \cite{Zucker2016, Mross2018, Wang2018, Lian2018, Zhu-Sheng}. On the other hand, if the topological order is known, much of the physics does not depend on the details of the wave function. Thus, it is useful to have simple representative wave functions for each of the 16 topological orders of the sixteenfold way. Such trial wave functions have been written for some of the orders, for example, Pfaffian \cite{MR1991}. As we will see, very similar trial wave functions emerge for all other composite fermion orders in half-integer FQHE. Of course, their construction sheds no light on which order is present in any given experimental system. This question can only be answered in a laboratory. In Secs.~\ref{sec:EXP} and~\ref{sec:MZ_experiment}, we address the relevant experimental signatures in detail.

In what follows, we will focus on the simplest setting of a single half-filled spin-polarized Landau level, $\nu=1/2$. Any disorder and LLM effects are neglected. The only exception will be the PH-Pfaffian wave function which greatly simplifies in a system with Landau-level mixing.

Wave functions of electrons in the lowest Landau level are significantly constrained by the analyticity requirement~\cite{Girvin1984}. They can be represented as
\begin{equation}
\label{dima2019-01}
\Psi(z_1,\ldots,z_N)=\Phi(z_1,\ldots,z_N)\exp\left(-\sum_{i=1}^{N}\frac{|z_i|^2}{4l_B^2}\right),
\end{equation}
where $z_k=x_k+i y_k$ are the positions of $N$ electrons, $l_B$ is the magnetic length, and $\Phi(z_1,\dots,z_N)$ is a holomorphic function. We will choose a polynomial $\Phi$. The polynomial is homogeneous since the degree of each monomial is proportional to the angular momentum. A single-electron wave function with $\Phi(z)\sim z^k$ describes charge density concentrated along a ring of radius $r\sim \sqrt k$~\cite{Girvin-notes}. Hence, the polynomial $\Phi(z_1,\ldots,z_N)$ describes electrons on a disk of radius $R\sim\sqrt{k_{\rm max}}$, where $k_{\rm max}$ is the highest power of $z_i$ in $\Phi$. The total degree $k=\sum k_i$ of each monomial $z_1^{k_1}\ldots z_N^{k_N}$ in $\Phi$ satisfies the relation $k\approx N^2/[2\nu]$.

This leads us to another constraint that the wave function should produce the correct charge density. At $\nu=1/2$, the simplest choice, compatible with the correct density, is $\Phi(z_1,\ldots,z_N)=\Phi_{\rm flux}=\Pi_{i<j}(z_i-z_j)^2$. This yields an acceptable wave function for bosons but not for electrons. To fix the statistics, $\Phi_{\rm flux}$  must be multiplied by an antisymmetric factor $\Phi_{\rm cf}(z_1,\ldots,z_N)$. To maintain the correct density in the thermodynamic limit $N\rightarrow \infty$,  the factor should change the degree $k$ of $\Phi$ by $o(N^2)$. 

The two factors $\Phi_{\rm flux}$ and $\Phi_{\rm cf}$ have a natural interpretation in the composite fermion picture~\cite{Jain_book}.
Note that experimental evidence exists for composite fermions at $\nu=5/2$ in GaAs \cite{polarized-CF, dima:W-CF}.
The former factor $\Phi_{\rm flux}$ describes two flux quanta attached to each electron. The resulting composite fermions move in a zero effective magnetic field. Their wave function is $\Phi_{\rm cf}$. It describes Cooper pairing of composite fermions.
%A fractional quantum Hall system with $\nu=5/2$ can be viewed as $\nu=2+1/2$. From numerical simulation, it was suggested that the lowest Landau level is completely filled by electrons %with both spins, whereas the remaining electrons at the first excited Landau level are spin polarized~\cite{Morf, Rezayi_Haldane}. When Landau level mixing is negligible, one tries to %neglect the completely filled Landau level and focus on the partially filled Landau level. Haplerin, Lee and Read developed a theory to describe the spin polarized half-filled electron %gas~\cite{HLR_theory}. Their theory was built on the concept of composite fermion, which attaches two flux quanta to each electron. Under the mean field approximation, the magnetic field from the flux quanta cancels the external magnetic field. In other words, the composite fermions feel a zero effective magnetic field, which allows them to form a well-defined %gapless Fermi liquid. If HLR theory is also valid for 
%$\nu=5/2$, then it will be the end of the story. No quantized Hall plateus is expected at this filling factor.
%However, the system with $\nu=5/2$ is gapped~\cite{Willet1987, Pan1999}. It is believed that the incompressible nature is originated from Cooper pairing of composite fermions near to %the Fermi surface. 
The most famous example is $p$-wave pairing in the Pfaffian state~\cite{Read-Green, Ivanov2001, Scarola_nature, Fisher_Nayak2007}. %which has one downstream Majorana mode at the edge. 
A trial wave function takes the form
\begin{eqnarray}
\text{Pf}\left(\frac{1}{z_i-z_j}\right)
\prod_{i<j}\left(z_i-z_j\right)^2
\exp{\left(-\sum_{i=1}^N \frac{\left|z_i\right|^2}{4l_B^2}\right)},
\end{eqnarray}
with $\Phi_{\rm cf}=\text{Pf}\left(\frac{1}{z_i-z_j}\right)$.
As was observed by Moore and Read \cite{MR1991}, the above choice of the bulk wave function determines the nature of the gapless edge modes. There are two of them: a charged boson and a neutral Majorana fermion.
%The Pfaffian comes from the ground state wavefunction of the BCS pairing, which is the origin of the non-Abelian nature for the topological order. The second term encodes the idea of flux attachment. 

In general, pairing between composite fermions can be described by an effective Bardeen-Cooper-Schrieffer (BCS) mean-field Hamiltonian. For spinless fermions, we have:
\begin{eqnarray} \label{eq:BCS_hamiltonian}
H_{\text{BCS}}
=
\sum_{\mathbf{k}}
\left[
\xi_{\mathbf{k}}c^{\dagger}_{\mathbf{k}}c_{\mathbf{k}}
+\frac{1}{2}
\left(\Delta^{*}_{\mathbf{k}}c_{-\mathbf{k}}c_{\mathbf{k}}
+\Delta_{\mathbf{k}}c^{\dagger}_{\mathbf{k}}c^{\dagger}_{-\mathbf{k}}\right)
\right].
\end{eqnarray} 
Here, $\xi_{\mathbf{k}}=\frac{k^2}{2m}-\mu$, with $m$ and $\mu$ labeling the effective mass of a composite fermion and the chemical potential of the system, respectively. The fermionic creation operator $c_{\mathbf{k}}^{\dagger}$ and destruction operator 
$c_{\mathbf{k}}$ satisfy the anti-commutation relation: 
$\{c_{\mathbf{k}}, c^{\dagger}_{\mathbf{k}'}\}=\delta_{\mathbf{k},\mathbf{k}'}$. The pairing function is denoted as $\Delta_{\mathbf{k}}$. When $\mu>0$, the system is in the weak-pairing phase~\cite{Read-Green}. In Ref.~\cite{Dubail_Read}, Dubail and Read modelled the gap function for the complex $l$-wave pairing as $\Delta_{\mathbf{k}}=\tilde{\Delta}(k_x\pm ik_y)^l$. By analyzing the entanglement spectrum in some specific cases, they showed that this kind of pairing between spinless fermions should lead to $l$ chiral Majorana fermions at the edge.

The edge structure establishes the connection of wave functions with Kitaev's sixteenfold way. As shown by Kitaev~\cite{Kitaev}, different topological orders in topological superconductors of composite fermions differ by the number of Majorana modes on the edge. Moreover, since the bulk is gapped, the universal low-energy physics is determined by the edge structure. 
As a consequence, all experimental probes, addressed in this paper, involve edge physics. 
Thus, we begin our discussion of topological orders with a review of their edge structures. This will uncover mother-daughter relations between various orders and will allow us to generate relevant wave functions in a straightforward way.

%From the previous discussion, we suggest that non-Abelian topological orders for $\nu=5/2$ FQH state with more Majorana modes at the edge can be constructed by generalizing the $p+ip$ %pairing to higher $l$-wave pairing. This is equivalent to increasing the number of neutral modes at the edge since a pair of Majorana modes can be combined into a conventional neutral %mode. In this section, we adopt another perspective by focusing on the edge structure. We demonstrate how the above topological orders can be generated systematically via particle-hole %conjugation and neutral-mode flipping. By doing so, we obtain the corresponding $K$ matrices and determine the scaling dimensions for the quasiparticles of the topological orders.

\subsection{Mother-daughter relations for non-Abelian orders}

As shown by Wen and Zee~\cite{Wen_Zee, Wen_book}, an Abelian topological order for a fractional quantum Hall state can be characterized by a $K$-matrix and a charge vector $t$. Although the Pfaffian order is non-Abelian, one can view it as a direct product between SU(2)$_2$ Ising anyon and an Abelian U(1) bosonic sector. The latter is characterized by a $1\times 1$ matrix $K=(2)$ and $t=1$, so that $\nu=t^T K^{-1} t =1/2$. The $K$-matrix determines the Abelian modes on the edge. In particular, their number equals the dimension of the matrix. Thus, the Pfaffian edge contains only one Abelian Bose-mode, in addition to a Majorana fermion from the Ising sector. The direction of those modes is determined by the sign of the magntic field and is called ``downstream". We will generally assume that downstream is counterclockwise \cite{foot-counterclockwise}.

To generate a chain of topological orders, we particle-hole conjugate \cite{APf_Lee2007, APf_Levin2007} the Pfaffian order. On the edge, this means reversing the direction of all Bose and Majorana edge modes (downstream $\rightarrow$ upstream) and adding an integer downstream Bose-mode. We obtain the anti-Pfaffian order with an upstream Majorana and a $2\times 2$ diagonal $K$-matrix encoding two Abelian modes: 
$K_{11}=1$, $K_{22}=-2$. The charge vector $t=(1,1)^T$. In terms of edge physics, the first element of the charge vector corresponds to the contribution from the $\nu=1$ edge and the second element comes from the reversed $\nu=1/2$ FQHE edge. We denote the two corresponding charged modes as $\left\{\phi_0,\phi_1\right\}$.  Disorder on the edge equilibrates the two modes. The appropriate language for edge physics involves then two linear combinations of $\phi_{0,1}$: an overall charged mode that propagates downstream and a neutral upstream boson~\cite{APf_Lee2007, APf_Levin2007}. Specifically, we consider a change of the basis $\left(\phi_\rho,\phi_n\right)^T=W\left(\phi_0,\phi_1\right)^T$. The $K$ matrix and the charge vector transform as
\begin{eqnarray} \label{eq:transform}
K\rightarrow (W^{-1})^T K W^{-1},~\quad
t\rightarrow (W^{-1})^T t.
\end{eqnarray}
One can easily check that the filling factor $\nu$ is invariant under the transformation. The $W$ matrix in this case is given by
\begin{eqnarray}
W
=\begin{pmatrix}
1 & 1 \\
1 & 2 \\
\end{pmatrix}.
\end{eqnarray}
The transformed $K$ matrix is diagonal with $K_{11}=2$, $K_{22}=-1$, and the charge vector becomes $t=(1,0)^T$. 

Our second trick is flipping the neutral modes.
Indeed, the SU(2)$_2$ edge structure can be obtained by flipping the directions of the neutral modes (Majorana fermion $\psi$ and the bosonic mode $\phi_n$). As a result, $\psi$ and $\phi_n$ get the same downstream chirality as $\phi_{\rho}$. The flipping of the bosonic neutral mode corresponds to changing the sign of  $K_{22}$: $-1\rightarrow 1$.
This trick can also be seen as negative-flux attachment \cite{Jolicoeur2007}.

\begin{figure}[htb]
\includegraphics[width=2.8 in]{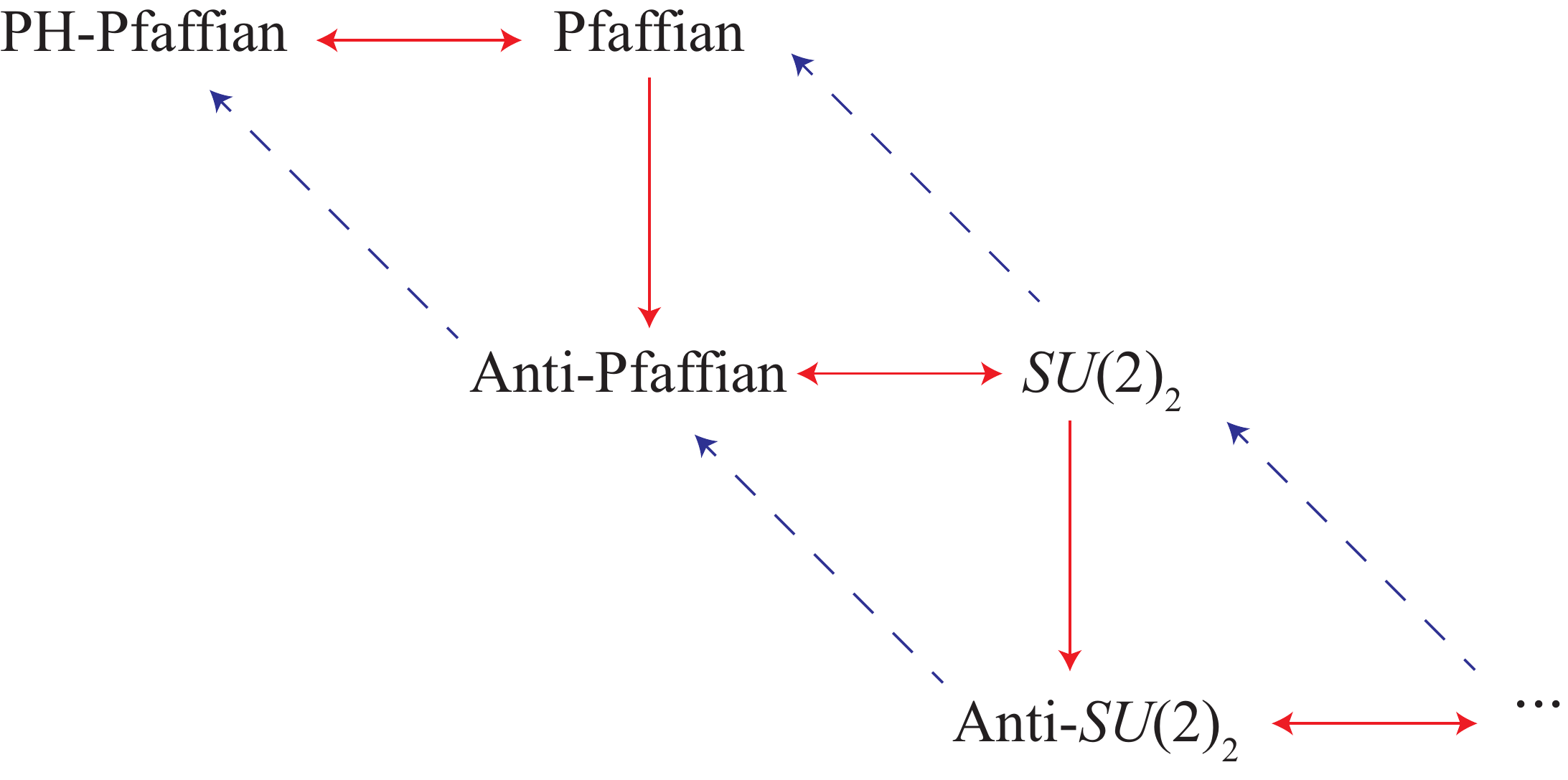}
\caption{(Color online) Illustration of generating non-Abelian topological orders for the $\nu=5/2$ fractional quantum Hall state by particle-hole conjugation (vertical arrows) and neutral-mode flipping (horizontal arrows). The topological orders can be related iteratively via a coupled-stripe construction in Sec.~\ref{sec:CW_construction}. The red solid lines correspond to the first construction (CW1) and the blue dashed lines correspond to the second construction (CW2).}
\label{fig:non_Abelian}
\end{figure}

By repeating the above processes, a chain of non-Abelian topological orders can be generated iteratively as shown in Fig.~\ref{fig:non_Abelian}. The $K$ matrices for the topological orders obtained with particle-hole conjugation are diagonal with $K_{11}=1$, $K_{22}=-2$ and $K_{ii}=-1~(i\geq 3)$, the charge vector being $t=(1,1,0,\cdots, 0)^T$. We rewrite the
$K$-matrices in the basis of a single downstream charged mode and multiple upstream neutral modes. This is achieved by using the following $W$ matrix:
\begin{eqnarray}
W=
\begin{pmatrix}
1 & 1 & 0 & 0 & \cdots & 0 & 0 \\
1 & 2 & 0 & 0 & \cdots & 0 & 0 \\
0 & 0 & 1 & 0 & \cdots & 0 & 0 \\
\vdots & \vdots & \vdots & \vdots & \ddots & \vdots & \vdots \\
0 & 0 & 0 & 0 & \cdots & 1 & 0 \\
0 & 0 & 0 & 0 & \cdots & 0 & 1 \\
\end{pmatrix}.
\end{eqnarray}
With the transformation from Eq.~\eqref{eq:transform}, the $K$ matrix transforms into 
\begin{eqnarray} \label{eq:diagonalized_K}
K=
\begin{pmatrix}
2 & 0 & 0 & \cdots & 0 \\
0 & -1 & 0 & \cdots & 0 \\
0 & 0 & - 1& \cdots & 0 \\
\vdots & \vdots & \vdots & \ddots &\vdots \\
0 & 0 & 0 &\cdots & -1 \\
\end{pmatrix}.
\end{eqnarray}
Here, the negative sign indicates that all neutral modes have opposite chirality with respect to the charged mode. This gives a natural description of a disorder-dominated phase~\cite{Kane-disorder-dominated}. % in which the system has a single charge mode and all remaining modes are neutral. 
To flip all the bosonic neutral modes, one simply changes all negative matrix elements from $-1$ to $1$.

\subsection{Mother-daughter relations for Abelian orders}

\begin{figure}[htb]
\includegraphics[width=2.5 in]{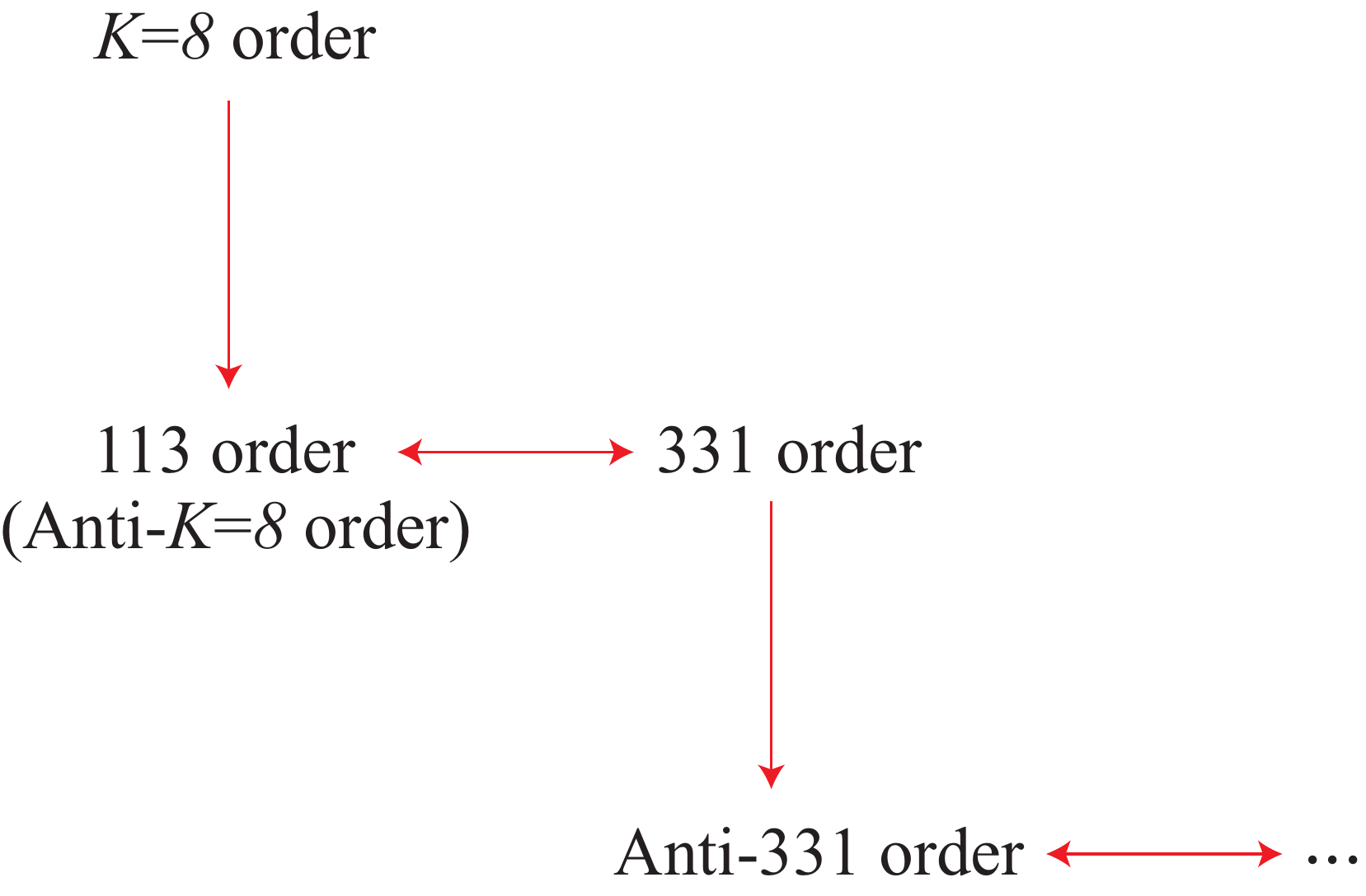}
\caption{(Color online) Illustration of generating Abelian topological orders  by particle-hole conjugation (vertical arrows) and neutral-mode flipping (horizontal arrows). It is reminded that the spin-polarized $113$ order is topologically equivalent to the anti-$K=8$ state 
%by a transformation $K'=S^T K S$
~\cite{Guang2014}. The topological orders can be related iteratively with the coupled-stripe construction of Sec.~\ref{sec:CW_construction}.}
\label{fig:Abelian}
\end{figure}

Using the techniques of the previous subsection, a chain of Abelian topological orders can also be generated. As illustrated in Fig.~\ref{fig:Abelian}, we start with the $K=8$ state to obtain the 
anti-$K=8$ state by particle-hole conjugation. Note that the edge modes of the anti-$K=8$ state cannot be equilibrated by weak disorder in the $T\rightarrow 0$ limit~\cite{Guang2014}.

The polarized version of the 113 order is topologically equivalent to the anti-$K=8$ state~\cite{Guang2014}. Neutral-mode flipping produces the 331 order from the 113 order. Then, the anti-331 order \cite{Guang2013} can be obtained by particle-hole conjugation. As in the non-Abelian case, the $K$ matrices can be diagonalized in the form ~\eqref{eq:diagonalized_K} with the corresponding charge vector $t=(1,0,\cdots,0)^T$. 

\subsection{Construction of wave functions}

One curious aspect of the existing proposals for quantum Hall states at half-integer filling factors is a great diversity of their names: Pfaffian, anti-Pfaffian, 331, SU(2)$_2$, $K=8$, and so on. This diversity reflects a great variety of the methods used to introduce those topological orders. The 331 state was discovered with a generalization of the Laughlin wave function for two flavors of electrons~\cite{Halperin331}; the Pfaffian state emerged from a connection with a conformal field theory (CFT)~\cite{MR1991}; SU(2)$_2$ was introduced with a parton construction~\cite{Jain-221}; $K=8$ can be understood as a quantum Hall state of bosons~\cite{Wen_Zee}; the anti-Pfaffian topological order was obtained with particle-hole conjugation~\cite{APf_Lee2007, APf_Levin2007}. Yet, the connection with the sixteenfold way shows that all those orders are close relatives. This is reflected by the mother-daughter relations between the orders. Below we will use those relations to generate a wave function for each order in a systematic way.

The structure of the wave functions will also motivate the prescription for finding allowed quasiparticle types and their mutual statistics. The prescription assumes that a CFT describes the edge of the system. An operator or operators are selected to describe electrons in CFT~\cite{Hansson-CFT}. All possible quasiparticles correspond to other CFT operators whose operator product expansions (OPE) with the electron operators are single-valued.

This prescription is broadly used, but its justification is not obvious. Indeed, the edges of realistic systems are never described by a CFT because different edge modes have different velocities, and numerous irrelevant and sometimes even relevant perturbations enter the Hamiltonian. In our case, the prescription will be placed on a firmer footing by the analysis of bulk wave functions for excited states. Of course, the best proof of the prescription consists in verifying that it reproduces the properties of the sixteenfold way. We confirm that in the next section.

We begin with non-Abelian states and briefly extend our arguments to the Abelian case. As is customary, we consider wave functions in the first Landau level.  Any such wave function is the product of an analytic function with the exponential factor $\exp(-\sum_i |z_i|^2/4l_B^2)$~\cite{Girvin1984}. We generate wave functions in an iterative way. The iterative procedure involves neutral-mode flipping and particle-hole conjugation. We illustrate these two tricks with constructions of the PH-Pfaffian and anti-Pfaffian liquids from the Pfaffian state.

The wave function of the Pfaffian state is well known:
\begin{equation}
\label{dima-1}
\Psi_{\rm Pf}(\{z_i\})
={\rm Pf}\{\frac{1}{z_i-z_j}\}
\prod_{i<j}(z_i-z_j)^2\exp(-\sum_{i} |z_i|^2/4l_B^2),
\end{equation}
where $z_k=x_k+iy_k$ is the position of an electron, and $l_B$ is the magnetic length. The complex analytic factor can be reinterpreted as a correlation function
$G=\langle\Pi \Psi_k \hat O\rangle$ of the electron operators 
$\Psi_k=\psi(z_k)\exp[2i\phi(z_k)]$ in the conformal field theory~\cite{MR1991} with the Lagrangian density
\begin{equation}
\label{dima-2}
\mathcal{L}
=\frac{2}{4\pi}[i\partial_y\phi\partial_x\phi+(\partial_x\phi)^2]+\psi(\partial_y-i\partial_x)\psi,
\end{equation}
where $y$ plays the role of the imaginary time; the operator $\hat O$ is localized far away from the system and compensates the electrical charge to ensure that the correlation function of the fields $\exp(2i\phi)$ is nonzero.

The CFT interpretation makes it easy to identify quasiparticles. We define wave functions of excited states as correlation functions $G_q=\langle\Pi \Psi_k \hat q(\xi_0) \rangle$, where $\hat q$ creates a quasiparticle (from now on we ignore the neutralizing operator $\hat O$). $\hat q$ is constructed from the operators of the CFT (\ref{dima-2}). For example, the twist field $\sigma$ of the Majorana part of the CFT corresponds to $\hat q=\sigma\exp(i\omega\phi)$. The parameter $\omega$ determines the charge of the excitation. It can be found from the requirement that the wave function is a single-valued function of the electron positions. This identifies $\omega=1/2+n$ and the quasiparticle charge is $e/4+ne/2$.

The PH-Pfaffian wave function is obtained with the help of complex conjugation of the Pfaffian factor in (\ref{dima-1}). This structure of the wave function reflects a close connection with the Pfaffian order. For example, all density-density correlations are exactly the same for the Pfaffian and PH-Pfaffian wave functions since the absolute values of the wave functions coincide.

The resulting wave function $\Psi_{\rm PH}$ is no longer holomorphic and hence does not describe electrons in a single Landau level. Given strong LLM in realistic systems, this does not create a problem. Nevertheless, it is important for us to discuss how wave functions  can be transformed into a holomorphic form. This involves projection to the lowest Landau level~\cite{Zucker2016}:
\begin{equation}
\label{dima-3}
\Psi_{\rm PH}\rightarrow\int\{d^2\xi_i\}\langle\{z_i\}|\{\xi_i\}\rangle\Psi_{\rm PH}(\{\xi_i\}),
\end{equation}
where $\langle\{z_i\}|\{\xi_i\}\rangle=\Pi_i\exp[-(|\xi_i|^2-2\bar\xi_iz_i+|z_i|^2)/4l_B^2]$, and the bar denotes complex conjugation. The wave function before projection can be understood as a correlation function of the CFT that differs from~\eqref{dima-2} by the opposite sign in front of 
$i\partial_x$ in the Majorana part of the action.
This corresponds to neutral-mode flipping, or, alternatively, negative-flux attachment. Excited states can again be represented in terms of the correlation functions 
with the insertion of quasiparticle operators. One can also multiply the wave function by a real rotationally-invariant function $R(\{\xi_i\})$ of 
the coordinates 
before the projection to the lowest Landau level. This is not expected to affect topological 
properties.
Physically allowed quasiparticles can be found from the single-valuedness of the wave function before the lowest Landau level projection. Indeed, the integral~\eqref{dima-3} is not well defined, if $\Psi_{\rm PH}(\{\xi_i\})$ is not single valued. 

Recent numerical work \cite{Mishmash} suggests that the simplest projection procedure generates a gapless state from $\Psi_{\rm PH}$. One possibility is that a factor  $R(\{\xi_i\})$ should be introduced before projection. Another possibility is that LLM is essential for maintaining a gap in the PH-Pfaffian liquid.

The CFT prescription ~\cite{Hansson-CFT} assumes that the ground-state wave functions it generates are separated by a gap from all excitations. This assumption is plausible and is supported by several well-understood examples. Nevertheless, it can only be proven by identifying a Hamiltonian for which the CFT-generated wave function is the gapped ground state. This explains the importance of Sec. VII below, where we use the same mother-daughter relations as in this Section to generate effective Hamiltonians for all topological orders of the sixteenfold way. We are working on an extension of our approach to translationally invariant Hamiltonians. 

The role of the particle-hole (PH) symmetry in the PH-Pfaffian state is another subtlety. The PH-Pfaffian topological order is consistent with the PH symmetry, but it is not protected by that symmetry, and so the corresponding ground-state wave functions do not have to be particle-hole symmetric. Moreover, it was argued \cite{Milovanovic_LLM} that a PH-symmetric wave function, which would naturally emerge in Son's picture of massless Dirac fermions, must be gapless. A gapped state with the PH-Pfaffian order requires massive Dirac fermions
 \cite{Milovanovic_LLM}. This means the absence of the microscopic PH symmetry.

The transition from the Pfaffian to PH-Pfaffian state is a template for neutral-mode flipping in our construction of wave functions for all topological orders. 
We start with a wave function of the form $\Psi_{\rm unflipped}=\Pi_{i<j}(z_i-z_j)^2
\Psi_{\rm neutral}(\{z_i\})$. The flipped wave function is obtained by the complex conjugation of the expression for $\Psi_{\rm neutral}$: $\Psi_{\rm flipped}=\Pi_{i<j}(z_i-z_j)^2\bar\Psi_{\rm neutral}(\{z_i\})$. If the resulting wave function remains non-analytic after the removal of the exponential factor $\exp(-\sum_i |z_i|^2/4l_B^2$), it has to be projected to the lowest Landau level. Particle-hole conjugation is somewhat trickier. We illustrate it with the transition from the Pfaffian to anti-Pfaffian order.

The particle-hole conjugate wave function~\cite{APf_Lee2007}
\begin{align}
\label{dima-4}
\nonumber
&\Psi_{\rm aP}(\{z_i\})
\\ 
=&\int\{d^2\xi_i\}\bar\Psi_{\rm Pf}(\{\xi_i\})
\prod_{i<j}(\xi_i-\xi_j)
\prod_{i<j}(z_i-z_j)
\prod_{i,j}(\xi_i-z_j)
\nonumber\\
&\times\exp(-\sum_i |\xi_i|^2/4l_B^2)\exp(-\sum_i |z_i|^2/4l_B^2),
\end{align}
where the Vandermonde factor $V=\Pi_{i<j}(\xi_i-\xi_j)\Pi_{i<j}(z_i-z_j)
\Pi_{i,j}(\xi_i-z_j)$ expresses the wave function of a filled Landau level. 
Since the filling factor is $1/2$, the numbers  of the $z_i$ and $\xi_i$ variables are the same.
We rewrite the Vandermonde factor as
\begin{align}
\label{dima-5}
V
&=\prod_{i<j}(\xi_i-\xi_j)^2\prod_{i<j}(z_i-z_j)^2
\times\frac{\displaystyle{\prod_{i,j}(\xi_i-z_j)}}
{\displaystyle{\prod_{i<j}(\xi_i-\xi_j)}\prod_{i<j}(z_i-z_j)}\nonumber\\
&=\prod_{i<j}(\xi_i-\xi_j)^2\prod_{i<j}(z_i-z_j)^2
|C|^2~\frac{\displaystyle{\prod_{i<j}(\bar\xi_i-\bar\xi_j)}
\prod_{i<j}(\bar z_i-\bar z_j)}
{\displaystyle{\prod_{i,j}(\bar \xi_i-\bar z_j)}},
\end{align}
where $C=\frac{\prod_{i,j}(\xi_i-z_j)}{\prod_{i<j}(\xi_i-\xi_j)\prod_{i<j}(z_i-z_j)}$. Thus,
\begin{align}
\label{dima-6}
\nonumber
\Psi_{\rm aP}
=&\int\{d^2\xi_i\}|R|^2\left[{\rm Pf}\{\frac{1}{\bar\xi_i-\bar\xi_j}\}\right]
\\
&\times\left[\frac{\prod_{i<j}(\bar\xi_i-\bar\xi_j)\prod_{i<j}(\bar z_i-\bar z_j)}
{\prod_{i,j}(\bar \xi_i-\bar z_j)}\right]
\left[\prod_{i<j}(z_i-z_j)^2\right],
\end{align}
where the real factor $|R^2|=|C|^2\Pi_{i<j} |\xi_i-\xi_j|^4 \exp(-\sum_i (2|\xi_i|^2+|z_i|^2)/4l_B^2)$ ensures convergence and is not expected to influence topological properties %\cite
of the wave function. Each term in the square brackets in Eq.~\eqref{dima-6} can be understood as a correlation function of a conformal field theory. The Pfaffian term in the first square brackets is a correlator of the antiholomorphic Majorana fermions $\psi$. The quadratic term in the third square brackets is a correlation function of the Bose fields $\exp(2i\phi)$ in the theory with 
$\mathcal{L}=\frac{2}{4\pi}[i\partial_y\phi\partial_x\phi+(\partial_x\phi)^2]$. The middle term is the correlation function 
$\langle\Pi\exp(i\theta(\xi_m))\Pi\exp(-i\theta(z_n))\rangle$ in the antiholomorphic theory with
$\mathcal{L}=\frac{1}{4\pi}[-i\partial_y\theta\partial_x\theta+(\partial_x\theta)^2]$. Thus, the topological properties of the wave function are encoded in the correlator
\begin{align} \label{dima-7}
\nonumber
G_{\rm aP}
=\Big\langle
&\prod_m \psi(\xi_m)\exp[i\theta(\xi_m)]
\\
&\times\prod_n\exp[-i\theta(z_n)]\exp[2i\phi(z_n)]\Big\rangle.
\end{align}
The insertion of a quasiparticle operator into the above correlation function must yield a single-valued function of $\xi_m$ and $z_n$. Consider an operator $\hat q=\sigma\exp(i\omega_1\theta)\exp(i\omega_2\phi)$. Single-valuedness with respect to $\xi$ fixes $\omega_1=1/2+n_1$. Hence, single-valuedness with respect to $z$ implies $\omega_2=1/2+n_2$. This, of course, agrees with the standard prescription for quasiparticles (cf. Sec. III).

Equations~\eqref{dima-4}-\eqref{dima-7} set a template for particle-hole conjugation in our construction. The key step is the transformation~\eqref{dima-5} which allows expressing the wave function via a correlator of the type~\eqref{dima-7}.

One can now repeat neutral-mode flipping and particle-hole conjugation  in turn a desired number of times to generate wave functions for the eight non-Abelian orders. In such iterative procedure, the projection to the lowest Landau level should only be performed once on the last step, if the wave-function does not become holomorphic after the removal \cite{footnote-dima-1} of 
$\exp(-\sum_i |z_i|^2/4l_B^2$). For example, an SU(2)$_2$ wave function is produced from $\Psi_{\rm aP}$ by neutral-mode flipping; an anti-SU(2)$_2$ wave function is produced by additional particle-hole conjugation; and so on.

The approach to the Abelian orders is the same. Thus, all that is left to do is to specify the wave function for the mother state. The $K=8$ state is best understood as a quantum Hall state of bosons~\cite{Wen_Zee}. For that reason, we use the $113$ state as the mother topological order. A convenient wave function for that order can be found in the Supplemental Material for Ref.~\cite{Guang2014}:
\begin{align}
\label{dima-8}
\nonumber
\Psi_{113}
=\int\prod\{d^2\xi_i\}&[~\prod_{i<j}(\bar\xi_i-\bar\xi_j)^4\prod_{i,j}(\xi_i-z_j)^2
\\
&\times\prod_{i<j}(z_i-z_j)
~|R_{113}|^2~],
\end{align}
where the real factor $|R_{113}|^2$ ensures convergence. This wave function corresponds to the hierarchical construction with the $K$ matrix 
$$
K=\left (
\begin{array}{cc}
1 & 2 \\
2 & -4
\end{array}
\right)
$$
We rewrite the complex factor in Eq.~\eqref{dima-8} as
\begin{align} \label{dima-9}
\nonumber
&\prod_{i<j}(\bar\xi_i-\bar\xi_j)^4\prod_{i,j}(\xi_i-z_j)^2\prod_{i<j} (z_i-z_j)
\\ \nonumber
=~&\frac{\prod_{i<j}(\bar\xi_i-\bar\xi_j)^4\prod_{i<j} (\bar z_i-\bar z_j)}
{\prod_{i,j}(\bar\xi_i-\bar z_j)^2}
\prod_{i<j} (z_i-z_j)^2
\\
&\times\prod_{i,j}|\xi_i-z_j|^4 \prod_{i<j}|z_i-z_j|^{-2}.
\end{align}
The real factor in the third line is not expected to affect topological properties. The second line can be understood as the correlation function 
$\langle\Pi \exp[2i\theta(\xi_m)]\Pi\exp[-i\theta(z_n)]\exp[2i\phi(z_n)]\rangle$ in the theory with the Lagrangian density
\begin{equation}
\mathcal{L}
=\frac{2}{4\pi}[i\partial_y\phi\partial_x\phi+(\partial_x\phi)^2]+\frac{1}{4\pi}[-i\partial_y\theta\partial_x\theta+(\partial_x\theta)^2].
\end{equation}
Quasiparticle operators $\hat q=\exp(i\omega_1\theta)\exp(i\omega_2\phi)$ must have single-valued OPE with $\exp(2i\theta)$ and $\exp(-i\theta)\exp(2i\phi)$. The first condition allows
$\omega_1=\pm 1/2$. The second condition then fixes $\omega_2=1/2+n$. This corresponds to the quasiparticle charge $e/4+ne/2$. Again, the results agree with the standard prescription.

The 331 order can next be obtained with neutral-mode flipping along the same lines as in the non-Abelian case; the anti-331 order can be obtained from the 331 wave function with particle-hole conjugation; and so on.

\subsection{Quasiparticle operators}

Based on the diagonal $K$ matrix in Eq.~\eqref{eq:diagonalized_K} and the $t$ vector, or, alternatively, on the edge theories of the preceding subsection, it is straightforward to determine the scaling dimensions for different types of quasiparticles. Here, we separately discuss the non-Abelian orders and the Abelian orders.

A note about notations. In the previous subsection, we did not explicitly consider edge theories with more than two Bose modes, and we used the notations $\theta$ and $\phi$ for the modes. In
this subsection, we will need multiple edge modes. We will denote all Bose fields as $\phi_{\rm index}$, where ${\rm index}=\rho$ for the charged mode, and ${\rm index= }~n_i$ or simply $i$ for a neutral mode, where $i=1,\dots,N$ numbers the Bose neutral modes. To label topological orders, we will use the Chern number $\nu_C=(2N+m_\psi)C$, where $m_{\psi}=0$ for the Abelian orders, $m_\psi=1$ for the non-Abelian orders, and $C=+1/-1$ for the orders with downstream/upstream neutral modes.

\subsubsection{Non-Abelian orders}

The simplest edge theory for the order with the Chern number $\nu_C$ has the Lagrangian density
\begin{align}
\label{eq:dima-action-non-Abelian}
\mathcal{L}_{\nu_C}
=&-\frac{2}{4\pi}\partial_x\phi_\rho(\partial_t+v_\rho\partial_x)\phi_\rho
\nonumber\\
&+i\psi(\partial_t+v_n{\rm sign}[\nu_C]\partial_x)\psi
\nonumber\\
&-\frac{1}{4\pi}\sum_{i=1}^N
\partial_x\phi_i({\rm sign}[\nu_C]\partial_t+v_n\partial_x)\phi_i.
\end{align}

For non-Abelian orders, the most relevant operator for electron takes the following form:
\begin{eqnarray}
\label{eq:dima-electrons}
\Psi_e=e^{2i\phi_\rho}e^{\pm i\phi_{n_j}}
~\text{or}~
\Psi_e=\psi e^{2i\phi_\rho}.
\end{eqnarray}
Here, the subscript $j$ runs from 1 to $N$, where $N$ is the number of neutral bosonic modes on the edge. For $e/2$ and $e/4$ quasiparticles, the operators are determined by requiring them to be local with all possible electronic operators. Hence, the most relevant operators for such quasiparticles are
\begin{eqnarray}
\Psi_{e/2}=e^{i\phi_\rho},\quad
\Psi_{e/4}=\sigma e^{i\phi_\rho/2}\prod_{i=1}^N e^{\pm i\phi_{n_j}/2}.
\end{eqnarray}
The twist field $\sigma$ has the conformal dimension \cite{foot-disorder} $h_\sigma=1/16$ and satisfies the fusion rule 
$\sigma\times\sigma=\psi+I$. Therefore, we determine the scaling dimensions for each type of quasiparticles \cite{Overbosch2008, Guang2013} as
\begin{eqnarray} \label{eq:g_5/2_nA}
\Delta_e=3/2,\quad
\Delta_{e/2}=1/4,\quad
\Delta_{e/4}=(N+1)/8.
\end{eqnarray}
From Eq.~\eqref{eq:g_5/2_nA}, it is noticed that the $e/4$ quasiparticle operator is the most relevant among all above operators at $N<1$ (PH-Pfaffian and Pfaffian orders). For $N=1$, the $e/4$ and $e/2$ quasiparticles are equally relevant  [anti-Pfaffian and SU(2)$_2$ orders]. For $N>1$, the $e/2$ quasiparticle becomes the most relevant [example: anti-SU(2)$_2$ order]. 

Note that different electron operators (\ref{eq:dima-electrons}) do not anticommute. This can be fixed by introducing Klein factors. We will not explicitly include them in the equations below since they are of little importance to our calculations.

%Both the analysis and the values obtained from equation~\eqref{eq:g_5/2_nA} agree with previous results published by different authors~\cite{Overbosch2008, Guang2013}. The above %universal exponents can be measured through tunneling experiment. Specifically, the zero-bias conductance scale with teperature $T$ as $G\sim T^{2g-2}$~\cite{Wen_book}.

\subsubsection{Abelian orders}

The simplest edge theory for the order with the Chern number $\nu_C$ has the Lagrangian density
\begin{align}
\label{eq:dima-action-Abelian}
\mathcal{L}_{\nu_C}
=&-\frac{2}{4\pi}\partial_x\phi_\rho(\partial_t+v_\rho\partial_x)\phi_\rho
\nonumber\\
&-\frac{1}{4\pi}\sum_{i=1}^N\partial_x\phi_i({\rm sign}[\nu_C]\partial_t+v_n\partial_x)\phi_i.
\end{align}

For Abelian orders, the Ising anyonic sector is absent. Therefore, the most relevant electronic operator (except in the $K=8$ state) takes the following form:
\begin{eqnarray}
\Psi_e=e^{2i\phi_\rho}e^{\pm i\phi_{n_j}}.
\end{eqnarray}
For $e/2$ and $e/4$ quasiparticles, the most relevant operators are now given by
\begin{eqnarray} \label{eq:qp_Abelian}
\Psi_{e/2}=e^{i\phi_\rho},\quad
\Psi_{e/4}=e^{i\phi_\rho/2}\prod_{i=1}^N e^{\pm i\phi_{n_j}/2}.
\end{eqnarray}
As a result, the scaling dimensions for each type of quasiparticles are determined \cite{Overbosch2008, Guang2013} as
\begin{eqnarray} \label{eq:g_5/2_A}
\Delta_e=3/2,\quad
\Delta_{e/2}=1/4,\quad
\Delta_{e/4}=N/8+1/16.
\end{eqnarray}
From Eq.~\eqref{eq:g_5/2_A}, we conclude that the $e/4$ quasiparticle is the most relevant for topological orders with $N\leq 1$ ($K=8$, $113$ and $331$ orders). For $N>1$, the $e/2$ quasiparticle is the most relevant (example: anti-331 order). Electrons are gapped in the $K=8$ state. The charge $q=ne/4$ excitation is described by $\Psi_q=\exp(i n\phi_\rho)$.

%----------------------- 16-fold way ------------------------------

\section{Fractional Statistics}
\label{sec:sixteen_fold}

After generating different topological orders for the $\nu=5/2$ FQH state in the previous section, we would like to check that all of them are connected 
with Kitaev's sixteenfold way~\cite{Kitaev}. Our present goal is twofold. First, we would like to describe quasiparticle statistics for all orders from the previous section. This is needed for the analysis of experimental probes. In the process, we achieve the second goal: explicitly observe a connection of all orders with Kitaev's classification.

In Kitaev's original proposal, all particles are neutral. Hence, we need to separate the neutral and charged sectors of the theory. 
Thanks to a simple form of the $K$ matrix in Eq.~\eqref{eq:diagonalized_K}, this task is not hard. 
The only charged field is $\phi_\rho$. As far as the contributions of the neutral fields to quasiparticle operators are concerned, there are three different classes of particles: the vacuum $(I)$ class of the particles whose operators contain only $\phi_\rho$, the fermion $(\varepsilon)$ class in which the charged part is multiplied by an operator with Fermi statistics, and the vortex $(\sigma)$ class. 
The products of charged fields and neutral Bose-operators are included in the $(I)$ class.
Every quasiparticle operator is a product of some exponent of the form $\exp(is\phi_\rho)$, and a ``neutralized" part. We identify the neutralized $e/4$ quasiparticles and the neutralized electron as the vortex and the fermion, respectively. Following Ref.~\cite{Kitaev}, the Chern number $\nu_C$ is defined as the net number of the Majorana fermions moving downstream. Since each neutral bosonic mode can be fermionized and split into two Majorana fermions, each downstream Bose mode contributes $2$ to the Chern number $\nu_C$, and each upstream Bose mode contributes $-2$.

\subsection{Sixteenfold way for Abelian topological orders}
\label{sec:16-fold_Abelian}

%To explicitly verify that all topological orders of Section II  can be identified with the orders of the 16-fold way, 
We introduce operators of neutral fermions
\begin{eqnarray} \label{eq:16-fold_fermion}
\varepsilon
=\prod_{i=1}^N
e^{i n_i\phi_i},
\end{eqnarray}
where we label neutral Bose fields as $\phi_i$.
Physically, these operators are the neutral parts of various electron operators.  Furthermore, $n_i$ is a set of integers which satisfies 
\begin{eqnarray}
\sum_{i=1}^N n_i \equiv 1~(\text{mod } 2).
\end{eqnarray}
We identify vortices $\sigma$ as the neutral parts of the $e/4$ quasiparticle operators: 
\begin{eqnarray} \label{eq:16-fold_vortex}
\sigma
=\prod_{i=1}^N
e^{i\left(\frac{1}{2}+l_i\right)\phi_i}.
\end{eqnarray}
Two vortices are said to be of the same type if they differ by an even number of fermions (equivalently, a boson). Otherwise, they are different types of vortices.

\subsubsection{Topological spin}

We start with computing the topological spin of the fermion and the vortex separately. Following the convention in Ref.~\cite{Kitaev}, we define the topological spin of a particle $a$ as
\begin{eqnarray} \label{eq:topological_spin}
\vartheta_a = e^{2\pi i (h_a-\bar{h}_a)}.
\end{eqnarray}
The symbols $h_a$ and $\bar{h}_a$ denote the holomorphic and anti-holomorphic conformal dimensions of the operator for the particle, respectively. Physically, the topological spin is directly related with the phase $\theta_a$ induced from exchanging two identical particles  as
\begin{eqnarray}  \label{eq:exchange_phase}
e^{i\theta_a}
=\vartheta _a.
\end{eqnarray}
Consider first the case of a positive Chern number $\nu_C$.
Since the $K$ matrix has been diagonalized in Sec.~\ref{sec:5/2}, the conformal dimension of a holomorphic vertex operator 
$V=e^{i\sum_i\alpha_i\phi_i}$ is 
\begin{eqnarray}
h
=\sum_i(K_{i+1~i+1})^{-1}(\frac{\alpha_i^2}{2})
=\sum_i \frac{\alpha_i^2}{2}.
\end{eqnarray}
Here $K_{i+1~i+1}=1$ is the diagonal matrix element, corresponding to the $i$-th neutral mode $\phi_i$. 
The same exactly scaling dimension $\bar h$ is obtained as a function of $\{\alpha_i\}$ for an anti-holomorphic vertex operator $V=e^{i\sum_i\alpha_i\phi_i}$ in a theory with a negative Chern number.

Based on the definition in Eq.~\eqref{eq:topological_spin}, the topological spin of $\varepsilon$ in Eq.~\eqref{eq:16-fold_fermion} is evaluated as
\begin{align} \label{eq:spin_fermion}
\nonumber
\vartheta_{\varepsilon}
&=\exp{\left[i~{\rm sgn}{(\nu_C)} \pi\sum_{i=1}^N n_i^2\right]}
\\
&=\exp{\left[i~{\rm sgn}{(\nu_C)} \pi \sum_{i=1}^N n_i\right]}
=-1.
\end{align}
 For $\sigma$, the topological spin is determined as
\begin{align} \label{eq:spin_vortex}
\nonumber
\vartheta_{\sigma}
=&~\exp{\left[i\pi~\text{sgn}(\nu_C) \sum_{i=1}^N \left(\frac{1}{2}+l_i\right)^2\right]}
\\ \nonumber
=&~e^{i\pi~\text{sgn}(\nu_C)N/4}
\exp{\left[i\pi~\text{sgn}(\nu_C) \sum_{i=1}^N l_i(l_i+1)\right]}
\\
=&~e^{i\pi\nu_C/8},
\end{align}
since $l_i(l_i+1)$ is even for any integer $l_i$.
Therefore, both $\vartheta_\varepsilon$ and $\vartheta_\sigma$ agree with the results by Kitaev~\cite{Kitaev}.

\subsubsection{Fusion rules}

Kitaev's fusion rules between a vortex and a fermion are satisfied automatically due to our definition of the two different types of vortices. Now, we show that the fusion rules for vortices can be grouped into two different cases. The result of fusing two vortices is given by
\begin{eqnarray} 
\sigma_1 \times \sigma_2
\sim\prod_{i=1}^N
e^{i\left(1+l_i+m_i\right)\phi_i}.
\end{eqnarray}

To determine the nature of the resulting particle, we evaluate its topological spin. From Eq.~\eqref{eq:topological_spin}, we have
\begin{align}
\nonumber
\vartheta
=&\exp{\left[i\pi \sum_{i=1}^N (1+l_i+m_i)^2 \right]}
\\
=&\exp{\left[i\pi \sum_{i=1}^N (1+l_i+m_i )\right]}.
\end{align}
If $\sigma_1$ differs from $\sigma_2$ by a boson, then 
$\sum_{i=1}^N (l_i+m_i)$ is an even integer. Hence, $\vartheta =e^{iN\pi}$. For odd $N$,  
$\vartheta=-1$ which indicates that $\sigma_1$ and $\sigma_2$ fuse to a fermion. On the other hand, $\vartheta=1$ when $N$ is even. Hence, the two vortices fuse to a boson. In summary, we have
\begin{align} \label{eq:fuse_same}
\nonumber
&\sigma\times\sigma
=\varepsilon
\quad\text{(when $N$ is odd)},
\\
&\sigma\times\sigma 
=I
\quad\text{(when $N$ is even)}.
\end{align}

When $\sigma_1$ and $\sigma_2$ are two different types of vortices, then 
$\sum_{i=1}^N (l_i+m_i)$ becomes an odd integer. In this case, we have the following fusion rules:
\begin{align} \label{eq:fuse_different}
\nonumber
&\sigma_1\times\sigma_2
=I
\quad\text{(when $N$ is odd)},
\\
&\sigma_1\times\sigma_2
=\varepsilon
\quad\text{(when $N$ is even)}.
\end{align} 
For the Abelian topological orders proposed in Sec.~\ref{sec:5/2}, all neutral modes have the same chirality. Hence, the Chern number satisfies $|\nu_C|=2N$. The cases of odd $N$ and even $N$ correspond to $\nu_C\equiv 2~\text{(mod 4)}$ and $\nu_C\equiv 0~\text{(mod 4)}$, respectively. One can easily check that the fusion rules in Eqs.~\eqref{eq:fuse_same} and~\eqref{eq:fuse_different} agree with Kitaev's results.

\subsubsection{Braiding rules}

The phase accumulated from exchanging two identical particles can be determined from Eqs.~\eqref{eq:exchange_phase},~\eqref{eq:spin_fermion}, and~\eqref{eq:spin_vortex}. In our discussion of interferometry, a slightly different phase is essential. We define $\phi^{ab}_c$ as the phase, accumulated by a particle of type $a$ making a full counterclockwise circle
\cite{foot-counterclockwise}
about a particle of type $b$. The two particles are in the fusion channel $c$. At $a=b$ one gets
\begin{align}
\label{eq:phase_encircle_fermion}
&\phi^{\varepsilon\varepsilon}_I
=2\theta_\varepsilon
=0,
\\
\label{eq:phase_encircle_vortex}
&\phi^{\sigma_1\sigma_1}_c
=\phi^{\sigma_2\sigma_2}_c
=2\theta_\sigma
=\frac{\pi\nu_C}{4}~(\text{mod }2\pi).
\end{align}

For non-identical particles, the exchange phase is not uniquely defined. For this reason, at $a\ne b$, we only consider the encircling phases $\phi^{ab}_c$. Let the neutral parts of the particles be described by the vertex operators $V_a=e^{i\sum_i l_i\phi_i}$ and $V_b=e^{i\sum_j m_j\phi_j}$. Since the $K$-matrix is diagonal, the correlation function for the two particles (and a distant additional vertex to ensure a nonzero answer) in the edge CFT is
\begin{eqnarray}
\langle V_a(z)V_b(w)\rangle
=(z-w)^{\displaystyle{\sum_{i=1}^N \frac{l_i m _i}{K_{i+1~i+1}}}}.
\end{eqnarray} 
Thus,
\begin{eqnarray}
\phi^{ab}_c
=2\pi \sum_{i=1}^N \frac{l_i m _i}{K_{i+1~i+1}}~(\text{mod } 2\pi).
\end{eqnarray}

\subsubsection*{3.1 Moving a fermion around a vortex}

Consider encircling a vortex by a fermion. This process induces the phase:
\begin{align} \label{eq:phase_vortex_fermion}
\nonumber
\phi^{\sigma\varepsilon}_c
&=2\pi~\text{sgn}(\nu_C)\sum_{i=1}^N n_i\left(\frac{1}{2}+l_i\right)
~(\text{mod } 2\pi)
\\ \nonumber
&=\pi~\text{sgn}(\nu_C)
\sum_{i=1}^N n_i
~(\text{mod } 2\pi)
\\
&=\pi
\end{align}
In the last step, we used the fact that $\sum_i n_i$ is odd since $\varepsilon$ is a fermion. Compare this with the rules from Tables 2 and 3 in Ref.~\cite{Kitaev}, which are summarized as follows:
\begin{align}
\nonumber
&\nu_C\equiv 0, 8~(\text{mod } 16):
R^{e\epsilon}_m=R^{\epsilon m}_e=1,~
R^{m\epsilon}_e=R^{\epsilon e}_m=-1
\\ \nonumber
&\nu_C\equiv \pm 4~(\text{mod } 16):
R^{e\epsilon}_m=R^{\epsilon e}_m
=R^{\epsilon m}_e=R^{m\epsilon}_e
=e^{i\pi\nu_C/8}
\\
&\nu_C\equiv \pm 2~(\text{mod } 4):
R^{a\epsilon}_{\bar{a}}=R^{\epsilon a}_{\bar{a}}
=R^{\bar{a}\epsilon}_a=R^{\epsilon\bar{a}}_a
=e^{-i\pi\nu_C/4}
\end{align}
For all three cases, the phase factor accumulated by a fermion on a complete circle around $\sigma$ equals
\begin{eqnarray}
R^{\sigma\epsilon}R^{\epsilon \sigma}=-1.
\end{eqnarray}
This is consisent with the $\pi$ phase ~\eqref{eq:phase_vortex_fermion}.

\subsubsection*{3.2 Moving vortices}

Since the topological spin for the vortex agrees with Ref.~\cite{Kitaev}, the phase (\ref{eq:spin_vortex}) induced from exchanging two identical vortices must be consistent with the braiding rules from
Ref.~\cite{Kitaev}. Furthermore, the same phase is induced if one of the vortices differs from the other by a boson. 

When the difference between the vortices is a fermion, the phase induced by moving one of them around the other is
\begin{align} \label{eq:phase_diff_vortices}
\nonumber
\phi^{\sigma_1\sigma_2}_c
&=2\pi~\text{sgn}(\nu_C)
\sum_{i=1}^N 
\left(\frac{1}{2}+l_i\right)\left(\frac{1}{2}+m_i\right)
~(\text{mod } 2\pi)
\\
&=\left(\frac{\pi\nu_C}{4}+\pi\right)~(\text{mod } 2\pi).
\end{align}
The corresponding braiding rules in Ref.~\cite{Kitaev} are
\begin{align}
\nonumber
&\nu_C\equiv 0, 8~(\text{mod } 16):
R^{em}_\epsilon=e^{i\pi\nu_C/8},~
R^{me}_\epsilon=-e^{i\pi\nu_C/8}
\\ \nonumber
&\nu_C\equiv \pm 4~(\text{mod } 16):
R^{em}_\epsilon=R^{me}_\epsilon=1
\\
&\nu_C\equiv \pm 2~(\text{mod } 4):
R^{a\bar{a}}_\epsilon=R^{\bar{a} a}_\epsilon
=e^{-i\pi\nu_C/8}
\end{align}
For all three cases, the phase, accumulated on a full circle, is
\begin{align}
R^{\sigma_1 \sigma_2}_c R^{\sigma_2 \sigma_1}_c
=e^{i(\pi+\pi\nu_C/4)},
\end{align}
which agrees with the phase from Eq.~\eqref{eq:phase_diff_vortices}. Thus, we have verified that all topological spins, fusion rules, and phases are consistent with Ref.~\cite{Kitaev}. Furthermore, the results in Eqs.~\eqref{eq:spin_fermion}, \eqref{eq:spin_vortex}, \eqref{eq:phase_vortex_fermion}, and \eqref{eq:phase_diff_vortices} are invariant under the change of $\nu_C\rightarrow \nu_C \pm 16$. Therefore, we conclude that the Abelian topological orders in Fig.~\ref{fig:Abelian} agree with Kitaev's sixteenfold way.

\subsection{Sixteenfold way for non-Abelian topological orders}

The non-Abelian topological orders introduced in Sec.~\ref{sec:5/2} can be viewed as direct products between an Ising conformal field theory and an Abelian U(1) sector. The Abelian sector is still described by the $K$ matrix in Eq.~\eqref{eq:diagonalized_K}. Here, we examine the extra contribution from the Ising CFT. To prevent confusion with the vortex $\sigma$, we change the notation for the spin field in the $e/4$ quasiparticle operator to $\chi$. As a reminder, the fusion rules for $\chi$ are $\chi\times\psi=\chi$ and $\chi\times\chi=\psi+I$~\cite{Das-Sarma}. Here, $\psi$ is the Majorana field with  the conformal dimension $1/2$. The phase induced from exchanging two $\chi$ is given by
\begin{eqnarray} \label{eq:braiding_twist}
\theta^{\chi\chi}_I
=-\frac{\pi}{8}\text{sgn}(\nu_C),~
\theta^{\chi\chi}_\psi 
=\frac{3\pi}{8}\text{sgn}(\nu_C).
\end{eqnarray}
Here, $\nu_C$ is the Chern number of the non-Abelian topological order which differs from Chern number of the associated U(1) Abelian sector by $\pm 1$.

\subsubsection{Topological spin}

The neutral fermion $\varepsilon$ is identified as
\begin{eqnarray} \label{eq:fermion_NA}
\varepsilon
=\prod_{i=1}^N
e^{i n_i\phi_i}
\quad\text{or}\quad
\varepsilon
=\psi\prod_{i=1}^N
e^{i m_i\phi_i}.
\end{eqnarray}
Here $\sum n_i$ is odd, whereas $\sum m_i$ is even. From this definition, we automatically have $\theta_\varepsilon=-1$. The non-Abelian vortex is
\begin{eqnarray} \label{eq:vortex_NA}
\sigma
=\chi\prod_{i=1}^N
e^{i\left(\frac{1}{2}+l_i\right)\phi_i}.
\end{eqnarray}
Based on the above definition, one can easily verify that $\sigma\times\sigma=\varepsilon+I$ and 
$\sigma\times\varepsilon=\sigma$. These fusion rules are consistent with Table 1 in Ref~\cite{Kitaev}. Although $N$ only counts the modes of the U(1) Abelian sector, $\vartheta_\sigma=e^{i\pi\nu_C/8}$ is still satisfied since the conformal dimension of $\chi$ is $1/16$. This contributes an additional factor of $e^{i\pi\text{sgn}(\nu_C)/8}$ to the topological spin. 

\subsubsection{Braiding rules}

For the fermion in Eq.~\eqref{eq:fermion_NA} and the vortex in Eq.~\eqref{eq:vortex_NA}, the phase induced by moving one of them around the other is
\begin{align}
\nonumber
\phi^{\sigma\varepsilon}_{\sigma}
&=2\pi~\text{sgn}(\nu_C)
\sum_{i=1}^N n_i (\frac{1}{2}+l_i)~(\text{mod }2\pi)
\\
\text{or}\quad
\phi^{\sigma\varepsilon}_{\sigma}
&=\text{sgn}(\nu_C)
\left[2\pi\sum_{i=1}^N m_i (\frac{1}{2}+l_i)-\pi\right]
~(\text{mod }2\pi)
\end{align}
In the second case, the additional $\pi$ phase comes from moving $\chi$ around $\psi$. In both cases, the results reduce to
\begin{eqnarray}
\phi^{\sigma\varepsilon}_{\sigma}=\pi.
\end{eqnarray}

Finally, the phase accumulated by exchanging a pair of non-Abelian vortices can be decomposed into two parts:
\begin{eqnarray}
\theta^{\sigma\sigma}_{\beta}
=\theta^{\chi\chi}_{\beta_1}
+\theta^{\sigma_A\sigma_A}_{\beta_2}.
\end{eqnarray}
In the above equation, $\sigma_A$ represents the Abelian vortices obtained from 
$\sigma$ by detaching $\chi$. Also, $\beta_1$ and $\beta_2$ should fuse into $\beta$, where $\beta$, $\beta_1$, and $\beta_2$ can be either $I$ or $\varepsilon=\psi$. 
Let us first assume that the two Abelian vortices $\sigma_A$ are described by identical operators. It is then meaningful to ask about the phase, 
accumulated when their positions are exchanged.

We start with the scenario of $\nu_C>0$. If $\nu_C\equiv 1~(\text{mod }4)$, then the two possible triplets for 
$(\beta,\beta_1,\beta_2)$ are $(I,I,I)$ and $(\varepsilon,\psi=\varepsilon,I)$. Then one has
\begin{align}
\nonumber
\theta^{\sigma\sigma}_I
&=\theta^{\chi\chi}_I+\theta^{\sigma_A \sigma_A}_I
=-\frac{\pi}{8}+\frac{\pi}{8}(\nu_C-1)
=\frac{\pi}{8}(\nu_C-2)
\\
\theta^{\sigma\sigma}_\varepsilon
&=\theta^{\chi\chi}_\psi+\theta^{\sigma_A \sigma_A}_I
=\frac{3\pi}{8}+\frac{\pi}{8}(\nu_C-1)
=\frac{\pi}{8}(\nu_C+2)
\end{align}

When $\nu_C\equiv 3~(\text{mod }4)$, the two possible triplets for 
$(\beta,\beta_1,\beta_2)$ become $(I,\psi,\varepsilon)$ and $(\varepsilon,I,\varepsilon)$. Thus,
\begin{align}
\nonumber
\theta^{\sigma\sigma}_I
&=\theta^{\chi\chi}_\psi+\theta^{\sigma_A \sigma_A}_\varepsilon
=\frac{3\pi}{8}+\frac{\pi}{8}(\nu_C-1)
=\frac{\pi}{8}(\nu_C+2)
\\
\theta^{\sigma\sigma}_\varepsilon
&=\theta^{\chi\chi}_I+\theta^{\sigma_A \sigma_A}_\varepsilon
=-\frac{\pi}{8}+\frac{\pi}{8}(\nu_C-1)
=\frac{\pi}{8}(\nu_C-2)
\end{align}
Similarly, one can also calculate $\theta^{\sigma\sigma}_I$ and $\theta^{\sigma\sigma}_\varepsilon$ for negative $\nu_C$. For all the four cases, the results can be rewritten as
\begin{align}
\label{eq:braid_sigma_I}
\theta^{\sigma\sigma}_I
&=\frac{\pi}{8}(\nu_C^2-\nu_C-1)~(\text{mod }2\pi),
\\
\label{eq:braid_sigma_fermion}
\theta^{\sigma\sigma}_\varepsilon
&=\frac{\pi}{8}(\nu_C^2+3\nu_C-1)~(\text{mod }2\pi),
\end{align}
which agree with the braiding rules listed in Table 1 in Ref.~\cite{Kitaev}. For encircling one vortex around another vortex, one has
\begin{align}
\label{eq:phase_enicrcle_sigma_I}
\phi^{\sigma\sigma}_I
&\equiv -\frac{\pi\nu_C}{4}~(\text{mod }2\pi),
\\
\label{eq:phase_encircle_sigma_fermion}
\phi^{\sigma\sigma}_\varepsilon
&\equiv \frac{3\pi\nu_C}{4}~(\text{mod }2\pi).
\end{align}

To finish our discussion we need to address the situation in which the Abelian parts $\sigma_A$ of the two vortices differ. Since we no longer consider identical operators for the two excitations, it is only meaningful to ask about the phase, accumulated when one anyon makes a complete circle around the other. The results turn out the same as in the above equations
(\ref{eq:phase_enicrcle_sigma_I}) and (\ref{eq:phase_encircle_sigma_fermion}).

To conclude, we have demonstrated that the sixteenfold way is satisfied for all the topological orders introduced in Sec.~\ref{sec:5/2}. This important feature will be useful when we discuss interferometry in Sec.~\ref{sec:MZ_experiment}.

%----------------------- Mach-Zehnder interferometer ------------------------------

\section{Experimental signatures}
\label{sec:EXP}
\subsection{Upstream modes}

\begin{figure}[htb]
\includegraphics[width=2.5 in]{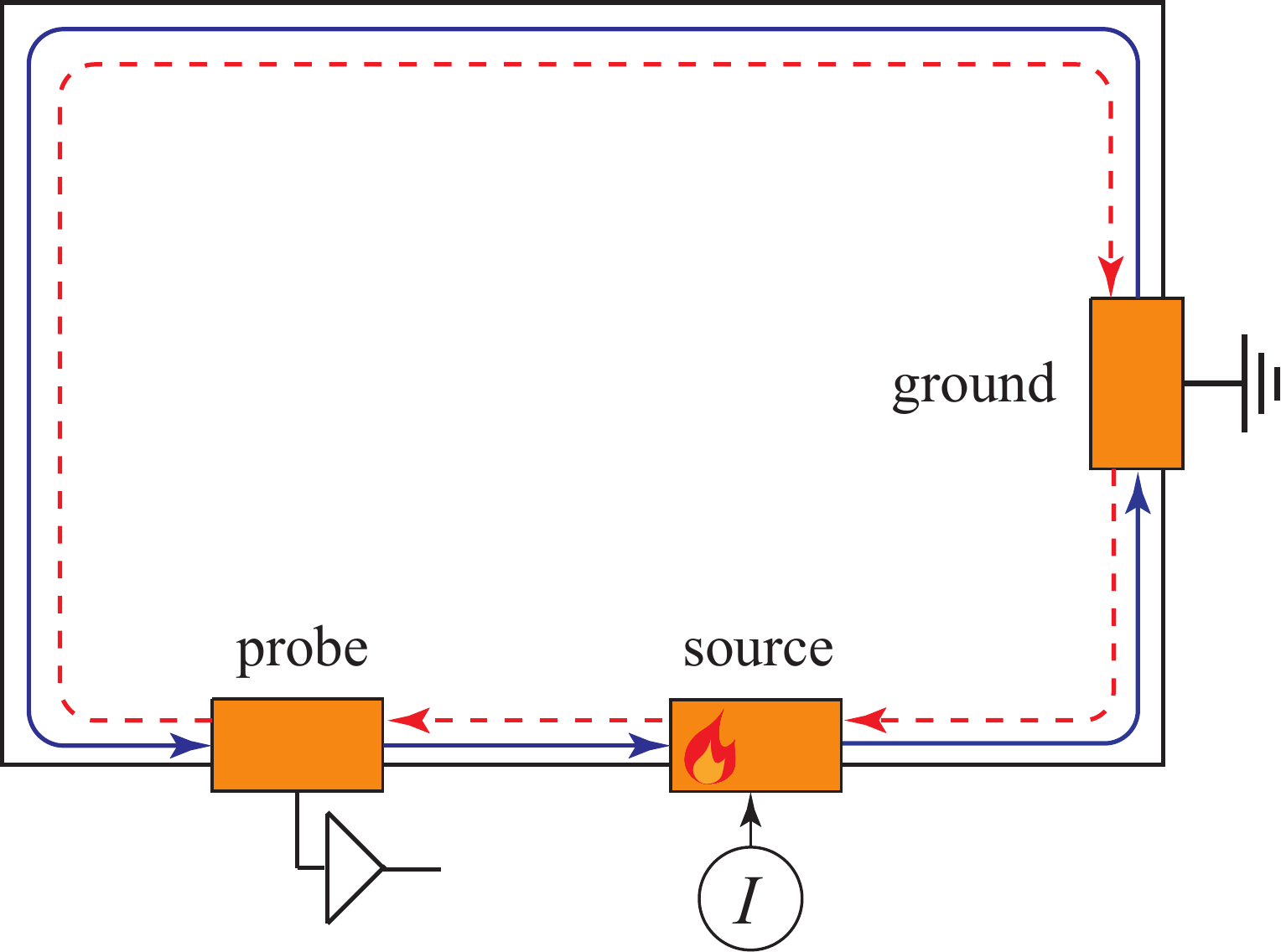}
\caption{(Color online) Dashed lines show upstream neutral modes. Solid lines show the charged mode. The neutral mode is excited at the hot spot in the source.}
\label{fig:upstream_neutral}
\end{figure}

The simplest experimental signature is the presence or absence of upstream neutral modes. It can be tested by probing upstream noise in the setup~\cite{Gross2012} of Fig. \ref{fig:upstream_neutral}. The source in Fig. \ref{fig:upstream_neutral} is maintained at a finite voltage, while the chiral charged mode enters it at zero voltage. Thus, a nonequilibrium hot spot \cite{hot-spot-1, hot-spot-2, hot-spot-3} forms at the point where the chiral charged mode enters the contact. Energy, dissipated in the hot spot, is carried by the upstream neutral mode towards the probe and heats it. This results in excess noise in the probe. Other related setups \cite{Bid2010, Dolev2011} were also proposed and used to observe upstream neutral modes.

Clearly, energy can only go upstream in the states with the negative Chern number $\nu_C$. A subtlety involves a possibility of upstream energy transport due to edge reconstruction~\cite{Overbosch2008}, if the edge is not long enough. This issue has been tackled experimentally by comparing the upstream noise at $\nu=5/2$ in GaAs with the upstream noise at $\nu=7/3$ and $\nu=8/3$~\cite{Dolev2011}. There is a topologically protected upstream mode at $\nu=8/3$ but not at $\nu=7/3$ (see Ref. \onlinecite{Ken2019} for a review of the $8/3$ and $7/3$ states in GaAs). Thus, if, in a given device, upstream noise is seen at $\nu=8/3$ but not at $\nu=7/3$, then the device can probe topologically protected upstream transport at other close filling factors.

\subsection{Thermal Hall conductance}

The thermal Hall conductance provides a complementary probe of the neutral modes. The existing thermal transport experiments cannot tell upstream modes from downstream modes \cite{Banerjee2017, Banerjee2018} since the experiments cannot determine the sign of the thermal conductance coefficient. Hence, to find the Chern number, one also has to test the presence of upstream modes. Thus, the thermal transport approach is most powerful if combined with the approach from the previous subsection. 

In this type of experiment, the Hall bar is connected with two heat reservoirs at different temperatures. One defines the thermal Hall conductance as 
$g_Q=dJ_Q/dT=\kappa T$ where $\kappa$ is the thermal conductance coefficient and $J_Q$ is the heat current. In an FQH system, the thermal energy is mainly carried by the edge modes. These edge modes are essentially one-dimensional ballistic channels. In the limit of a long propagation length, it was shown that 
$\kappa=c\kappa_0$, where $\kappa_0=\pi^2 k_B^2/(3h)$ and $c$ denotes the central charge of the topological order which is related to the net number of the 
downstream modes~\cite{Read-Green, Kane_thermal, Cappelli_thermal}. A negative $c$ corresponds to upstream heat transport.

For the $\nu=5/2$ FQH system, there are two downstream bosonic modes from the filled lowest Landau level.
 Also, an additional downstream charged bosonic mode exists for the second Landau level with 
$\nu=1/2$. 
Finally, each topological order has its unique neutral sector. In other words, the central charges for different topological orders are different. For Abelian orders, all neutral modes are bosonic. Each contributes $\pm 1$ to the central charge, depending on the propagation direction. Hence, the thermal conductance coefficient is given by
\begin{eqnarray}
\label{eq:long-edge-1}
\kappa
=(2+1\pm N)\frac{\pi^2 k_B^2}{3h}
=(3\pm N)\frac{\pi^2 k_B^2}{3h},
\end{eqnarray}
where the minus sign corresponds to upstream neutral modes.
On the other hand, a single Majorana mode exists at the edge of a non-Abelian system. The central charge of the Majorana mode is $\pm 1/2$. Therefore, one has
\begin{eqnarray}
\label{eq:long-edge-2}
\kappa
=\left[3\pm\left(N+\frac{1}{2}\right)\right]\frac{\pi^2 k_B^2}{3h}.
\end{eqnarray}
The positive (negative) sign corresponds to topological orders with downstream (upstream) neutral modes. Recently, the thermal conductance of $\kappa=2.5\kappa_0$ 
was reported in a $\nu=5/2$ FQH system in GaAs~\cite{Banerjee2018} in agreement with the predictions~\cite{Zucker2016} for the PH-Pfaffian state. 
Equations (\ref{eq:long-edge-1}) and (\ref{eq:long-edge-2}) apply to long edges in thermal equilibrium. See Refs.~\cite{Banerjee2018, Ken2019, Banerjee2017, comment2018} for a discussion of finite-size effects in some of the states.

%This experimental result suggests that the PH-Pfaffian order is the only consistent choice for the system.

\subsection{Tunneling}
\label{sec:dima-tunneling}

\begin{figure}[htb]
\includegraphics[width=2.5 in]{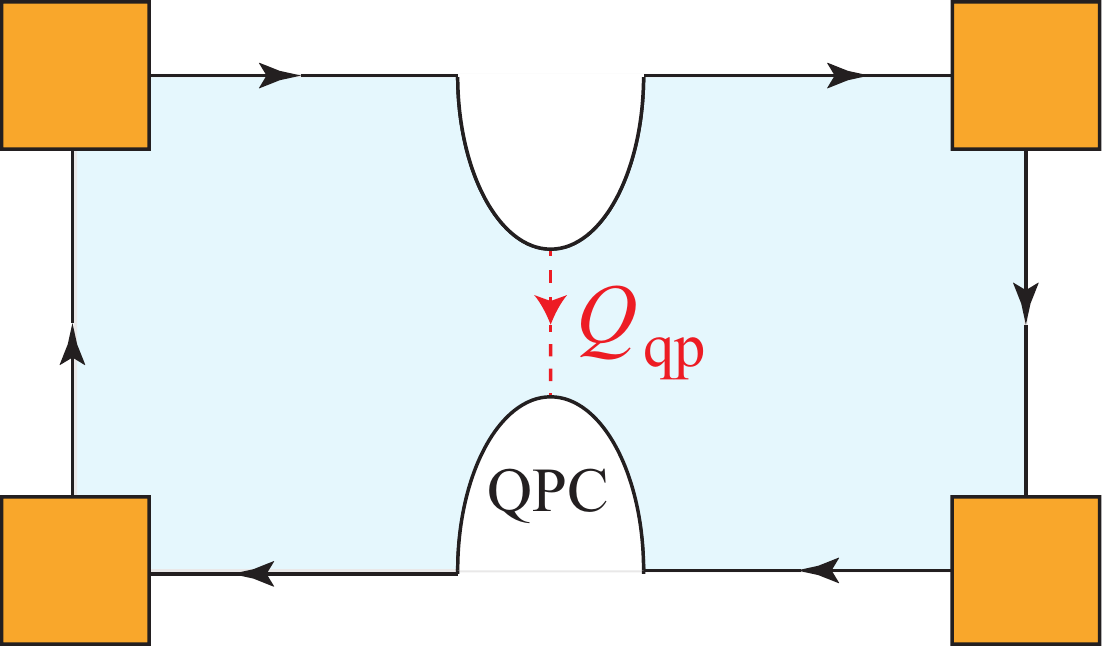}
\caption{(Color online) Quasiparticle tunneling in the quantum point contact (QPC) is shown by the dashed line.}
\label{fig:tunneling}
\end{figure}

A very different approach to probe topological order is tunneling transport \cite{Radu2008, Lin2012, Baer2014, Fu2016}. Imagine that a constriction is created in an FQH liquid (Fig. \ref{fig:tunneling}). Quasiparticles can then tunnel through the constriction. To estimate the tunneling conductance, one uses the scaling dimensions 
(\ref{eq:g_5/2_nA},\ref{eq:g_5/2_A}) of the quasiparticle operators $\Delta_q$, where $q$ stands for the quasiparticle type. At low temperatures, the linear conductance can be found from the renormalization group (RG) and is determined by the scaling dimension of the most relevant tunneling operator $\Gamma\Psi^\dagger_{qu}\Psi_{qd}$, where $\Psi_{qu,qd}$ denote the quasiparticle operators on the upper and lower edges on the two sides of the constriction QPC.  Under the action of RG, $\Gamma$ grows as $E^{2\Delta_q-1}$, where $E$ is the energy cutoff. Thus, $\Gamma_{\rm eff}(T)\sim T^{2\Delta_q-1}$ at the energy scale set by the temperature. The conductance $G\sim |\Gamma_{\rm eff}(T)|^2\sim T^{4\Delta_q-2}=T^{2g-2}$~\cite{Wen_book}. 

The tunneling exponents $g$ are listed in Table VI. The most relevant quasiparticle is the $e/2$-particle in most states, and hence $g=1/2$ for a majority of the states. Smaller values of $g$ correspond to $\nu_C=0,\pm 1,$ and $\pm 2$, that is, the $K=8$, Pfaffian, PH-Pfaffian, 331, and 113 states. We generally expect that the tunneling amplitudes $\Gamma_{1,2}$ are higher for lower-charge particles. Thus, {\it unrenormalized} tunneling amplitudes are expected to be higher for $e/4$-quasiparticles than for $e/2$-quasiparticles. The dominant low-energy tunneling process depends on the renormalized amplitudes. At $|\nu_C|<3$, the most relevant tunneling operator is that of $e/4$-particles and hence they dominate tunneling. At $\nu_C=\pm 3$, the $e/4$ and $e/2$ tunneling operators have the same scaling dimension, so it is plausible that 
the $e/4$ tunneling dominates. On the other hand, at $|\nu_C|\ge 7$, the $e/4$ tunneling operator is marginal or irrelevant. Hence, the $e/2$ tunneling is more important. The case of 
$3<|\nu_C|<7$ is subtle. Both the $e/4$ and $e/2$ tunneling are relevant in the RG sense, yet, the $e/2$ tunneling operator has  a lower scaling dimension. What sort of particles dominates depends then on the ratio of their unrenormalized tunneling amplitudes.

The above discussion tacitly assumed that the neutral modes do not interact with the charged mode. As we explain below, the results for the tunneling exponents do not depend on this assumption. This point is well known for positive Chern numbers~\cite{Wen_book}. For negative Chern numbers and in the absence of disorder, the exponents are non-universal~\cite{Wen_book}.  The PH-Pfaffian state ($\nu_C=-1$) is an exception to this rule since no RG-relevant interaction between a single upstream Majorana mode and the charged mode exists in that case~\cite{Zucker2016}. For $\nu_C<-2$, the universality of the tunneling exponents is guaranteed by disorder~\cite{APf_Lee2007, APf_Levin2007, Guang2013}. Thus, only the 113 state with $\nu_C=-2$ should show a dependence of tunneling exponents on the interaction of the upstream neutral and downstream charged modes. Even in that case, the interaction effect is weak~\cite{Guang2014} and will be neglected below.

The predicted scaling is only observed in the absence of Coulomb interaction across the constriction~\cite{Guang2013, Papa}, edge reconstruction~\cite{Rosenow-edge, Yang-edge}, and dissipation \cite{Carrega-noise}. Otherwise, one expects a nonuniversal $g$ that exceeds the ideal theoretical value. In a very clean sample, momentum-resolved tunneling \cite{Yang-momentum, Chenjie-momentum} would give detailed information about the structure of the edge.

\subsection{Fabry-P\'{e}rot interferometry}

\label{app:FP_experiment}

In order to directly probe the fractional statistics of anyons in the fractional quantum Hall system, it is necessary to braid quasiparticles and examine the consequences. An experimental technique based on a Fabry-P\'{e}rot interferometer was proposed by Chamon \textit{et al.} for Abelian states~\cite{Chamon1997}. Later, the same technique was generalized to study 
$\nu=5/2$ fractional quantum Hall systems~\cite{Das-Sarma, Stern2006, Bonderson2006} and many other FQH states~\cite{Chung2006, Bonderson2007, Bishara2009, Ilan2008, Ilan2009, Bonderson_PRB2010}. In this subsection, we review Fabry-P\'{e}rot interferometry for all states introduced in Sec.~\ref{sec:5/2}. The key feature is the topological even-odd effect
 \cite{Stern2006, Bonderson2006} which was originally predicted for the Pfaffian state, but can easily be seen to occur in all non-Abelian states. Depending on the details, it can also be mimicked by Abelian orders \cite{Stern_PRB2010}.

\begin{figure}[htb]
\includegraphics[width=3.0 in]{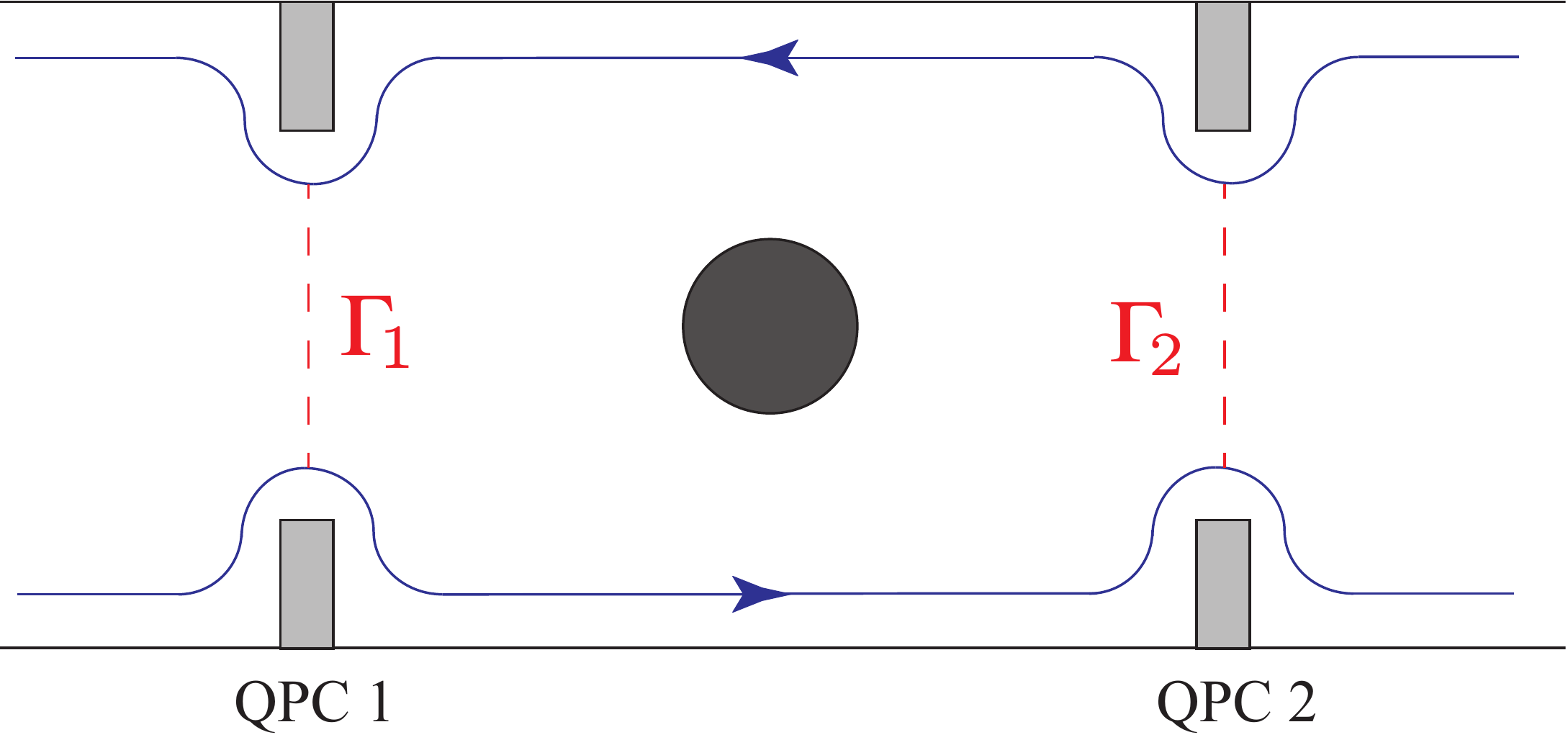}
\caption{(Color online) A Fabry-P\'{e}rot interferometer with two quantum point contacts. Quasiparticles traveling along the edge tunnel at the two contacts with the amplitudes $\Gamma_1$ and $\Gamma_2$. The number of the quasiparticles in the antidot (dark circle in the middle) is controlled by the gate voltage.}
\label{fig:Fabry_setup}
\end{figure}

In Fig.~\ref{fig:Fabry_setup}, we sketch a Fabry-P\'{e}rot interferometer with two quantum point contacts (QPC). Quasiparticles traveling along an edge can tunnel to the opposite edge at the contacts, with the corresponding tunneling amplitudes $\Gamma_1$ and $\Gamma_2$. These values are controlled by tuning the voltage on the gates that define the QPCs. In the middle of the interferometer, an antidot is created by applying a voltage to the central gate. By tuning the voltage there, the number of quasiparticles in the antidot can be adjusted. In the experiment, an interference pattern in the tunneling current due to two possible tunneling paths is measured.

In the following discussion, we will only focus on the weak-tunneling regime. We assume that both $\Gamma_1$ and $\Gamma_2$ are small, such that the backscattering current between the upper and lower edges of the interferometer is determined by the single-particle tunneling probability. To the lowest order in $\Gamma_1$ and $\Gamma_2$, the tunneling probability is given by~\cite{Das-Sarma, Bonderson2006}:
\begin{eqnarray}
\label{dima:FP1}
p=
r[\left|\Gamma_1\right|^2+\left|\Gamma_2\right|^2
+2u\left|\Gamma_1\right| \left|\Gamma_2\right|
\cos{\left(\phi_{\rm{AB}}+\phi_s+\delta\right)}],
\end{eqnarray}
where $\phi_{\rm{AB}}$ denotes the Aharonov-Bohm phase and $\phi_s$ is the statistical phase, accumulated by a quasiparticle that makes a full circle around the interferometer.  
We define $\delta=\text{arg}(\Gamma_2/\Gamma_1)$. $r=r(V,T)$ and $u=u(V,T)$ depend on microscopic details. They satisfy one important constraint. 
Indeed, the current must flow from higher voltage to lower voltage irrespective of the Aharonov-Bohm phase $\phi_{\rm{AB}}$. In other words, the current cannot change sign as a function of 
$\phi_{\rm{AB}}$. This means that the combination

\begin{equation}
\label{eq:dima-s}
s=\frac{2|u\Gamma_1\Gamma_2|}{|\Gamma_1|^2+|\Gamma_2|^2}
\end{equation}
must satisfy the inequality

\begin{equation}
\label{eq:dima-s-2}
s\le 1.
\end{equation}

%\subsection{$e/4$-quasiparticle tunneling}

\subsubsection{Tunneling operators}

Equation (\ref{dima:FP1}) tacitly assumes that only one type of quasiparticles is allowed to tunnel. This is never the case and the Hamiltonian of an interferometer assumes the form

\begin{equation}
\label{eq:dima-H-FP}
H=H_{\rm edges}+\sum_{\alpha}(\Gamma_1^\alpha T_1^\alpha+\Gamma_2^\alpha T_2^\alpha+{\rm H.c.}),
\end{equation}
where $H_{\rm edges}$  describes the edges [see Eqs. (\ref{eq:dima-action-non-Abelian}) and (\ref{eq:dima-action-Abelian})]; $\Gamma_{1,2}^\alpha$ and $T_{1,2}^\alpha$ are the tunneling amplitudes and the tunneling operators for quasiparticle type $\alpha$ at QPC1 and QPC2. The index $\alpha$ covers both electric and topological charges. 

We have argued in Sec.~\ref{sec:dima-tunneling} that quasiparticles of only one electric charge can be expected to dominate tunneling. This charge is either $e/4$ or $e/2$. 
The $e/2$ case is easy since there is only one allowed most relevant tunneling operator $T=\exp(i[\phi_\rho^u-\phi_\rho^d]/2)$, where the indices $u$ and $d$ refer to the upper and lower edges. Thus, we come back to Eq.~(\ref{dima:FP1}). The situation is more complex for $e/4$ particles, provided that $\nu_C\ne 0$.

One complication is a possibility that $e/4$ and $e/2$ particles dominate tunneling at the two different QPCs. To avoid that issue, we will assume that QPC1 and QPC2 are identical. In particular $|\Gamma_1^\alpha|=|\Gamma_2^\alpha|$. This assumption will also be used in our discussion of 
Mach-Zehnder interferometry below. Second, in all Abelian orders with $\nu_C\ne 0$,
there are two topologically distinct $e/4$ quasiparticles. Thus, two different tunneling operators must be included at each QPC. This will be of great importance in subsequent sections.

We observe that one tunneling operator is sufficient in Eq.~(\ref{eq:dima-H-FP}) for $e/2$ tunneling in all states and for $e/4$ tunneling in all non-Abelian states and in the 
$K=8$ state ($\nu_C=0$). All other Abelian orders (integer $\nu_C\ne 0$) should be described by Hamiltonians that include tunneling of two sorts of $e/4$ quasiparticles.

One more subtlety involves a possibility of several equally relevant quasiparticle operators for $e/4$ particles. For example, at $\nu_C=4$, such operators are 
$Q_+=\exp(i\phi_\rho/2)\exp(i[\phi_1+\phi_2]/2)$ and $Q_-=\exp(i\phi_\rho/2)\exp(-i[\phi_1+\phi_2]/2)$. A tunneling operator $T_i^\alpha$ can include contributions from all such quasiparticle operators, consistent with label $\alpha$. This point is of little consequence at $|\nu_C|\le 2$, but affects possible values of $s$ [Eqs. (\ref{eq:dima-s} and (\ref{eq:dima-s-2})] at $|\nu_C|>2$. Naively, any value of $0\le s\le 1$ is allowed and $s\rightarrow 1$ at $V,T\rightarrow 0$. The argument is based on the renormalization group treatment of the Hamiltonian (\ref{eq:dima-H-FP}). Indeed, the renormalization group procedure decreases the distance between any two points on each step. When the distance becomes shorter than the ultra-violet cutoff, the points can be seen as merging. Hence, if the thermal and voltage lengths $\hbar v_{\rho,n}/T$ and $\hbar v_{\rho,n}/eV$ exceed the interferometer size, the renormalization group procedure stops after the two tunneling contacts end up in the same spatial point. For identical $T_1$ and $T_2$ this implies $u(V,T)=1$. Hence, $s=1$ at $|\Gamma_1|=|\Gamma_2|$.

The above argument works, provided that the edge actions are given by equations of the type (\ref{eq:dima-action-non-Abelian}) and (\ref{eq:dima-action-Abelian}). A realistic system may well not be described by this type of an action even in the scaling limit, where all irrelevant operators can be ignored. Indeed, relevant perturbations are missing in our simplest equations for the edge actions. One such perturbation is present at any $\nu_C$. It is the random potential that couples to the charged mode: $w(x)\partial_x\phi_\rho$. Such perturbation can be eliminated from the Hamiltonian density $\frac{v_\rho}{2\pi}(\partial_x\phi_\rho)^2+w(x)\partial_x\phi_\rho$ by the variable shift $\phi_\rho\rightarrow\phi_\rho+\pi\int w(x)dx/v_\rho$.
The shift changes the relative phases of $\Gamma_1$ and $\Gamma_2$ and has no effect on the range of $s$. Similar perturbations are among various relevant perturbations that involve neutral modes. For example, the perturbation $P_n=w_n(x)\partial_x\phi_1$ is allowed. Such perturbations do not affect the range of $s$ at $|\nu_C|\le 2$. This changes at $|\nu_C|>2$.
Indeed, $P_n$ can be eliminated by a shift of $\phi_1$. This changes the relative phases of the contributions, containing $Q_1$ and $Q_2$, in the tunneling operators. As a result, $T_1$ and $T_2$ cease being identical. This undermines the argument for the possibility to reach $s=1$.

We now turn to the analysis of the current through the interferometer.
First, we consider the situation, in which the tunneling process is dominated by the $e/4$ quasiparticles.

\subsubsection{Non-Abelian topological orders}

Suppose an $e/4$ quasiparticle is sent to the interferometer as a probe particle. The braiding phase it accumulates around the antidot is given by 
%equation~\eqref{eq:phase_interferometer_e/4}. 
%Furthermore, the statistical phase $\phi_s$ is given by
\begin{eqnarray} \label{eq:phase_interferometer_e/4}
\phi_s
=\frac{n\pi}{4}+\phi^{\sigma\alpha}_{\beta}.
\end{eqnarray} 
Here, $ne/4$ is the total charge inside the interferometer (i.e., $n$ is the number of $e/4$ quasiparticles), $\alpha$ denotes the topological charge inside the interferometer,
and $\beta$ is the fusion outcome between $\sigma$ and $\alpha$. The phase $\phi^{\sigma\alpha}_\beta$  comes from the neutral degrees of freedom. 
The first term $n\pi/4$ comes from the Abelian charged sector which is the same in all 16 states. 
As a reminder, we quote the results for 
$\phi^{\sigma\alpha}_\beta$ from Sec.~\ref{sec:sixteen_fold}:
\begin{align}
\phi^{\sigma\sigma}_{I}
&\equiv -\frac{\pi\nu_C}{4}~(\text{mod }2\pi),
\\
\phi^{\sigma\sigma}_{\psi}
&\equiv\frac{3\pi\nu_C}{4}~(\text{mod }2\pi), 
\\
\phi^{\sigma\psi}_{\sigma}
&=\pi
\\
\phi^{\sigma I}_{\sigma}
&=0
\end{align}
%The phase $\phi^{\sigma\alpha}_\beta$ is resulted from encircling the neutral modes in the probe particles and the neutral modes in the quasiparticles at the antidot (with topological %charge $\alpha$).

%We denote the number of quasiparticles in the antidot as $n$.

When the number $n$ of trapped quasiparticles is odd, then
$\alpha=\sigma$. Since $\sigma\times\sigma=\psi+I$, there are two possible fusion channels for the vortices. Both channels contribute to the measured backscattering current. Moreover, the probabilities of having $\beta=\psi$ and $I$ are the same. From Eqs.~\eqref{eq:phase_enicrcle_sigma_I} and~\eqref{eq:phase_encircle_sigma_fermion}, the phase difference between the two cases is determined as 
\begin{align}
\Delta\phi
=\phi^{\sigma\sigma}_\psi-\phi^{\sigma\sigma}_I
=\frac{3\pi\nu_C}{4}-(-\frac{\pi\nu_C}{4})
\equiv\pi~(\text{mod } 2\pi)
\end{align}
Therefore, the two fusion channels correspond to the opposite values of the cosine term in the probability (\ref{dima:FP1}).  Hence, the tunneling current does not depend on the magnetic flux enclosed by the two QPCs. 

On the other hand, $\alpha$ can be either $I$ or $\psi$ when $n$ is even. Nevertheless, the antidot must be in one of the superselection states, but not in their superposition. In Sec.~\ref{sec:sixteen_fold}, we found that $\phi^{\sigma\psi}_\sigma=\pi$ and 
$\phi^{\sigma I}_\sigma=0$ (the second equation is, of course, trivial). Furthermore, these two values are independent of $\nu_C$. Therefore, we conclude that for all non-Abelian topological orders satisfying the sixteenfold way, the flux-dependent term in the tunneling current is given by
\begin{align}
&\text{$n$ is odd}: I_{\Phi}=0.
\\
&\text{$n$ is even}: 
I_{\Phi}
=\frac{er}{2}|\Gamma_1| |\Gamma_2|\cos{(\gamma+\frac{n\pi}{4}+N_{\psi}\pi)}.
\end{align}
Here, $N_{\psi}=1$ if the antidot has the topological charge $\psi$, $N_{\psi}=0$ otherwise. Also, we have defined $\gamma=\phi_{\rm{AB}}+\delta$. The above expresses the celebrated even-odd effect.

\subsubsection{Abelian topological orders with flavor symmetry}

It has been argued that the even-odd effect was observed experimentally at
$\nu=5/2$~\cite{Willett2009, Willett2010}. At this time, the interpretation of the experiment remains ambiguous \cite{5/2-review-2019}, in part, because the even-odd effect may also be observed ~\cite{Stern_PRB2010} in the Abelian 331 state. Below, we argue that all the Abelian orders in the $16$-fold way can demonstrate the same effect if they have the exact flavor symmetry for the two species of quasiparticles $\sigma_1$ and $\sigma_2$. The $K=8$ order is an exception since it has only one quasiparticle type. The flavor symmetry is defined as the equivalence of the two quasiparticle types. This implies two properties: (i) the two species of quasiparticles have the same tunneling amplitudes at the QPC; and (ii) the probabilities
of their presence in the antidot are the same. For us, only (i) matters.

Suppose the antidot contains a total number of $n$ quasiparticles, such that $n_1$ of them are the first species of vortex and $n_2$ of them are the second species of vortex. Then, the condition $n=n_1+n_2$ must hold. Due to the exact flavor symmetry, the topological charge of the probe particle can be either $\sigma_1$ or $\sigma_2$ with the same probability. Depending on the species of the probe particle, the phase from encircling the antidot is given by
\begin{align}
\phi_1
&=\frac{n\pi}{4}
+n_1\phi^{\sigma_1\sigma_1}_{c_1}
+n_2\phi^{\sigma_1\sigma_2}_{c_2},
\\
\phi_2
&=\frac{n\pi}{4}
+n_1\phi^{\sigma_2\sigma_1}_{c_2}
+n_2\phi^{\sigma_2\sigma_2}_{c_1}.
\end{align}
In Sec.~\ref{sec:sixteen_fold}, we proved that 
$\phi^{\sigma_1 \sigma_1}_{c_1}=\phi^{\sigma_2 \sigma_2}_{c_1}=\pi\nu_C/4$ and
$\phi^{\sigma_1 \sigma_2}_{c_2}=\phi^{\sigma_2 \sigma_1}_{c_2}=\pi\nu_C/4+\pi$. Thus, we obtain
\begin{align}
\label{eq:phase1_FP}
\phi_1
&=\frac{n\pi}{4}
+(n_1+n_2)\frac{\pi\nu_C}{4}
+n_2\pi,
\\
\label{eq:phase2_FP}
\phi_2
&=\frac{n\pi}{4}
+(n_1+n_2)\frac{\pi\nu_C}{4}
+n_1\pi.
\end{align}

The phase difference between the two cases is given by 
$\Delta\phi=\phi_1-\phi_2=(n_2-n_1)\pi$. When $n$ is odd, $\Delta\phi\equiv\pi~(\text{mod } 2\pi)$ which implies that the measured backscattering current would have no oscillating pattern. On the other hand, $\Delta\phi\equiv 0~(\text{mod } 2\pi)$ when $n$ is even. Hence, constructive interference is present. Now, the evenness of $n$ means that $n_1$ and $n_2$ can be either both even or both odd. From Eqs.~\eqref{eq:phase1_FP} and~\eqref{eq:phase2_FP}, we see that a change in the parity of $n_1$ and $n_2$ shifts both $\phi_1$ and $\phi_2$ by a phase of $\pi$. This phenomenon is identical to the result for non-Abelian topological orders where two different topological charges of the antidot are possible at each even $n$, and correspond to two phases that differ by $\pi$. Therefore, we conclude that all topological orders satisfying the sixteenfold way can demonstrate the even-odd effect if the Abelian orders have an exact symmetry for the $e/4$ quasiparticles.

\subsubsection{$e/2$-quasiparticle tunneling}

We complete our discussion of Fabry-P\'{e}rot interferometery by examining the tunneling current when the tunneling process is dominated by the $(e/2, I)$ quasiparticles. In this scenario, the braiding phase from moving an $(e/2, I)$ quasiparticle around an $e/4$ particle is $\pi n/2$. Hence, the periodic term for the backscattering current is given by~\cite{Bishara2009}: 
\begin{eqnarray}
I_{e/2}
\propto\cos{\left(\frac{2\pi\Phi}{2\Phi_0}+\frac{n\pi}{2}\right)},
\end{eqnarray}
where $\Phi_0=hc/e$ and $\Phi$ is the magnetic flux. In other words, the backscattering current can tell nothing about the nature of the topological order. 

%To distinguish different topological orders, a more sophisticated experimental technique, such as the Mach-Zehnder interferometer is required.

\section{Mach-Zehnder interferometry}
\label{sec:MZ_experiment}

%In Appendix~\ref{app:FP_experiment}, we argue that the Fabry-P\'{e}rot interferometer cannot distinguish different topological orders effectively. Another drawback for the %Fabry-P\'{e}rot experiment is its sensitivity to the number fluctuation of quasiparticles in the interferometer. As a result, 
In this section, we consider a more complicated setup than a Fabry-P\'{e}rot interferometer. A Mach-Zehnder interferometer \cite{KT2006, MZ-2003} is harder to fabricate, but it offers two advantages over other approaches to interferometry. First, it produces substantially different signatures for different topological orders of the sixteenfold way. Second, this approach is immune to complications from fluctuations of the quasiparticle charge inside the interferometer~\cite{Kang-FPI}. If such fluctuations happen on a shorter time- scale than a typical time interval between tunneling events at the point contacts in the interferometer, then the fluctuations would destroy or greatly modify the interference picture in any device. Slow fluctuations still greatly affect the behavior of a Fabry-P\'{e}rot interferometer~\cite{Simon-noise}, while a Mach-Zehnder device is not sensitive to them.

The physics of a Mach-Zehnder interferometer is considerably more involved than in the experimental setups from the previous section. It was addressed for some topological orders before
~\cite{Feldman2006, KT2006, KT_noise, Ponomarenko2007, Ponomarenko2010, Zucker2016, Chenjie2010, Guang2015, KT2008}. Our present goal is to review the expected signatures in all states of the sixteenfold way.
We will consider not only the tunneling current, but also the low-frequency noise in the interferometer. At weak tunneling, the noise and the current are not independent probes in the Fabry-P\'{e}rot setup. Indeed, at $T=0$, the noise $S=\int dt \langle I(0)I(t)+I(t)I(0)\rangle$ reduces to the Schottky formula $S=2qI$, where $q$ is the charge of tunneling quasiparticles~\cite{KT_noise}. Interestingly, the noise exhibits a much more complicated behavior in the Mach-Zehnder setup. This happens due to the memory of the previous tunneling events.

%In this section, we will evaluate the tunneling current and the Fano factor in a shot-noise experiment for all topological orders introduced in Section~\ref{sec:5/2}. 
Below, we focus on zero temperature, so that quasiparticles can only tunnel from the edge with the higher electrochemical potential (edge 1) to the edge with the lower electrochemical potential (edge 2). A systematic treatment for systems at a finite temperature \cite{KT_noise, Zucker-thesis} can be found in Appendix~\ref{app:Finite_T_MZ}.

\begin{figure}[htb]
\includegraphics[width=3.0 in]{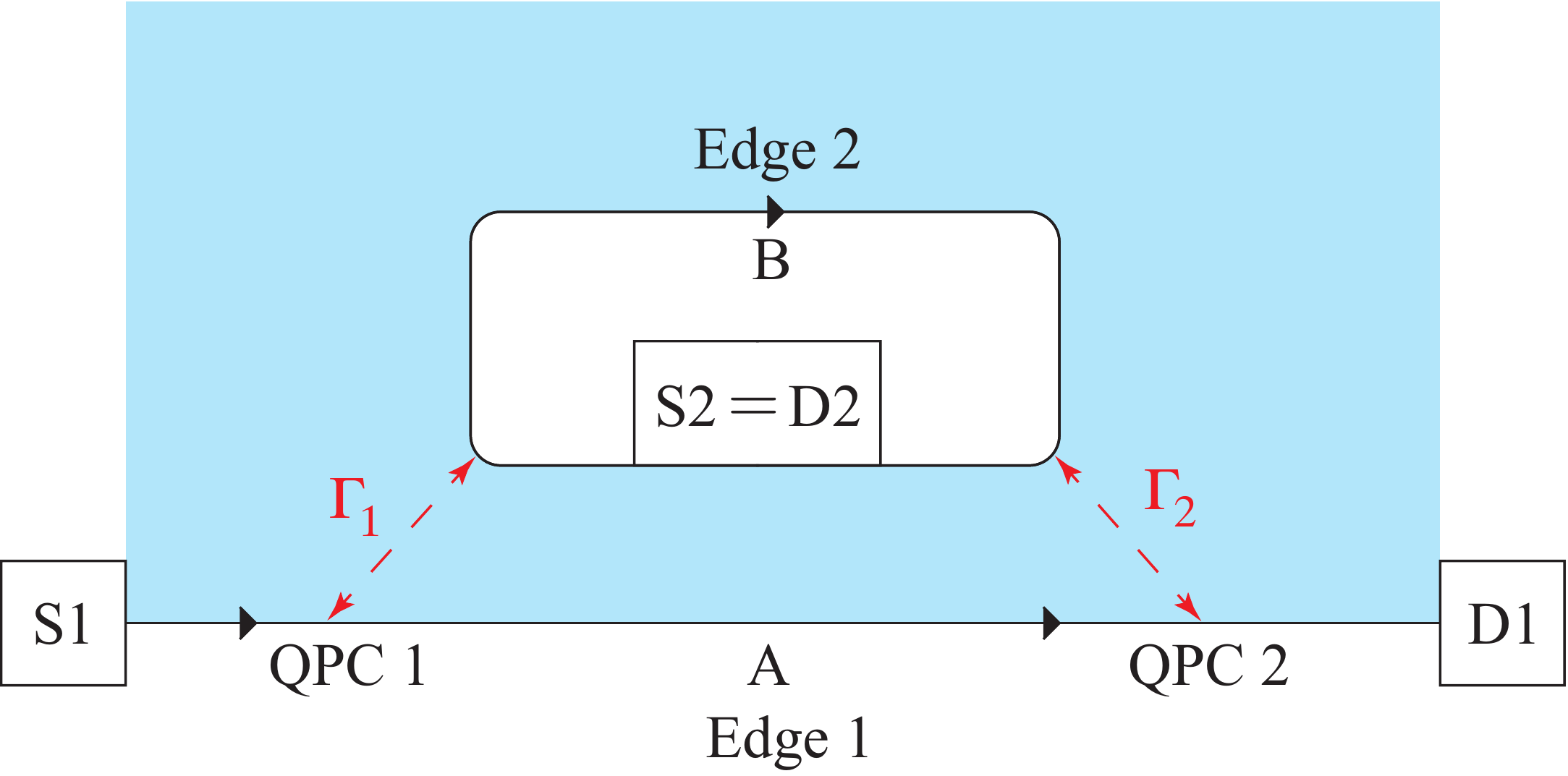}
\caption{(Color online) A schematic picture of an electronic Mach-Zehnder interferometer. Charges propagate from source S1 to drain D1 and source S2 to drain D2. Tunneling is possible at the two quantum point contacts, QPC1 and QPC2.}
\label{fig:MZ}
\end{figure}

A typical setup for a Mach-Zehnder interferometer is illustrated in Fig.~\ref{fig:MZ}. In the figure, the arrows show the propagation of charged modes along the quantum Hall edges. Quasiparticles are allowed to tunnel between the edges at the two quantum point contacts, QPC1 and QPC2. Source S1 is biased so that the electrochemical potential of edge 1 is higher than that of edge 2 by  $eV$. We are interested in the tunneling current from source S1 to drain D2 and the corresponding noise, which depends on $V$ and the magnetic flux enclosed by the loop QPC1-A-QPC2-B-QPC1.

The key piece of physics is the memory effect. Each quasiparticle, absorbed by drain D2, remains forever inside the loop QPC1-A-QPC2-B-QPC1. The probability of each subsequent tunneling event is affected by the mutual statistical phase $\phi_s$ of the tunneling quasiparticle and drain D2.

\subsection{Tunneling current for non-Abelian orders}

Since the bulk excitations are gapped, the system can be described by a low-energy edge theory. The tunneling process in Fig.~\ref{fig:MZ} is modeled by the following Hamiltonian:
\begin{eqnarray}
\hat{H}
=\hat{H}_{\text{edge}}
+\left[\left(\Gamma_1\hat{T}_1+\Gamma_2\hat{T}_2\right)+\text{H.c.}\right],
\end{eqnarray}
where $\hat{H}_{\text{edge}}$ is the Hamiltonian for the two edges of the FQH liquid. The tunneling amplitudes for particles at the two quantum point contacts are labeled as 
$\Gamma_1$ and $\Gamma_2$, with the corresponding tunneling operators denoted as 
$\hat{T}_1$ and $\hat{T}_2$. Here, we choose a gauge such that both the Aharonov-Bohm phase $\phi_{\text AB}$ and the statistical phase $\phi_s$ are absorbed in $\hat{T}_2$. Depending on the number of neutral bosonic modes on the edge  and the tunneling amplitudes for different types of quasiparticles, the tunneling process can be dominated by either $e/2-$ or 
$e/4-$quasiparticles [see Eqs.~\eqref{eq:g_5/2_nA} and~\eqref{eq:g_5/2_A}].
In the following, we will calculate the tunneling current for each case separately.

\subsubsection{Case 1: $e/4$-quasiparticle tunneling}

\begin{figure}[htb]
\includegraphics[width=3.0 in]{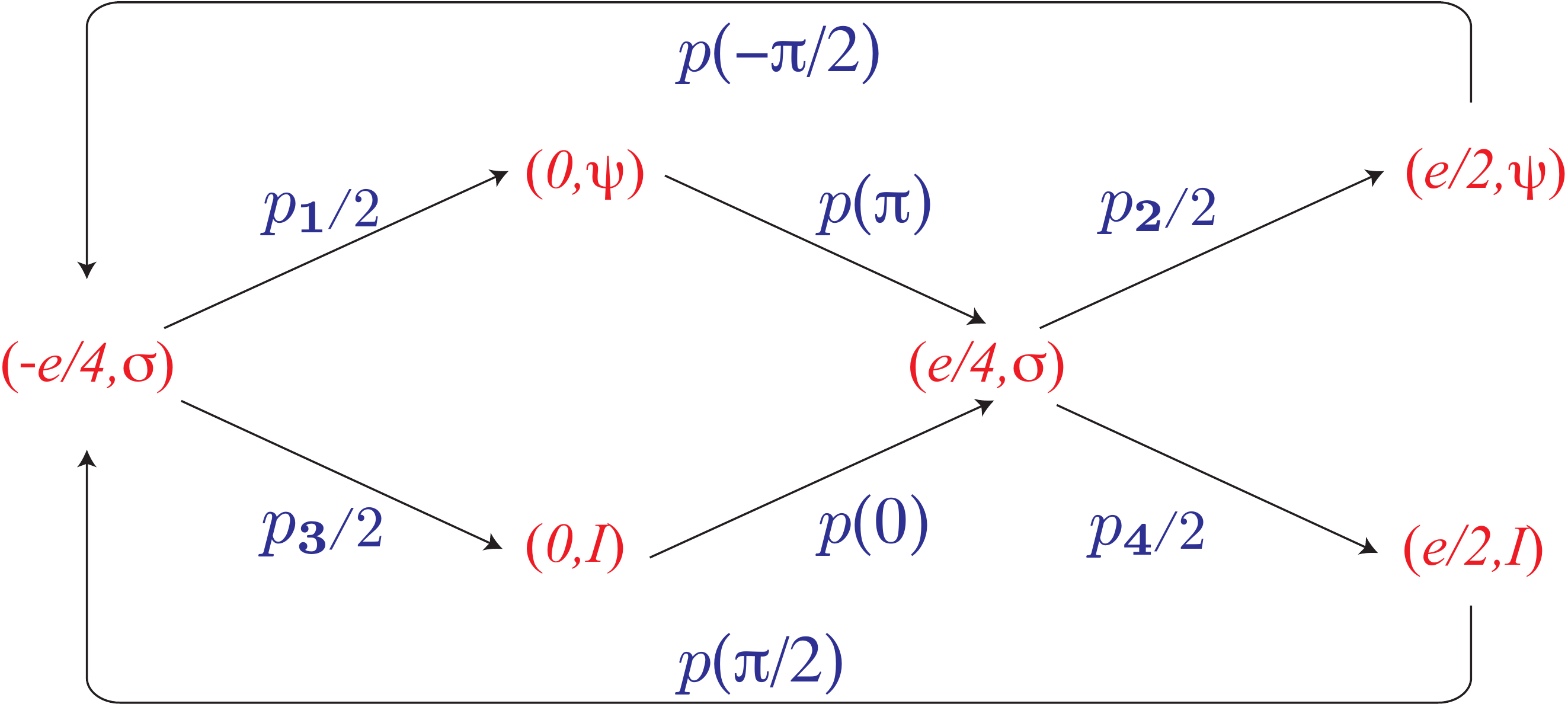}
\caption{(Color online) The six possible superselection sectors for drain D2 when the tunneling is dominated by charge-$e/4$ quasiparticles. The arrows show all possible transitions between different sectors at zero temperature. The corresponding transition probabilities and statistical phases are shown in blue (four phases are listed in Table~\ref{tab:phase_MZ_non-Abelian}).}
\label{fig:MZ_e/4}
\end{figure}

For all proposed non-Abelian topological orders in Sec.~\ref{sec:5/2}, the fundamental excitations are quasiparticles with charge $e/4$. Suppose the tunneling process is dominated by $e/4$ quasiparticles. Then, there are six possible superselection sectors for drain D2 as shown in Fig.~\ref{fig:MZ_e/4}. Each sector is labeled by the electric and topological charges in parentheses. The electric charge is always $ne/4$, where $n=-1,0,1,2$, since changing $n$ by 4 amounts to adding the charge of a topologically trivial electron. Thus,
$(-e/4,\sigma)$ can be considered to be in the same sector as $(3e/4,\sigma)$ \cite{footnote-03-20}. 
Since the temperature is assumed to be zero, all transitions between different sectors are unidirectional.

When both $\Gamma_1$ and $\Gamma_2$ are small, and assuming that the fusion channel of the tunneling particle with the topological charge in D2 is known, a general expression for the transition rate between two sectors can be written as~\cite{KT2006}:
\begin{align} \label{eq:transition_rate}
\nonumber
p(\phi_s)
=r[&(\left|\Gamma_1\right|^2+\left|\Gamma_2\right|^2)
\\
&+2u\left|\Gamma_1\right| \left|\Gamma_2\right|
\cos{\left(\phi_{\text{AB}}+\phi_s+\delta\right)}],
\end{align}
with $\delta=\text{arg}~(\Gamma_2/\Gamma_1)$ and $u\le 1$. Here, $\phi_{\text{AB}}=2\pi\Phi/(4\Phi_0)$ is the Aharonov-Bohm phase accumulated by an $e/4$ quasiparticle moving around the interferometer loop QPC1-A-QPC2-B-QPC1. Four probabilities $p(n\pi/2)$ in Fig. \ref{fig:MZ_e/4} are given by the above expression with $\phi_s =n\pi/2$. The remaining  probabilities are $p_{\bf i}/2$, where $p_{\bf i}$
is given by Eq. (\ref{eq:transition_rate}) with $\phi_s$ from Table \ref{tab:phase_MZ_non-Abelian}. The factor of $1/2$ in each probability comes from two possible fusion channels,
$\sigma\times\sigma=I$ or $\psi$, and reflects the equal probabilities of the two fusion outcomes.

%Furthermore, the statistical phase $\phi_s$ is given by
%\begin{eqnarray} \label{eq:phase_interferometer_e/4}
%\phi_s
%=\frac{n\pi}{4}+\phi^{\sigma\alpha}_{\beta}.
%\end{eqnarray} 
%Here, $\alpha$ denotes the topological charge of drain D2 before tunneling and $\beta$ is the fusion outcome between $\sigma$ and $\alpha$. As a reminder, we quote the results for 
%$\phi^{\sigma\alpha}_\beta$ from Section~\ref{sec:sixteen_fold}:
%\begin{align}
%\phi^{\sigma\sigma}_{I}
%&\equiv -\frac{\pi\nu_C}{4}~(\text{mod }2\pi),
%\\
%\phi^{\sigma\sigma}_{\psi}
%&\equiv\frac{3\pi\nu_C}{4}~(\text{mod }2\pi), 
%\\
%\phi^{\sigma\psi}_{\sigma}
%&=\pi
%\\
%\phi^{\sigma I}_{\sigma}
%&=0
%\end{align}

%The corresponding values of $\phi_s$ in for transitions between different  superselection sectors are summarized in Table~\ref{fig:MZ_e/4}.

\begin{table} [htb] 
\begin{center}
\begin{tabular}{| c | c ||  c | c  |}
\hline
~~index ~~ & ~~ $\phi_s$ ~~ & ~~ index ~~ & ~~ $\phi_s$ ~~ 
\\ \hline\hline
$\mathbf{1}$  & ~$\pi(3\nu_C-1)/4$~ & 
~$\mathbf{2}$~ & ~ $\pi(3\nu_C+1)/4$~
\\ \hline
$\mathbf{3}$  & ~$-\pi(\nu_C+1)/4$~ & 
~$\mathbf{4}$~ & ~$-\pi(\nu_C-1)/4$~
\\ \hline
\end{tabular}
\caption{Statistical phases $\phi_s$ for transitions between different superselection sectors as shown in Fig.~\ref{fig:MZ_e/4}. Notice that 
$p_{\mathbf{1}}+p_{\mathbf{3}}=p_{\mathbf{2}}+p_{\mathbf{4}}
=2r\left(\left|\Gamma_1\right|^2+\left|\Gamma_2\right|^2\right)$.}
\label{tab:phase_MZ_non-Abelian}
\end{center}
\end{table}

As shown in Fig.~\ref{fig:MZ_e/4}, there are four possible ways for drain D2 to absorb one electron charge from source S1 and return back to the original sector $(-e/4,\sigma)$. They correspond to four paths  $\mathcal{P}_i$, $i=1,\dots,4$, on the oriented graph in the figure. 
For example, one path $\mathcal{P}_1$ is $(-e/4,\sigma)\rightarrow(0,\psi)\rightarrow(e/4,\sigma)\rightarrow(e/2,\psi)\rightarrow(-e/4,\sigma)$. 
To compute the average current detected in drain D2, we need to know the average time $\bar t_{e/4}$ to transfer four successive quasiparticles: $I=e/\bar t_{e/4}$.
The average time is a weighted sum of the average times $\bar{t}_{\mathcal{P}_i}$ to travel along each of the paths $\mathcal{P}_i$. For example, the probability $q_1$ that the system chooses path $\mathcal{P}_1$ equals

\begin{eqnarray}
q_1=
\left(\frac{p_{\mathbf{1}}}{p_{\mathbf{1}}+p_{\mathbf{3}}}\right)
\left(\frac{p_{\mathbf{2}}}{p_{\mathbf{2}}+p_{\mathbf{4}}}\right).
\end{eqnarray}
The average time is given by 

\begin{equation}
\bar{t}_{e/4}=\sum_{i=1}^4 q_i \bar{t}_{\mathcal{P}_i},
\label{eq:dima_te4}
\end{equation}
where $q_i$ are the probabilities of the four paths.

The expressions for $q_i$ and $\mathcal{P}_i$ are similar for all $i$.
We only show the contribution from the first path:
\begin{align}
\nonumber
q_1 \bar{t}_{\mathcal{P}_1}=
&\int_0^{\infty}\left[e^{-(p_\mathbf{1}+p_\mathbf{3})t_1/2}\cdot
\left(\frac{p_{\mathbf{1}}}{2}\right)\right]
\left[e^{-p(\pi)t_2}\cdot p(\pi)\right]
\\ \nonumber
&\times\left[e^{-(p_\mathbf{2}+p_\mathbf{4})t_3/2}\cdot
\left(\frac{p_{\mathbf{2}}}{2}\right)\right]
\left[e^{-p(-\frac{\pi}{2})t_4}\cdot p(-\frac{\pi}{2})\right]
\\
&\times(t_1+t_2+t_3+t_4)~dt_1 dt_2 dt_3 dt_4,
\end{align}
%From the result of the integral, a diagrammatic approach for evaluating $\bar{t}_{e/4}$ is obtained. The average time is given by $\bar{t}_{e/4}=\sum_{i=1}^4 q_i %\bar{t}_{\mathcal{P}_i}$, where $q_i$ and $\bar{t}_{\mathcal{P}_i}$ denote the weight and avergae time for four successive tunneling via path $\mathcal{P}_i$. For $\mathcal{P}_1$, the %weight is assigned as
so that
\begin{eqnarray}
\bar{t}_{\mathcal{P}_1}=
\frac{2}{p_{\mathbf{1}}+p_{\mathbf{3}}}
+\frac{1}{p(\pi)}
+\frac{2}{p_{\mathbf{2}}+p_{\mathbf{4}}}
+\frac{1}{p(-\pi/2)}.
\end{eqnarray}
\\
For convenience in the later discussion, we define
\begin{eqnarray}
A=
\left(4+\sum_{j=1}^4 \frac{p_{\mathbf{j}}}{p[(j+1)\pi/2]}\right).
\end{eqnarray}
After summing over all four paths with the weights $q_i$, we have
\begin{eqnarray}
\bar{t}_{e/4}
=\frac{A}{2r(\left|\Gamma_1\right|^2+\left|\Gamma_2\right|^2)},
\end{eqnarray}
The same result can also be derived with the kinetic equation approach~\cite{KT_noise, KT2006}.

The tunneling current  $I_{e/4}=e/\bar{t}_{e/4}$ 
takes four different values for different $\nu_C$. Indeed, $\phi_s$ in Table \ref{tab:phase_MZ_non-Abelian} is invariant under $\nu_C\rightarrow\nu_C+8$, and $\nu_C$ is an odd number for non-Abelian orders. In terms of the parameter $s=2u\left|\Gamma_1\right| \left|\Gamma_2\right|/
(\left|\Gamma_1\right|^2+\left|\Gamma_2\right|^2)$, Eqs.~(\ref{eq:dima-s}) and (\ref{eq:dima-s-2}), one has
\begin{widetext}
\begin{align}
\label{eq:case1_I_e/4}
&\text{when } \nu_C=1:\quad
I_{e/4}
=\frac{er}{4}\left(\left|\Gamma_1\right|^2+\left|\Gamma_2\right|^2\right)
\left[
\frac{1-s^2+\frac{s^4}{4}\sin^2{2\gamma}}
{1-\frac{3s^2}{4}
+\frac{s^4}{16}
\left(1-\cos{4\gamma}
-\sin{4\gamma}\right)}
\right],
\\
\label{eq:case4_I_e/4}
&\text{when } \nu_C=-1:\quad
I_{e/4}
=\frac{er}{4}\left(\left|\Gamma_1\right|^2+\left|\Gamma_2\right|^2\right),
\\
\label{eq:case2_I_e/4}
&\text{when } \nu_C=3~\text{or}-5:\quad
I_{e/4}
=\frac{er}{4}\left(\left|\Gamma_1\right|^2+\left|\Gamma_2\right|^2\right)
\left[
\frac{1-s^2+\frac{s^4}{4}\sin^2{2\gamma}}
{1-\frac{s^2}{2}}\right],
\\
\label{eq:case3_I_e/4}
&\text{when } \nu_C=5~\text{or}-3:\quad 
I_{e/4}
=\frac{er}{4}\left(\left|\Gamma_1\right|^2+\left|\Gamma_2\right|^2\right)
\left[
\frac{1-s^2+\frac{s^4}{4}\sin^2{2\gamma}}
{1-\frac{3s^2}{4}
+\frac{s^4}{16}
\left(1-\cos{4\gamma}+\sin{4\gamma}\right)}
\right].
\end{align}
\end{widetext}
In the above equations, we have defined $\gamma=\phi_{\text{AB}}+\delta$. We remark that Eqs.~\eqref{eq:case1_I_e/4} and~\eqref{eq:case4_I_e/4} reproduce the results for the Pfaffian order~\cite{Feldman2006} and the PH-Pfaffian order~\cite{Zucker2016}, respectively. The PH-Pfaffian case is strikingly different from all others since the current 
(\ref{eq:case4_I_e/4}) exhibits no flux dependence. $\nu_C=\pm 7$ are not included in the above equations since $e/4$-particles are not expected to dominate tunneling at those Chern numbers.

\subsubsection{Case 2: $e/2$ quasiparticle tunneling}

As discussed in Sec.~\ref{sec:dima-tunneling}, $e/4$-particles dominate tunneling at 
$|\nu_C|<4$. At $|\nu_C|\geq 7$, the most important tunneling process involves $e/2$-particles. Both quasiparticle types can dominate tunneling at the intermediate values of the Chern number.
Thus, it is essential to address the interference of both $e/4$ and $e/2$ charges. Below we 
investigate the tunneling of the particles from the $(e/2,I)$ sector. In a striking contrast with the $e/4$ case, the results do not depend on statistics, at least, in the simplest model.
In fact, the tunneling current is the same for the Abelian and non-Abelian orders.

\begin{figure}[htb]
\includegraphics[width=3.3 in]{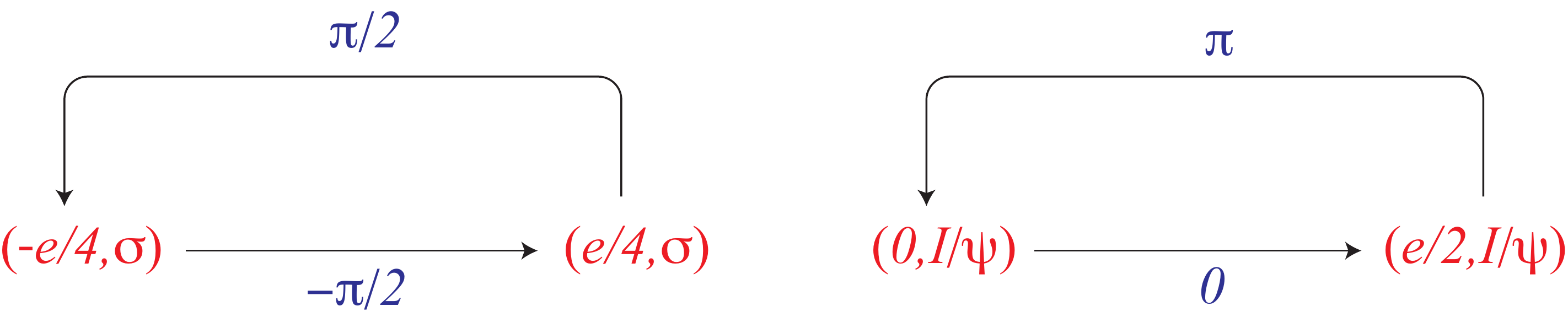}
\caption{(Color online) Possible superselection sectors for drain D2 when the tunneling is dominated by charge-$e/2$ quasiparticles. The left (right) panel illustrates the scenario when the number of charge-$e/4$ quasiparticles in D2 is odd (even). When there are even number of quasiparticles in drain D2, the topological charge of the drain can be either $I$ or $\psi$. However, the topological charge does not change after tunneling events since $e/2$ quasiparticles carry the trivial topological charge $I$. The arrows show all possible transitions between different sectors at zero temperature.}
\label{fig:MZ_e/2}
\end{figure}

%When $\nu_C=\pm 7$ or larger,  the tunneling process will be dominated by the charge-$e/2$ quasiparticles. In this case, the most relevant operator is given by $I e^{i\phi_\rho}$. 
As before, we denote the number of $e/4$-quasiparticles in D2 as $n$. Depending on the parity of $n$, possible superselection sectors for the drain are shown in Fig.~\ref{fig:MZ_e/2}. From the figure, one sees that the tunneling current depends on the parity of $n$. The Aharonov-Bohm phase becomes $\phi_{\text{AB}}'=2\pi\Phi/(2\Phi_0)$. The statistical phase is $\phi_s'=n\pi/2$ irrespectively of $\nu_C$. 

When $n$ is odd, the topological charge for the drain can be $\sigma$ only. The average time required for D2 to absorb one electron is then given by
$\bar{t}_{e/2}=1/p(-\pi/2)+1/p(\pi/2)$. On the other hand, the topological charge of D2 can be either $I$ or $\psi$, when $n$ is even. 
%Since the tunneling quasiparticles have the trivial topological charge $I$, the topological charge for the drain does not change. 
In both cases, the time for D2 to absorb an electron is $\bar{t}_{e/2}=1/p(0)+1/p(\pi)$. Therefore, we determine the tunneling current $I_{e/2}=e/\bar{t}_{e/2}$ as
\begin{align} 
\label{eq:tunneling_e/2: case 1}
I_{e/2} 
&=\frac{er}{2}(\left|\Gamma_1\right|^2+\left|\Gamma_2\right|^2)
\left(1-s^2\sin^2{\gamma'}\right)
~\text{for odd $n$},
\\ \label{eq:tunneling_e/2: case 2}
I_{e/2} 
&=\frac{er}{2}(\left|\Gamma_1\right|^2+\left|\Gamma_2\right|^2)
\left(1-s^2\cos^2{\gamma'}\right)
~\text{for even $n$}.
\end{align}
Here, $\gamma'=\phi'_{\text{AB}}+\delta$ and $s$ has a similar definition to the definition in Eqs. (\ref{eq:case1_I_e/4})-(\ref{eq:case3_I_e/4}). This result resembles the even-odd effect in the Fabry-P\'{e}rot interferometry.
 
A more general analysis should incorporate rare tunneling events of charge-$e/4$ particles. Such tunneling events switch the system between the two sides of Fig. \ref{fig:MZ_e/2}.
In turn, the tunneling of $e/4$ particles is sensitive to possible tunneling of neutral fermions $\psi$. Fermion tunneling is marginal in the RG sense~\cite{Fisher_Nayak2007} and hence likely more important than the tunneling of $e/4$ charges at $|\nu_C|\ge 7$.
%We should point out that the most general consideration for the tunneling process is incorporating both charge-$e/2$ and charge-$e/4$ quasiparticle tunneling. 
To include such effects, it is necessary to set up a full set of kinetic equations. This cumbersome general procedure is beyond the scope of this paper. On the other hand, one does not need to include rare tunneling events of $e/2$ particles and neutral fermions in the analysis of the previous subsection, where we assumed that $e/4$ particles dominate. 
The difference between this subsection and the previous subsection is due to the fact that $e/4$ charge tunneling cycles the system through all superselection sectors. Any additional tunneling events just occasionally change the phase of the cycle. When the dominant tunneling process is due to $e/2$ particles, some superselection sectors are available only through rare tunneling events of other charges.

%\newpage

\subsection{Fano factor in shot-noise experiment for non-Abelian orders}

The Fano factor in a shot-noise experiment is another useful parameter to differentiate topological orders~\cite{KT_noise}. The non-equilibrium noise is defined as the Fourier transform of the current-current correlation function:
\begin{eqnarray}
S(\omega)
=\frac{1}{2}
\int_{-\infty}^{\infty}
\langle\hat{I}(0)\hat{I}(t)+\hat{I}(t)\hat{I}(0)\rangle e^{i\omega t}dt.
\end{eqnarray}
This definition differs by a factor of $1/2$ from a definition, frequently found in the literature.
We focus on the low-frequency limit. In this case, the shot noise can be written as 
$S=\overline{\delta Q^2(t)}/t$, where $Q(t)$ is the charge, transmitted through the interferometer over the time $t$, and $\delta Q(t)$ is its fluctuation. 
The Fano factor $e^*$ is the ratio between the noise and the current: 
\begin{eqnarray}
\label{eq:dima-noise-def}
e^*
=S/ I
=\overline{\delta Q^2(t)}/Q(t)
=e(\overline{\delta t^2}/\overline{t}^2),
\end{eqnarray}
where $\overline{t}$ is the average time needed to transfer the total charge $e$ through the interferometer, and $\overline{\delta t^2}$ is the mean square fluctuation of that time.
The last equality in Eq. (\ref{eq:dima-noise-def}) was derived in Ref.~\cite{KT_noise}.

We first assume that tunneling is dominated by $e/4$-particles.
Now, we proceed to evaluate $\overline{t^2}_{e/4}$. 
One easily verifies that 

\begin{equation}
\label{eq:dima-new-noise-1}
\overline{t^2}_{e/4}=\sum q_i \overline{t^2}_{\mathcal{P}_i},
\end{equation}
where $\overline{t^2}_{\mathcal{P}_i}$ are the fluctuations of the times, corresponding to the four paths through the diagram in Fig. \ref{fig:MZ_e/4}.
For the path $\mathcal{P}_1$, the contribution $q_1\overline{t^2}_{\mathcal{P}_1}$ is given by
\begin{align}
\nonumber
\int_0^{\infty}
&\left[e^{-(p_\mathbf{1}+p_\mathbf{3})t_1/2}\cdot
\left(\frac{p_{\mathbf{1}}}{2}\right)\right]
\left[e^{-p(\pi)t_2}\cdot p(\pi)\right]
\\ \nonumber
&\times\left[e^{-(p_\mathbf{2}+p_\mathbf{4})t_3/2}\cdot
\left(\frac{p_{\mathbf{2}}}{2}\right)\right]
\left[e^{-p(-\frac{\pi}{2})t_4}\cdot p(-\frac{\pi}{2})\right]
\\
&\times(t_1+t_2+t_3+t_4)^2~dt_1 dt_2 dt_3 dt_4.
\end{align}
%Again, one can associate a diagrammatic rule to the above integral. 
This yields 
%$\overline{t^2}_{e/4}$ from $\mathcal{P}_1$ is 

\begin{align}
\nonumber
\overline{t^2}_{\mathcal{P}_1}
=&~\frac{8}{(p_{\mathbf{1}}+p_{\mathbf{3}})^2}
+\frac{8}{(p_{\mathbf{2}}+p_{\mathbf{4}})^2}
+\frac{8}{(p_{\mathbf{1}}+p_{\mathbf{3}})(p_{\mathbf{2}}+p_{\mathbf{4}})}
\\ \nonumber
&+\frac{4}{(p_{\mathbf{1}}+p_{\mathbf{3}})p(\pi)}
+\frac{4}{(p_{\mathbf{1}}+p_{\mathbf{3}})p(-\pi/2)}
\\ \nonumber
&+\frac{4}{(p_{\mathbf{2}}+p_{\mathbf{4}})p(\pi)}
+\frac{4}{(p_{\mathbf{2}}+p_{\mathbf{4}})p(-\pi/2)}
\\
&+\frac{2}{[p(\pi)]^2}+\frac{2}{[p(-\pi/2)]^2}+\frac{2}{p(\pi)p(-\pi/2)}.
\end{align}
%Mathematically, the diagrammatic rule follows from 
%$\langle (\sum t_i)^2\rangle=\sum\langle t_i^2\rangle+\sum\langle t_i\rangle \langle t_j\rangle$ and $\langle t_i^2\rangle=2/p_i^2$. Using this, $\overline{t^2}_{e/4}$ can be evaluated %as 
%$\overline{t^2}_{e/4}=\sum q_i \overline{t^2}_{\mathcal{P}_i}$. 
A lengthy but straightforward calculation for all four paths gives the following Fano factor:
\begin{align} \label{eq: Fano_nA}
\nonumber
\frac{e^*}{e}
=&~\frac{p_{\mathbf{1}}p_{\mathbf{3}}}{A^2}
\left[\frac{1}{p(0)}-\frac{1}{p(\pi)}\right]^2
+\frac{p_{\mathbf{2}}p_{\mathbf{4}}}{A^2}
\left[\frac{1}{p(-\frac{\pi}{2})}-\frac{1}{p(\frac{\pi}{2})}\right]^2
\\
&+\frac{p_{\mathbf{1}}+p_{\mathbf{3}}}
{A^2}
\sum_{j=1}^4 \frac{p_{\mathbf{j}}} {(p[(j+1)\pi/2])^2}
+\frac{8}{A^2}.
\end{align}
By substituting the probabilities at different $\nu_C$, one gets 
$e^*/e$ as
\begin{widetext}
\begin{align}
&\text{when } \nu_C=1:\quad
\frac{e^*}{e}
=\frac{1}{4}
\left[
\frac{1-s^2+\frac{s^4}{4}\left(2+\cos{4\gamma}-\sin{4\gamma}\right)
+\frac{s^6}{8}\sin{4\gamma}-\frac{s^8}{128}\left(1-\cos{8\gamma}\right)}
{\left[1-\frac{3s^2}{4}+\frac{s^4}{16}
\left(1-\cos{4\gamma}-\sin{4\gamma}\right)\right]^2}
\right],
\\
\label{eq:dima-PH-F}
&\text{when } \nu_C=-1:\quad
\frac{e^*}{e}
=\frac{1}{4}
\left[
\frac{1-\frac{s^2}{2}}
{1-s^2+\frac{s^4}{8}\left(1-\cos{4\gamma}\right)}
\right],
\\
&\text{when } \nu_C=3~\text{or}-5:\quad
\frac{e^*}{e}
=\frac{1-\frac{s^2}{2}+\frac{s^4}{8}\left(1+3\cos{4\gamma}\right)
+\frac{s^6}{16}\left(1-\cos{4\gamma}\right)}
{\left(2-s^2\right)^2},
\\
&\text{when } \nu_C=5~\text{or}-3:\quad
\frac{e^*}{e}
=\frac{1}{4}
\left[
\frac{1-s^2+\frac{s^4}{4}\left(2+\cos{4\gamma}+\sin{4\gamma}\right)
-\frac{s^6}{8}\sin{4\gamma}-\frac{s^8}{128}\left(1-\cos{8\gamma}\right)}
{\left[1-\frac{3s^2}{4}+\frac{s^4}{16}
\left(1-\cos{4\gamma}+\sin{4\gamma}\right)\right]^2}
\right].
\end{align}
\end{widetext}
From these equations, we extract the maximal and minimal possible values of $e^*/e$ in the limit of the maximal possible $s=1$. Those values and the corresponding values of 
$\gamma=\phi_{\text{AB}}+\delta$ are summarized in Table~\ref{tab:MZ_table_NA}.

\begin{table} [htb] 
\begin{center}
\begin{tabular}{| c | c | c | c | c |}
\hline
~~$\nu_C~(\text{mod }8)$ ~~  & ~~ $e^*_{\text{max}}/e$ ~~ 
& ~~ $\gamma_{\text{max}}$~~ & ~~ $e^*_{\text{min}}/e$ ~~ 
& ~~ $\gamma_{\text{min}} ~~$
\\ \hline\hline
$1$  &  ~$3.20$~ & ~$0.09$~ &  ~$0.44$~ & ~$-0.82$~
\\ \hline
$3$  &  ~$1$~ & ~$0$~ &  ~$3/8$~ & ~$\pm\pi/4$~
\\ \hline
$5$  &  ~$3.20$~ & ~$-0.09$~ &  ~$0.44$~ & ~$0.82$~
\\ \hline
$7$  &  ~$\infty$~ & ~$0$ & ~ $1/2$~ & ~$\pm\pi/4$~
\\ \hline
\end{tabular}
\caption{Maximal and minimal values of the Fano factor in a Mach-Zehnder interferometer experiment when $s=1$. Here, we focus on non-Abelian orders for the  $\nu=5/2$ FQH state and assume that the tunneling process is dominated by $e/4$-quasiparticles. The third and last columns provide the values for $\gamma=\phi_{\text{AB}}+\delta$ at which $e^*=e^*_{\text{max}}$ and $e^*=e^*_{\text{min}}$, respectively. Notice that these values of $\gamma$ are modulo $\pi/2$.}
\label{tab:MZ_table_NA}
\end{center}
\end{table}

When the tunneling process is dominated by $e/2$ quasiparticles, the physics is similar to that of  a Laughlin state \cite{KT_noise} as can be seen from Fig.~\ref{fig:MZ_e/2}. The Fano factor is simply given by
\begin{eqnarray}
e^*
=e\left[\frac{1+s^2\sin^2{(\phi'_{\text{AB}}+\delta)}}{2}\right],
\end{eqnarray}
or a similar expression with a cosine in place of the sine.
Hence, the maximal value of the Fano factor is $e$ in the limit of $\Gamma_1\approx\Gamma_2$ and $u\approx 1$. The minimal Fano factor is always $e/2$.

\subsection{Abelian topological orders with flavor symmetry}

\begin{figure}[htb]
\includegraphics[width=3.3 in]{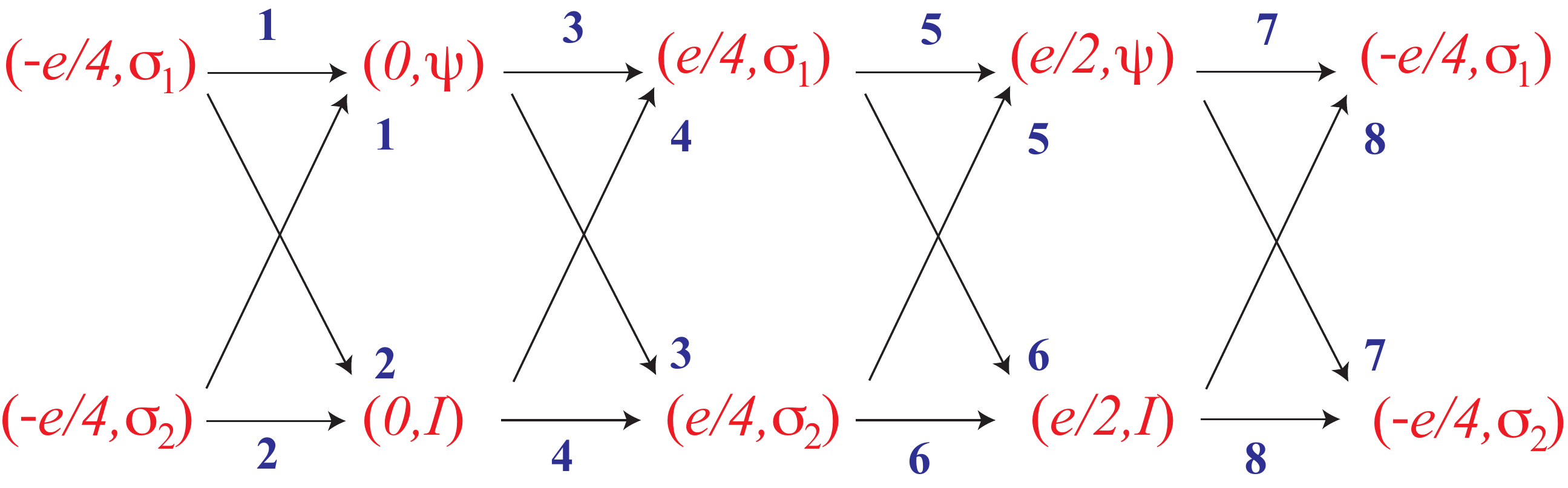}
\caption{(Color online) For Abelian topological orders of the $\nu=5/2$ FQH state, there are eight superselection sectors for drain D2. The arrows show possible transitions between different sectors at zero temperature due to the tunneling of $e/4$ particles. The corresponding statistical phases are listed in Tables~\ref{tab:phase_MZ_Abelian_0} and~\ref{tab:phase_MZ_Abelian_2}.} 
\label{fig:MZ_Abelian_e/4}
\end{figure}

Similar calculations can be performed for Abelian topological orders. However, there are two different species of $e/4$-quasiparticles due to the 
two different types of vortices, $\sigma_1$ and 
$\sigma_2$, as shown in Sec.~\ref{sec:sixteen_fold}. Consequently, there are eight distinct superselection sectors for drain D2 as shown in Fig.~\ref{fig:MZ_Abelian_e/4}. 
Generally, the two types of quasiparticles can have different tunneling amplitudes at the quantum point contacts. Thus, one has to consider many more transition rates than in the 
non-Abelian case. The calculations become very cumbersome. In the past, they were performed numerically for some of the proposed topological orders~\cite{Chenjie2010, Guang2015}.

The statistical phase, accumulated after one $e/4$-quasiparticle encircles another, is still given by Eq.~\eqref{eq:phase_interferometer_e/4}. Since the sixteenfold way is also satisfied by Abelian vortices, $\phi_s$ can be evaluated easily. Depending on the Chern number of the topological order, the results are shown in Tables~\ref{tab:phase_MZ_Abelian_0} and~\ref{tab:phase_MZ_Abelian_2}.

\begin{table} [htb] 
\begin{center}
\begin{tabular}{| c | c ||  c | c  |}
\hline
~~index ~~ & ~~ $\phi_s$ ~~ & ~~ index ~~ & ~~ $\phi_s$ ~~ 
\\ \hline\hline
$\mathbf{1}$  & ~$\pi(\nu_C+3)/4$~ & ~$\mathbf{5}$~ & ~$\pi(\nu_C-3)/4$~
\\ \hline
$\mathbf{2}$  & ~$\pi(\nu_C-1)/4$~ & ~$\mathbf{6}$~ & ~$\pi(\nu_C+1)/4$~
\\ \hline
$\mathbf{3}$  & ~$\pi$~ & ~$\mathbf{7}$~ & ~$-\pi/2$~
\\ \hline
$\mathbf{4}$  & ~$0$~ & ~$\mathbf{8}$~ & ~$\pi/2$~
\\ \hline
\end{tabular}
\caption{Statistical phase $\phi_s$ for the transitions between different superselection sectors as shown in Fig.~\ref{fig:MZ_Abelian_e/4}. Here, 
$\nu_C\equiv 0~(\text{mod }4)$.}
\label{tab:phase_MZ_Abelian_0}
\end{center}
\end{table}

\begin{table} [htb] 
\begin{center}
\begin{tabular}{| c | c ||  c | c  |}
\hline
~~index ~~ & ~~ $\phi_s$ ~~ & ~~ index ~~ & ~~ $\phi_s$ ~~ 
\\ \hline\hline
$\mathbf{1}$  & ~$\pi(\nu_C-1)/4$~ & ~$\mathbf{5}$~ & ~$\pi(\nu_C+1)/4$~
\\ \hline
$\mathbf{2}$  & ~$\pi(\nu_C+3)/4$~ & ~$\mathbf{6}$~ & ~$\pi(\nu_C-3)/4$~
\\ \hline
$\mathbf{3}$  & ~$\pi$~ & ~$\mathbf{7}$~ & ~$-\pi/2$~
\\ \hline
$\mathbf{4}$  & ~$0$~ & ~$\mathbf{8}$~ & ~$\pi/2$~
\\ \hline
\end{tabular}
\caption{Statistical phase $\phi_s$ for the transitions between different superselection sectors as shown in Fig.~\ref{fig:MZ_Abelian_e/4}. Here, 
$\nu_C\equiv 2~(\text{mod }4)$.}
\label{tab:phase_MZ_Abelian_2}
\end{center}
\end{table}

%\newpage

\begin{figure}[htb]
\includegraphics[width=3.0 in]{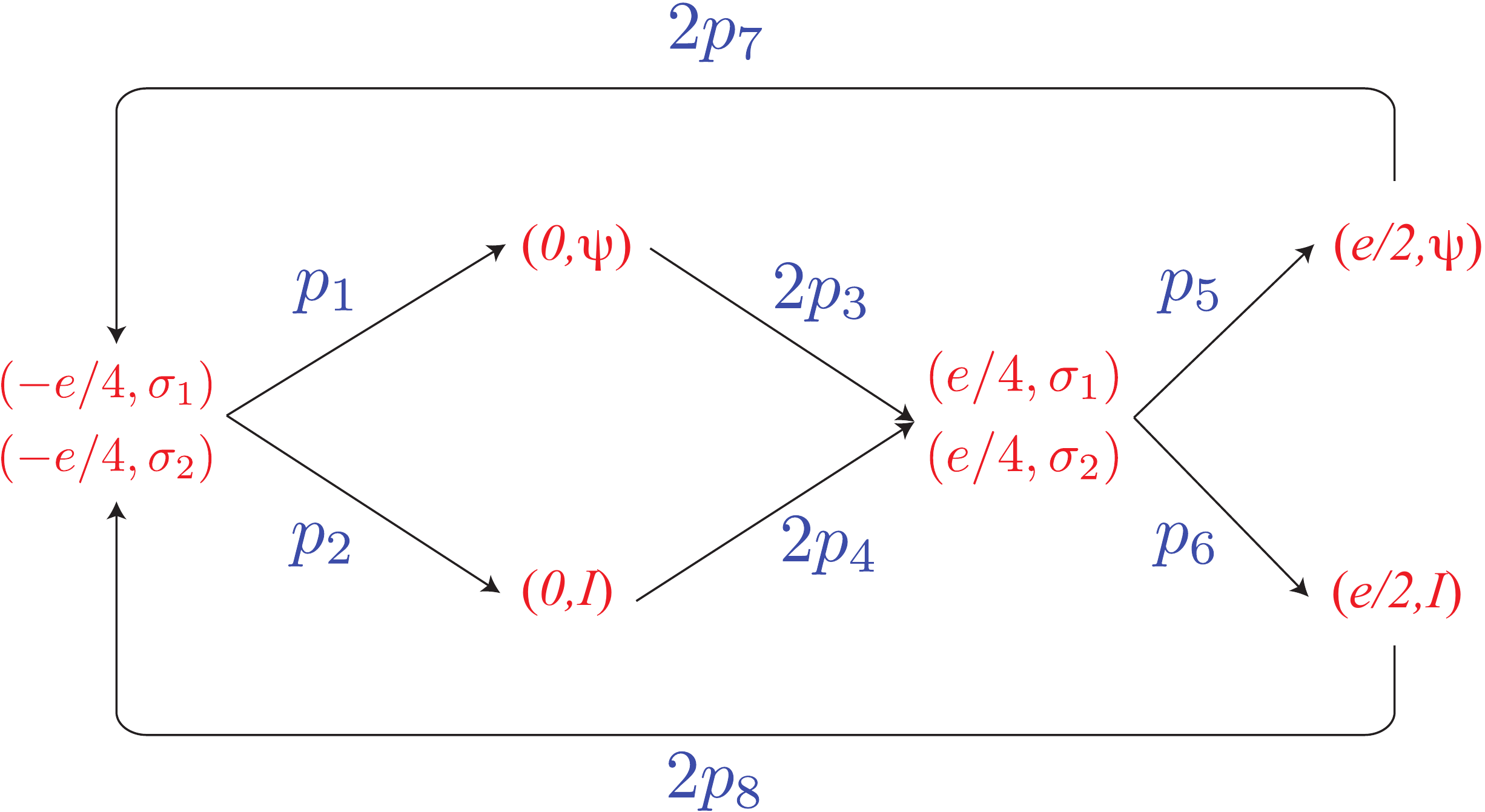}
\caption{(Color online) Transitions between superselection sectors in an Abelian system with flavor symmetry. Note a coefficient of 2 in some of the probabilities. For example, the transition rate from the $(0,I)$ sector is $2p_4$ since this is the combined tunneling rate for two quasiparticle types.} 
\label{fig:MZ-Abelian(revised)}
\end{figure}

In principle, the tunneling current and the Fano factor can be evaluated in essentially the same way as in the above subsection. To avoid unwieldy expressions, we focus on the situation with flavor symmetry of the quasiparticles. In other words, the tunneling amplitudes for the two types of $e/4$ quasiparticles at the QPCs are the same. Under this assumption, Fig.~\ref{fig:MZ_Abelian_e/4} reduces to a version of Fig.~\ref{fig:MZ_e/4}, as shown in Fig.~\ref{fig:MZ-Abelian(revised)}. Using the same technique as before, we determine the tunneling current as

\begin{widetext}
\begin{align}
\label{eq:case2_I_Abelian_e/4}
&\text{when } \nu_C=2~\text{or}-6:\quad
I_{e/4}
=\frac{er}{2}
\left[
\frac{\left(\left|\Gamma_1\right|^2+\left|\Gamma_2\right|^2\right)
\left(
1-s^2+\frac{s^4}{4}\sin^2{2\gamma}
\right)}
{1-(\frac{3}{4}-\frac{1}{4\sqrt{2}})s^2
+\frac{s^4}{16}
\left[
\left(1-\frac{1}{\sqrt{2}}\right)
\left(1-\cos{4\gamma}\right)
-\frac{1}{\sqrt{2}}\sin{4\gamma}\right]}
\right],
\\
\label{eq:case3_I_Abelian_e/4}
&\text{when } \nu_C=4~\text{or}-4:\quad
I_{e/4}
=\frac{er}{2}
\left[
\frac{\left(\left|\Gamma_1\right|^2+\left|\Gamma_2\right|^2\right)
\left(
1-s^2+\frac{s^4}{4}\sin^2{2\gamma}
\right)}
{1-(\frac{3}{4}-\frac{1}{4\sqrt{2}})s^2
+\frac{s^4}{16}
\left[
\left(1-\frac{1}{\sqrt{2}}\right)
\left(1-\cos{4\gamma}\right)
+\frac{1}{\sqrt{2}}\sin{4\gamma}\right]}
\right],
\\
\label{eq:case4_I_Abelian_e/4}
&\text{when } \nu_C=6~\text{or}-2:\quad
I_{e/4}
=\frac{er}{2}
\left[
\frac{\left(\left|\Gamma_1\right|^2+\left|\Gamma_2\right|^2\right)
\left(
1-s^2+\frac{s^4}{4}\sin^2{2\gamma}
\right)}
{1-(\frac{3}{4}+\frac{1}{4\sqrt{2}})s^2
+\frac{s^4}{16}
\left[
\left(1+\frac{1}{\sqrt{2}}\right)
\left(1-\cos{4\gamma}\right)
+\frac{1}{\sqrt{2}}\sin{4\gamma}\right]}
\right].
\end{align}
\end{widetext}
Just as in the non-Abelian case, the results are grouped into four different classes (notice the sign differences in the denominator). It is easy to verify that the cases with $\nu_C=2$ and $-2$ recover the expressions for the 331 order~\cite{Chenjie2010} and the 113 order~\cite{Guang2015}, respectively. Finally, we remark that the tunneling current retains the structure of Eqs.~\eqref{eq:tunneling_e/2: case 1} and~\eqref{eq:tunneling_e/2: case 2}, if the tunneling process is dominated by $e/2$ quasiparticles. 

The Fano factor can be calculated in the same way as before. Since the expressions are too lengthy, we do not display them here. The maximal and minimal values of the Fano factors for different Chern numbers are found numerically and are summarized in Table~\ref{tab:MZ_table_Abelian}.

\begin{table} [htb] 
\begin{center}
\begin{tabular}{| c | c |  c | c | c |}
\hline
~~$\nu_C~(\text{mod }8)$ ~~ & ~~ $e^*_{\text{max}}/e$ ~~ 
& ~~ $\gamma_{\text{max}}$~~ & ~~ $e^*_{\text{min}}/e$ ~~ 
& ~~ $\gamma_{\text{min}}$~~
\\ \hline\hline
$2$  & ~$1.39$~ & ~$0.10$~ & ~$0.393$~ & ~$0.80$~
\\ \hline
$4$  & ~$1.39$~ & ~$-0.10$~ & ~$0.393$~ & ~$-0.80$~
\\ \hline
$6$  & ~$13.5$~ & ~$-0.05$~ & ~$0.381$~ & ~$0.98$~
\\ \hline
\end{tabular}
\caption{Maximal and minimal values of the Fano factor in a Mach-Zehnder interferometer experiment when $s=1$. Here, we focus on Abelian orders for the $\nu=5/2$ FQH state with flavor symmetry and assume that the tunneling process is dominated by $e/4$-quasiparticles. The third and last columns of the table show the values of $\gamma=\phi_{\text{AB}}+\delta$ at which $e^*=e^*_{\text{max}}$ and $e^*_{\text{min}}$, respectively. Notice that these values are modulo $\pi/2$. }
\label{tab:MZ_table_Abelian}
\end{center}
\end{table}

\subsection{A special topological order: the  $K=8$ state} \label{sec:K8_state}

In contrast to other Abelian orders, the $K=8$ state is obtained by pairing two electrons into a charge-$2e$ boson. Then, the bosons condense into a Laughlin state with the filling factor of $1/8$~\cite{Wen_Zee}. In this state, single-electron excitations are gapped. There are no neutral modes, and the vertex operator for the charge-$e/4$ quasiparticle is $e^{i\phi_\rho/2}$, where 
$\phi_\rho$ is the charged mode. In contrast to all other Abelian orders in the sixteenfold way, there is only one type of $e/4$-quasiparticles in the $K=8$ state. Here, we examine its tunneling current and the Fano factor in a Mach-Zehnder experiment. 

\begin{figure}[htb]
\includegraphics[width=2.5 in]{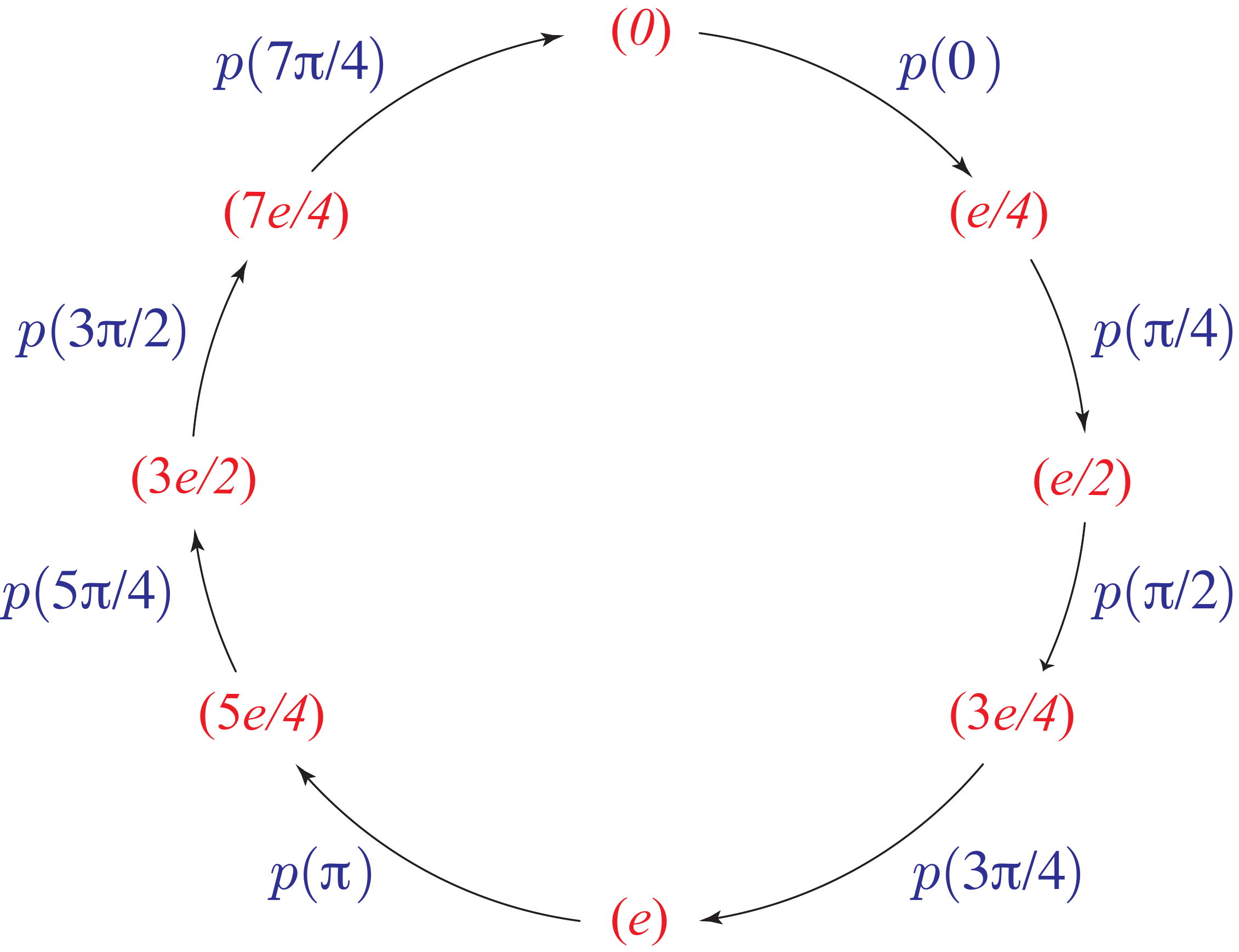}
\caption{(Color online) In the $K=8$ state,  the superselection sectors of drain D2 are described by the drain's charge modulo $2e$. The drain makes a full circle through the diagram  
after eight tunneling events of $e/4$-quasiparticles.} 
\label{fig:K8}
\end{figure}

In Fig.~\ref{fig:K8}, we show all eight possible superselection sectors for drain D2, with the corresponding transition probabilities.	 The phase accumulated by a quasiparticle, encircling the drain, equals $\phi_s=n\pi/4$, where the drain charge is $ne/4$ modulo $2e$, that is, $n=0,1,\cdots,7$. In order for drain D2 to return to its initial superselection sector, it is necessary to transfer a total charge of $2e$. The structure of the diagram resembles the simple diagram of a Laughlin state~\cite{KT2006}.

From the figure, the average time required for eight successive tunneling events of charge-$e/4$ quasiparticles is given by $\bar{t}=\sum_{n=0}^7 \frac{1}{p(n\pi/4)}$. The tunneling current is determined as $I=2e/\bar{t}$. This leads to
\begin{align}
\nonumber
I=
\frac{er}{4}(\left|\Gamma_1\right|^2+\left|\Gamma_2\right|^2)
\left[
\frac{1-2s^2+\frac{5s^4}{4}-\frac{s^6}{4}+\frac{s^8}{64}\sin^2{4\gamma}}
{\left(1-\frac{s^2}{2}\right)\left(1-s^2+\frac{s^4}{8}\right)}\right].
\\
\label{eq:current_K8}
\end{align}

The variance of $t$ can be evaluated as 
$\overline{\delta t^2}=\sum_{n=0}^7 [1/p(n\pi/4)]^2$. This yields the following Fano factor:
\begin{align}
\nonumber
\frac{e^*}{e}
=&~\frac{64-160s^2+152s^4-68s^6+\frac{29s^8}{2}
-\frac{5s^{10}}{4}+\frac{s^{12}}{16}+\frac{s^{14}}{32}}
{\left(2-s^2\right)^2\left(8-8s^2+s^4\right)^2}
\\
&+\frac{s^8\left(\frac{7}{2}-\frac{15s^2}{4}+\frac{15s^4}{16}-\frac{s^6}{32}\right)}
{\left(2-s^2\right)^2\left(8-8s^2+s^4\right)^2}
\cos{8\gamma}
\label{eq:Fano_K8}
\end{align}
When $s=1$, the Fano factor takes its maximum value, $e_{\text{max}}^*=2e$~\cite{Chenjie2010} at $\gamma=m\pi/4$. On the other hand, it assumes the minimum value
$e_{\text{min}}^*/e=11/16$ at $\gamma=\pi/8+m\pi/4$, where $m$ is an integer. Equations~\eqref{eq:current_K8} and~\eqref{eq:Fano_K8} suggest that the tunneling current and 
the Fano factor are periodic in $\Phi$ with the period of $\Phi_0/2=h/2e$. This agrees with the formation of Cooper pairs of electrons in the system~\cite{Byers-Yang}. 

\section{Summary of experimental signatures} 
\label{sec:summary}

Experimental signatures of the topological orders of the sixteenfold way are summarized in Table~\ref{tab:summary}.  The PH-Pfaffian order seems to agree best with the existing data for the $5/2$ plateau in GaAs at the electron densities $n\sim 2-3\times 10^{11}~{\rm cm}^{-2}.$   Indeed, this order possesses an upstream Majorana mode, has a tunneling exponent of $g_{e/4}=1/4<1/2$, demonstrates the even-odd effect in a Fabry-P\'{e}rot experiment, and  shows the thermal Hall conductance coefficient of $\kappa_H=2.5$, i.e. $\kappa=2.5\kappa_0$.

\begin{table*}[t]
\centering
\begin{tabular}{| l | c | c | c | c | c | c | c | c | c | c | c |}
\hline
   ~ Name ~ & ~ $\nu_C$ ~ & ~ Non-Abelian? ~ & ~$q$~ & ~~$g_{e/4}$~~ & ~~$g_{e/2}$~~ 
   & ~~$g_e$~~ & ~$\kappa_H$~ & ~Even-odd effect?~ & 
   ~$(e^*/e)_{\text{max}}$~ & ~$(e^*/e)_{\text{min}}$~
    \\ \hline\hline
    ~$K=8$~ & ~$0$~ & ~No~ & ~~$e/4$~~ & ~$\bf{1/8}$~ 
    & ~$1/2$~ & ~$\infty$~ & ~$3$~ & ~No~ & ~$2$~ & ~$11/16$~
    \\  \hline
    ~Pfaffian~ & ~$1$~ & ~Yes~ & ~$e/4$~ & ~$\bf{1/4}$~ 
    & ~$1/2$~ & ~$3$~ & ~$3.5$~ & ~Yes~ & ~$3.20$~ & ~$0.44$~
    \\  \hline
    ~PH-Pfaffian~ & ~$-1$~ & ~Yes~ & ~$e/4$~ & ~$\bf{1/4}$~ 
    & ~$1/2$~ & ~$3$~ & ~$2.5$~ & ~Yes~ & ~$\infty$~ & ~$1/2$~
    \\  \hline
    ~$331$~ & ~$2$~ & ~No~ & ~$e/4$~ & ~$\bf{3/8}$~ 
    & ~$1/2$~ & ~$3$~ & ~$4$~ & ~Maybe~ & ~$1.39$~ & ~$0.39$~
    \\  \hline
    ~$113$~ & ~$-2$~ & ~No~ & ~$e/4$~ & ~$\bf{3/8}$~ 
    & ~$1/2$~ & ~$3$~ & ~$2$~ & ~Maybe~ & ~$13.5$~ & ~$0.38$~
    \\  \hline
    ~SU(2)$_2$~ & ~$3$~ & ~Yes~ & ~$e/4$~ & ~$\bf{1/2}$~ 
    & ~$\bf{1/2}$~ & ~$3$~ & ~$4.5$~ & ~Yes~ & ~$1$~ & ~$3/8$~
    \\  \hline
    ~Anti-Pfaffian~~ & ~$-3$~ & ~Yes~ & ~$e/4$~ & ~$\bf{1/2}$~ 
    & ~$\bf{1/2}$~ & ~$3$~ & ~$1.5$~ & ~Yes~ & ~$3.20$~ & ~$0.44$~
    \\  \hline
    ~$\nu_C=4$~ & ~$4$~ & ~No~ & ~$e/4$~ & ~$5/8$~ 
    & ~$\bf{1/2}$~ & ~$3$~ & ~$5$~ & ~Maybe~ & ~$1.39$~ & ~$0.39$~
    \\  \hline
    ~Anti-331~ & ~$-4$~ & ~No~ & ~$e/4$~ & ~$5/8$~ 
    & ~$\bf{1/2}$~ & ~$3$~ & ~$1$~ & ~Maybe~ & ~$1.39$~ & ~$0.39$~
    \\ \hline
    ~$\nu_C=5$~ & ~$5$~ & ~Yes~ & ~$e/4$~ & ~$3/4$~ 
    & ~$\bf{1/2}$~ & ~$3$~ & ~$5.5$~& ~Yes~ & ~$3.20$~ & ~$0.44$~
    \\ \hline
    ~Anti-SU(2)$_2$~ & ~$-5$~ & ~Yes~ & ~$e/4$~ & ~$3/4$~ 
    & ~$\bf{1/2}$~ & ~$3$~ & ~$0.5$~ & ~Yes~ & ~$1$~ & ~$3/8$~
    \\ \hline
    ~$\nu_C=6$~ & ~$6$~ & ~No~ & ~$e/4$~ & ~$7/8$~ 
    & ~$\bf{1/2}$~ & ~$3$~ & ~$6$~ & ~Maybe~ & ~$13.5$~ & ~$0.38$~
    \\ \hline
     ~$\nu_C=-6$~ & ~$-6$~ & ~No~ & ~$e/4$~ & ~$7/8$~ 
    & ~$\bf{1/2}$~ & ~$3$~ & ~$0$~ & ~Maybe~ & ~$1.39$~ & ~$0.39$~
    \\ \hline
    ~$\nu_C=7$~ & ~$7$~ & ~Yes~ & ~$e/4$~ & ~$1$~ 
    & ~$\bf{1/2}$~ & ~$3$~ & ~$6.5$~ & ~Yes~ & ~$1$~ & ~$1/2$~
    \\ \hline
    ~$\nu_C=-7$~ & ~$-7$~ & ~Yes~ & ~$e/4$~ & ~$1$~ 
    & ~$\bf{1/2}$~ & ~$3$~ & ~$-0.5$~ & ~Yes~ & ~$1$~ & ~$1/2$~
    \\ \hline
     ~$\nu_C=8$~ & ~$8$~ & ~No~ & ~$e/4$~ & ~$9/8$~ 
    & ~$\bf{1/2}$~ & ~$3$~ & ~$7$~ & ~Maybe~ & ~$1$~ & ~$1/2$~
    \\ \hline
\end{tabular}
\caption{Experimental signatures of different topological orders in the sixteenfold way. The second column provides the Chern number of the edge, which should equal the Chern number of the bulk. A non-Abelian (Abelian) order has an odd (even) Chern number. The states with negative $\nu_C$ have upstream neutral modes. For all topological orders, the most fundamental quasiparticle has the charge $q=e/4$. The fifth to seventh columns give the universal tunneling exponents for different types of quasiparticles, with the most relevant one being boldfaced. The eighth column provides the thermal Hall conductance coefficients, which are half-integers (integers) for non-Abelian orders (Abelian orders). In the last three columns, we list the expected results from interferometry.
The bottom three entries in the last two columns refer to the situation in which the dominant process is $e/2$-tunneling. Everywhere else, we assume that the dominant quasiparticle has the charge $e/4$. All non-Abelian orders should demonstrate even-odd effect in a Fabry-P\'{e}rot interferometer. Abelian orders (except the $K=8$ state) may also show the same effect, if they possess flavor symmetry. The last two columns list the maximal and minimal values of the Fano factor in a shot-noise experiment with a symmetric Mach-Zehnder interferometer 
($s=1$). 
%Renormalization group analysis predicts that the tunneling amplitude sacles with temperature as $T^{1-g}$, so the $e/4$-quasiparticle tunneling becomes marginal at $\nu_C=\pm 7$ and %irrelevant for $\nu_C=8$. For these three cases, the tunneling process across the quantum-point contacts should be dominated by the $e/2$ quasiparticles.
}
\label{tab:summary}
\end{table*}

%----------------------- Coupled-stripe construction ------------------------------

\section{Iterative Coupled quantum-Hall-stripes construction} 
\label{sec:CW_construction}

Effective Hamiltonians for different fractional quantum Hall states have been designed with coupled-wire constructions in Refs.~\cite{Sondhi_Yang, Kane_CW, Teo-Kane, Kane-Stern-Halperin, Fuji-Furusaki}. Motivated by the mother-daughter relations from Sec.~\ref{sec:5/2}, we propose an iterative construction of effective Hamiltonians for all orders in the sixteenfold way. In contrast to the previous work, we start with a collection of quantum Hall stripes and not wires (cf. Refs. \cite{BMF2015, MRF2016, WY2016}). We choose one of the 16 orders and assume that the ground state of the bulk Hamiltonian of each stripe has the chosen order. Such Hamiltonian is well known for the Pfaffian order
 \cite{Hamiltonian1, Greiter1992}. Thus, we choose the Pfaffian order as our starting point in the following discussion. At the same time, all other orders can be used as a starting point. 

We consider a large number of parallel stripes.
The stripes host gapped QHE liquids in the bulk. In the absence of interaction between the stripes, they have gapless edge modes: charged and Majorana. We choose the inter-stripe interaction that gaps those edge modes out. Indeed, our goal is to generate a system, in which gapless edge modes are confined to its uppermost and downmost parts. 

%The charged modes and the Majorana modes in the stripes are gapped out by introducing suitable coupling between the stripes. The end result is that only a gapless charged mode and a %brunch of gapless Majorana modes are left at the edge of the system. 

To demonstrate our idea, we start with constructing the effective Hamiltonian of the PH-Pfaffian state. This example provides a template for neutral-mode flipping in our coupled-stripe construction.

\subsection{From Pfaffian to PH-Pfaffian}
\label{sec:Pf_PH-Pf}

\begin{figure}[htb]
\includegraphics[width=3.2 in]{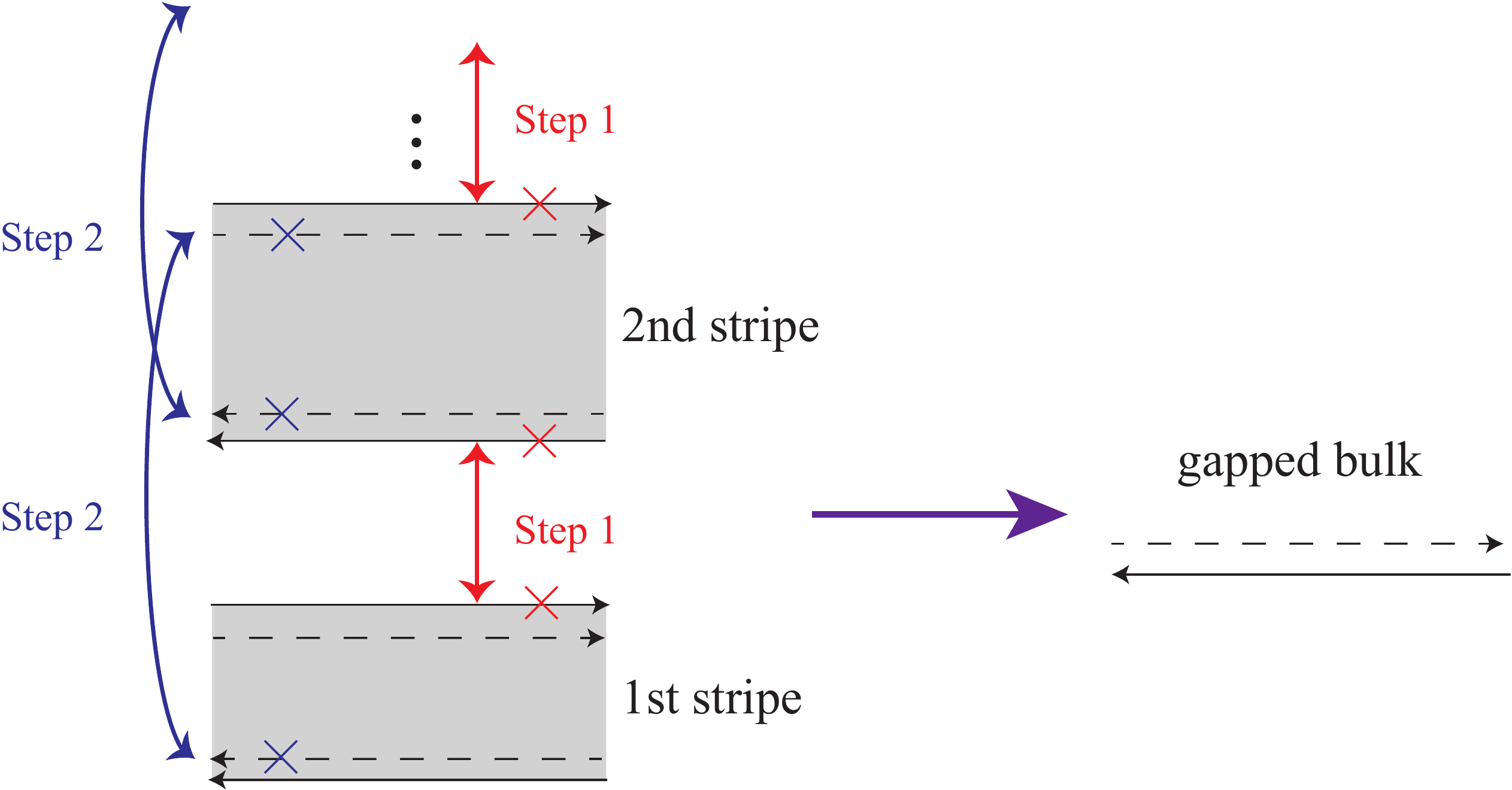}
\caption{(Color online) Coupled-stripe construction of the effective Hamiltonian for the PH-Pfaffian order from a collection of quantum Hall stripes in the Pfaffian state. The solid lines (the dashed lines) refer to the charged bosonic modes (Majorana fermions). The lines with arrows on both ends represent inter-stripe interactions which gap out 
the pairs of the modes, connected by those lines.}
\label{fig:PH-Pf_Pf}
\end{figure}

Consider a system of quantum Hall stripes in the Pfaffian state as illustrated in Fig.~\ref{fig:PH-Pf_Pf}. First, assume no inter-stripe interaction. The Hamiltonian density of the gapless edge channels of the  decoupled system is given by
\begin{align}
\nonumber
\mathcal{H}_0
=\frac{2v_{\rho}}{4\pi}\sum_{j=1}^{\infty}
\left[\left(\partial_x\phi_{\rho,j,L}\right)^2+\left(\partial_x\phi_{\rho,j,R}\right)^2\right]
\\
+iv_{\psi}\sum_{j=1}^{\infty}
\left(\psi_{j,L}\partial_x \psi_{j,L}
-\psi_{j,R}\partial_x \psi_{j,R}\right).
\end{align}
Here, $v_\rho$ and $v_\psi$ denote the speeds of the charged mode $\phi_{\rho}$ and the Majorana mode $\psi$, respectively. The subscripts $L$ and $R$ label the left and right chiralities of the modes. The integer index $j$ labels the quantum Hall stripes. The commutation relations of the Majorana fermions are 
$\{\psi_{j,D}(x),\psi_{i,D'}(x')\}=\frac{1}{2}\delta(x-x')\delta_{ij}\delta_{DD'}$, where $D$ and $D'$ can be $R$ and $L$.

Let us summarize the idea of the construction. Recall that the PH-Pfaffian and Pfaffian states are related by neutral-mode flipping. As shown in Fig.~\ref{fig:PH-Pf_Pf}, the couplings between the stripes gap out pairs of modes and leave a single gapless charged mode and a single gapless Majorana mode at the edge of the system. Furthermore, this Majorana mode has the opposite chirality to that of the gapless boson. Thus, the gapless edge acquires the structure demanded by the PH-Pfaffian order. Hence, an effective Hamiltonian for the PH-Pfaffian
order is constructed.

Explicitly, we first gap out charged modes by introducing {\it electron-pair} tunneling between neighboring stripes (step 1 in Fig.~\ref{fig:PH-Pf_Pf}). The coupling is described by the following Hamiltonian density:
\begin{align} \label{eq:gap_charge}
\nonumber
\mathcal{H}_1
&=\Gamma_1
\sum_{j=1}^{\infty}
\left(\psi_{j,R} e^{2i\phi_{\rho, j, R}}\right)^2
\left(\psi_{j+1,L} e^{-2i\phi_{\rho, j+1, L}}\right)^2
+\text{H.c.}
\\
&=~2\Gamma_1
\sum_{j=1}^{\infty}
\cos{\left(4\phi_{\rho, j, R}-4\phi_{\rho, j+1, L}\right)}.
\end{align}
Here, we have used the property that $\psi^2$ is a $c$-number. This number is dimensional, so, strictly speaking, the constants $\Gamma_1$ are not identical on the two sides of Eq. 
(\ref{eq:gap_charge}). This minor point is of no importance below. Note that we can add a density-density interaction $w\partial_x\phi_{\rho, j, R}\partial_x\phi_{\rho, j+1, L}$ that makes the tunneling (\ref{eq:gap_charge}) relevant in the renormalization group sense. As always with coupled-wire constructions, it is essential that the arguments commute for any two cosines (or any one cosine in different points) in Eq. (\ref{eq:gap_charge}):

\begin{align}
[4(\phi_{\rho, j, R}(x)-\phi_{\rho, j+1, L}(x)),4(\phi_{\rho, i, R}(y)-\phi_{\rho, i+1, L}(y))]\nonumber\\
=0
\end{align}
for any $i,~j,~x,$ and $y$. As a consequence, it may be legitimate to treat the arguments of the cosines as $c$-numbers.

When a negative $\Gamma_1$ is sufficiently large, the combination $4\phi_{\rho, j, R}-4\phi_{\rho, j+1, L}$  is pinned to a multiple of $2\pi$. This gaps out the modes $\phi_{\rho, j, R}$  and $\phi_{\rho, j+1, L}$. 
Only $\phi_{\rho,1, L}$ is not coupled with a right-moving mode by the above tunneling operator and hence $\phi_{\rho,1, L}$ remains gapless. Next, the Majorana modes are gapped out by the following coupling (step 2 in Fig.~\ref{fig:PH-Pf_Pf}):
\begin{align} \label{eq:gap_Majorana}
\nonumber
\mathcal{H}_2
&=\Gamma_2
\sum_{j=1}^{\infty}
\left(\psi_{j,L} e^{2i\phi_{\rho, j, R}}\right)
\left(\psi_{j+1,R} e^{-2i\phi_{\rho, j+1, L}}\right)
+\text{H.c.}
\\
&=\Gamma_2
\sum_{j=1}^{\infty}
\psi_{j,L}\psi_{j+1,R}
e^{2i\left(\phi_{\rho, j, R}-\phi_{\rho, j+1, L}\right)}
+\text{H.c.}
\end{align}
We must explain why such coupling is legitimate. Two requirements must be satisfied. First, the interaction must conserve the electric charge as it obviously does. Second, it should conserve the topological charge. To understand why the second condition is satisfied, observe that the above tunneling interaction consists of products of operators of the type
$\psi_{j,L} \exp(2i\phi_{\rho, j, R})=\hat A\hat B$ with $\hat A=\psi_{j,L}\exp(2i\phi_{\rho, j, L})$ and $\hat B=\exp(2i\phi_{\rho, j, R}-2i\phi_{\rho, j, L})$. $\hat A$ is a topologically trivial electron operator. $\hat B$ transfers one electron charge between the two sides of a single stripe and hence is also topologically trivial. Hence, the product of
$\hat A$ and $\hat B$ also conserves the topological charge, as does the interaction (\ref{eq:gap_Majorana}).

%Here, we combined a charged mode and a Majorana mode with opposite chirality to form an electron operator. This is legitimate as the topological spins for $\psi_L$ and $\psi_R$ satisfy %$e^{2\pi i/2}=e^{-2\pi i/2}$, which indicates that they have the same topological property. 

At this point, we observe that the combination of the charged modes $4\phi_{\rho, j, R}-4\phi_{\rho, j+1, L}$ was fixed to be a multiple of $2\pi$ at the first step. Hence, the exponential factor $\exp(2i\left(\phi_{\rho, j, R}-\phi_{\rho, j+1, L}\right))$ in Eq.~\eqref{eq:gap_Majorana} is $\pm 1$. As a consequence, $\mathcal{H}_2$ can be simplified into
\begin{eqnarray}
\mathcal{H}_2
=\tilde{\Gamma}_2
\sum_{j=1}^{\infty}
\psi_{j,L}\psi_{j+1,R}
+\text{H.c.},
\end{eqnarray}
where the $\pm$ sign factor is absorbed into $\tilde{\Gamma}_2$. To make sure that $\tilde{\Gamma}_2$ is the same for all $j$, one may also need to redefine 
$\psi_{j,L}\rightarrow -\psi_{j,L}$.

The overall Hamiltonian density of the coupled system 
$\mathcal{H}=\mathcal{H}_0+\mathcal{H}_1+\mathcal{H}_2$ can be separated into the bulk and edge parts: $\mathcal{H}=\mathcal{H}_b+\mathcal{H}_e$. The gapped bulk contribution is
\begin{align} \label{eq:bulk_H}
\nonumber
\mathcal{H}_b
=&\frac{4v_\rho}{4\pi}\sum_{j=1}^{\infty} 
\left[\left(\partial_x\theta_j\right)^2+\left(\partial_x\varphi_j\right)^2\right]
+2\Gamma_1\sum_{j=1}^\infty \cos{(8\varphi_j)}
\\ \nonumber
+&iv_{\psi}\sum_{j=1}^{\infty}
\left(\psi_{j,L}\partial_x \psi_{j,L}-\psi_{j+1,R}\partial_x \psi_{j+1,R}\right)
\\
+&\tilde{\Gamma}_2\sum_{j=1}^\infty \psi_{j,L}\psi_{j+1,R}+\text{H.c.},
\end{align}
where
\begin{align}
\theta_j=(\phi_{\rho,j,R}+\phi_{\rho,j+1,L})/2,
\\ 
\varphi_j=(\phi_{\rho,j,R}-\phi_{\rho,j+1,L})/2.
\end{align}
The edge contribution $\mathcal{H}_e$ is gapless.

To verify that the bulk is gapped, we need to check that the Majorana modes in Eq. (\ref{eq:bulk_H}) are gapped out. We expand the Majorana operators as superpositions of plane waves:
\begin{align}
\psi_{j,L}(x)
=\frac{1}{\sqrt{L}}\sum_k a_{j,k} e^{ikx},
\\
\psi_{j,R}(x)
=\frac{1}{\sqrt{L}}\sum_k \tilde{a}_{j,k} e^{ikx},
\end{align}  
where $L$ is the length of the stripes.
The condition $\psi(x)=\psi^{\dagger}(x)$ implies that 
$a_{j,-k}=a^{\dagger}_{j,k}$.
The anti-commutation relations for $a_{j,k}$ and 
$\tilde{a}_{j,k}$ are 
\begin{align}
\{a_{i,k}, a^{\dagger}_{j,k'}\}
&=\frac{1}{2}\delta_{i,j}\delta_{k,k'}
\\
\{\tilde{a}_{i,k}, \tilde{a}^{\dagger}_{j,k'}\}
&=\frac{1}{2}\delta_{i,j}\delta_{k,k'}
\\
\{a_{i,k}, \tilde{a}^{\dagger}_{j,k'}\}
&=0
\end{align}
The Hamiltonian of the bulk Majorana degrees of freedom is given by the integral 
$H_\psi=\int_0^L dx \mathcal{H}_{b\psi}$, where $\mathcal{H}_{b\psi}$ is the sum of the last two rows in Eq.~\eqref{eq:bulk_H}. With the new notation $a_{j,k}$, $\tilde a_{j,k}$, the  Hamiltonian $H_{\psi}$ can be rewritten as 
\begin{align} 
\nonumber
H_{\psi}
=&~2v_{\psi}\sum_{j=1}^{\infty}
\sum_{k>0} k \left(\tilde{a}^{\dagger}_{j+1,k}\tilde{a}_{j+1,k}
-a_{j,k}^{\dagger}a_{j,k}\right)
\\
&+\left[\tilde{\Gamma}_2\sum_{j=1}^\infty
\sum_{k>0} \left(a^{\dagger}_{j,k}\tilde{a}_{j+1,k} 
-\tilde{a}^{\dagger}_{j+1,k} a_{j,k}\right)
+\text{H.c.}\right].
\end{align}
Then, $H_\psi$ can be diagonalized by the following transformation:
\begin{align}
c_{j,k}
&=\frac{(\lambda-v_\psi k)a_{j,k}+i~\text{Im}(\tilde{\Gamma}_2)\tilde{a}_{j+1,k}}
{\sqrt{[\text{Im}(\tilde{\Gamma}_2)]^2+(v_\psi k-\lambda)^2}},
\\
d_{j,k}
&=\frac{i~\text{Im}(\tilde{\Gamma}_2)a_{j,k}+(\lambda-v_\psi k)\tilde{a}_{j+1,k}}
{\sqrt{[\text{Im}(\tilde{\Gamma}_2)]^2+(v_\psi k-\lambda)^2}},
\end{align}
where $\lambda=\sqrt{(v_\psi k)^2+[\text{Im}(\tilde{\Gamma}_2)]^2}$. The anti-commutation relations for $c_{j,k}$ and $d_{j,k}$ are the same as the relations for $a_{j,k}$ and 
$\tilde{a}_{j,k}$. The above transformation leads to the following diagonalized $H_\psi$:
\begin{align} 
H_\psi
=2\sum_{j=1}^\infty \sum_{k>0}
\sqrt{v_\psi^2 k^2+[\text{Im}(\tilde{\Gamma}_2)]^2}
~(c^{\dagger}_{j,k}c_{j,k}-d^{\dagger}_{j,k}d_{j,k}).
\end{align}
It is now evident that as long as $\text{Im}(\tilde{\Gamma}_2)\neq 0$, the Majorana modes are gapped with the gap of $|\text{Im}(\tilde{\Gamma}_2)|$.

The bulk Hamiltonian is thus gapped: 
\begin{align}
\nonumber
&\int dx \mathcal{H}_b
\\ \nonumber
=&~ \frac{4v_\rho}{4\pi}\int dx\sum_{j=1}^{\infty} 
\left[\left(\partial_x\theta_j\right)^2+\left(\partial_x\varphi_j\right)^2\right]
\\
\nonumber
&+ 2\Gamma_1\int dx\sum_{j=1}^\infty \cos{(8\varphi_j)}
\\
&+2\sum_{j=1}^\infty \sum_{k>0}
\sqrt{v_\psi^2 k^2+[\text{Im}(\tilde{\Gamma}_2)]^2}
~(c^{\dagger}_{j,k}c_{j,k}-d^{\dagger}_{j,k}d_{j,k}).
\end{align}
At the same time, the modes $\phi_{\rho, 1, L}$ and $\psi_{1,R}$ remain gapless and are described by the edge Hamiltonian  $\int dx\mathcal{H}_e$,
\begin{align}
\mathcal{H}_e
=\frac{2v_\rho}{4\pi}\left(\partial_x\phi_{\rho,1,L}\right)^2
-iv_\psi \psi_{1,R}\partial_x \psi_{1,R}.
\end{align}
This is the edge theory of the PH-Pfaffian order. The electron operator

\begin{align}
\Psi_e\sim&\exp(2i\phi_{\rho,1,L})\psi_{1,L}[\psi_{1,L}\psi_{1,R}]
\nonumber\\
&\sim\exp(2i\phi_{\rho,1,L})\psi_{1,R}.
\end{align}

\subsection{First coupled-stripe construction (CW1) for non-Abelian topological orders} \label{sec:nA_neutral}

The previous construction can be generalized to relate other non-Abelian topological orders which possess neutral bosonic modes, 
or, equivalently, more than one Majorana mode at the edge. Below, we will use the language of a single Majorana edge mode.
%The Hall stripes are assumed to be in the disorder-dominated phase~\cite{Kane-disorder-dominated}. 
The $K$ matrices, describing the Abelian edge modes, take the form \eqref{eq:diagonalized_K} with the corresponding charge vector $t=(1,0,\cdots,0)^T$. 

We are going to introduce coupled-stripe constructions of two types. The first construction describes neutral-mode flipping. The second construction describes particle-hole conjugation. 
We will call these two constructions CW1. A different approach CW2 to the coupled-stripe construction will be considered in the next subsection. 

\subsubsection{Effective coupled-stripe construction for neutral-mode flipping}

Our goal is to transform a system with the Chern number $-\nu_C<0$ into a system with the opposite Chern number $\nu_C$.

\begin{figure}[htb]
\includegraphics[width=3.3 in]{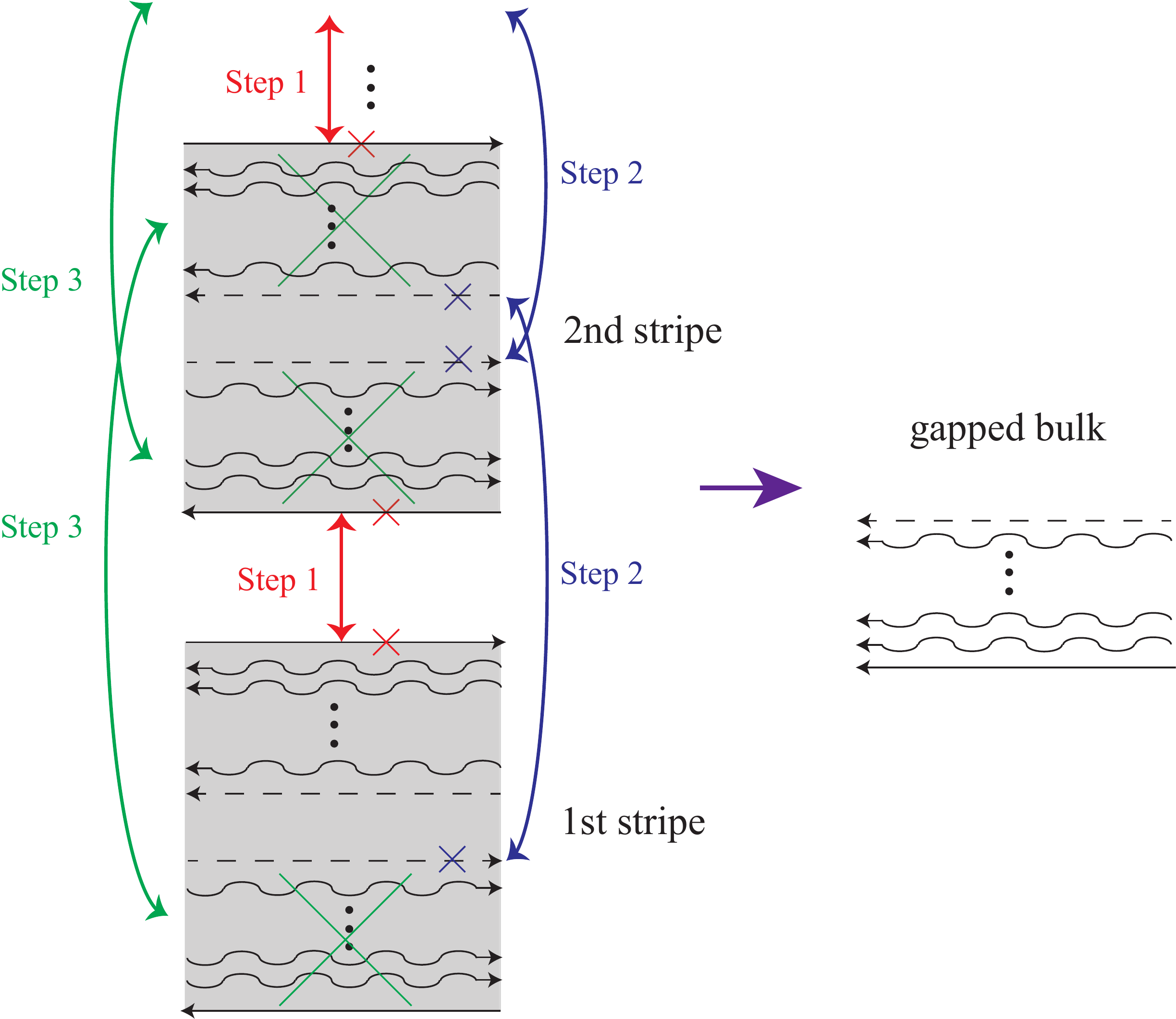}
\caption{(Color online)  Coupled-stripe construction for neutral-mode flipping in non-Abelian topological orders. The charged modes, the Majorana modes, and the bosonic neutral modes are denoted by solid lines, dashed lines, and wavy lines, respectively. Three different tunneling processes between neighboring stripes are introduced to gap out the modes in the bulk of the system.  The remaining gapless edge modes are shown in the right panel. %The final gapless edge structure has a Chern number $\nu_C$.
}
\label{fig:CW1_non_Abelian}
\end{figure}

As shown in Fig.~\ref{fig:CW1_non_Abelian}, we start with a system of decoupled quantum Hall stripes. Each edge of each stripe contains a single downstream charged mode, one upstream Majorana mode, and $N$ upstream bosonic neutral modes so that $-\nu_C=-(2N+1)$. The velocities of all upstream modes are the same.  By introducing electron tunneling processes between neighboring stripes, we gap out pairs of the modes. At the end, gapless modes remain only at the topmost and bottommost edges of the system of the stripes. The structure of the gapless modes corresponds to the desired Chern number $\nu_C$. 

The Hamiltonian density of decoupled stripes with no interstripe tunneling is given by
\begin{align} 
\label{eq:dima-decoupled}
\nonumber
\mathcal{H}_0
=&\frac{2v_{\rho}}{4\pi}\sum_{j=1}^{\infty}
\left[\left(\partial_x\phi_{\rho,j,L}\right)^2+\left(\partial_x\phi_{\rho,j,R}\right)^2\right]
\\ \nonumber
&+\frac{{v}_n}{4\pi}\sum_{j=1}^{\infty}\sum_{i=1}^N
\left[\left(\partial_x\phi_{n_i,j,L}\right)^2+\left(\partial_x\phi_{n_i,j,R}\right)^2\right]
\\
&+i{v}_n\sum_{j=1}^{\infty}
\left(\psi_{j,L}\partial_x \psi_{j,L}
-\psi_{j,R}\partial_x \psi_{j,R}\right),
\end{align}
where $v_\rho$ labels the speed of the charged mode, ${v}_n$ is the speed of the neutral modes. The sub-subscript $i$ in 
$\phi_{n_i}$ ranges from $1$ to $N$ and enumerates the $N$ neutral bosonic modes at the edge of each stripe. 
%They have the same speed $\bar{v}_n$ in the disorder-dominated phase~\cite{Kane-disorder-dominated}. 

As in Sec.~\ref{sec:Pf_PH-Pf}, the charged modes are gapped out by introducing electron-pair tunneling between neighboring stripes (step 1 in Fig.~\ref{fig:CW1_non_Abelian}):
\begin{align}
\nonumber
\mathcal{H}_1
&=\Gamma_1
\sum_{j=1}^{\infty}
\left(\psi_{j,L} e^{2i\phi_{\rho, j, R}}\right)^2
\left(\psi_{j+1,R}e^{-2i\phi_{\rho, j+1, L}}\right)^2
+\text{H.c.}
\\
&=2\Gamma_1
\sum_{j=1}^{\infty}
\cos{\left(4\phi_{\rho, j, R}-4\phi_{\rho, j+1, L}\right)}.
\end{align}
Notice that the Majorana mode and the charged mode in the electron operator have opposite chiralities as the topological order has a negative Chern number (for example, PH-Pfaffian or anti-Pfaffian). The coupling constant $\Gamma_1$ is set to a sufficiently large negative number to make sure that the charged modes are gapped in the bulk of the system. 

Next, we proceed to gap out the Majorana modes in the bulk by adding single-electron tunneling  (step 2 in Fig.~\ref{fig:CW1_non_Abelian}):
\begin{align}
\label{eq:dima-Majorana-gap}
\nonumber
\mathcal{H}_2
&=\Gamma_2
\sum_{j=1}^{\infty}
\left(\psi_{j,R} e^{2i\phi_{\rho, j, R}}\right)
\left(\psi_{j+1,L} e^{-2i\phi_{\rho, j+1, L}}\right)
+\text{H.c.}
\\
&=\tilde{\Gamma}_2
\sum_{j=1}^{\infty}
\psi_{j,R}\psi_{j+1,L} 
+\text{H.c.}
\end{align}

The neutral bosonic modes in the bulk can be gapped by an additional inter-stripe tunneling as shown as step 3 in Fig.~\ref{fig:CW1_non_Abelian}. Recall that $\Psi_e=e^{-i\phi_{n_i}}e^{2i\phi_{\rho}}$ is a legitimate electron operator for any $i=1,2,\cdots, N$. 
%Furthermore, the left-moving and the right-moving neutral modes have the same topological property because their topological spins satisfy $e^{2\pi i(1/2)}=e^{2\pi i(-1/2)}$.
Thus, by analogy with Eq. (\ref{eq:dima-Majorana-gap}), one can consider the following electron tunneling process:
\begin{align}
\nonumber
\mathcal{H}_3
&=\Gamma_3
\sum_{j=1}^{\infty}
\sum_{i=1}^N
\left(e^{-i\phi_{n_i, j, R}} e^{2i\phi_{\rho, j, R}}\right)
\\ \nonumber
&\quad\quad\quad\quad\quad\quad
\times\left(e^{i\phi_{n_i, j+1, L}} e^{-2i\phi_{\rho, j+1, L}}\right)
+\text{H.c.}
\\
&=2\tilde{\Gamma}_3
\sum_{j=1}^{\infty}
\sum_{i=1}^N
\cos{(\phi_{n_i, j+1, L}-\phi_{n_i, j, R})}.
\label{eq:gap_neutral_modes}
\end{align}

All modes in the coupled stripes are completely gapped out by the above three tunneling processes, except for the modes which do not appear in $\mathcal{H}_1$, $\mathcal{H}_2$, and 
$\mathcal{H}_3$. As a result, the effective Hamiltonian for the gapped bulk is given by
\begin{align}
\nonumber
H_b
&=\int \mathcal{H}_b~dx
\\ \nonumber
&=~\frac{4v_{\rho}}{4\pi}
\int \sum_{j=1}^{\infty}
\left[\left(\partial_x \theta_{\rho, j}\right)^2
+\left(\partial_x \varphi_{\rho, j}\right)^2\right]~dx
\\ \nonumber
&~~ +2\Gamma_1 \int \sum_{j=1}^\infty \cos{(8\varphi_{\rho,j})}~dx
\\ \nonumber
&~~ +\frac{2v_n}{4\pi}
\int \sum_{j=1}^{\infty} \sum_{i=1}^N
\left[\left(\partial_x \theta_{n_i, j}\right)^2
+\left(\partial_x \varphi_{n_i, j}\right)^2\right]~dx
\\ \nonumber
&~~ +2\tilde{\Gamma}_3 
\int\sum_{j=1}^\infty\sum_{i=1}^N \cos{(2\varphi_{n_i,j})}~dx
\\
&~~ +2\sum_{j=1}^{\infty}\sum_{k>0}
\sqrt{k^2 v_n^2+\text{Im}(\tilde{\Gamma}_2)^2}
\left(c_{j,k}^{\dagger}c_{j,k}-
d_{j,k}^{\dagger}d_{j,k}\right),
\end{align}
where
\begin{align}
\varphi_{n_i,j}=(\phi_{n_i,j+1,L}-\phi_{n_i,j,R})/2;\\
\theta_{n_i,j}=(\phi_{n_i,j+1,L}+\phi_{n_i,j,R})/2;\\
\varphi_{\rho,j}=(\phi_{\rho,j+1,L}-\phi_{\rho,j,R})/2;\\
\theta_{\rho,j}=(\phi_{\rho,j+1,L}+\phi_{\rho,j,R})/2.
\end{align}

The Hamiltonian density of the gapless edge at the bottom of the system of the stripes is
\begin{align}
\nonumber
\mathcal{H}_e
=&\frac{2v_{\rho}}{4\pi}
\left(\partial_x \phi_{\rho, 1, L}\right)^2
+\frac{v_n}{4\pi}
\sum_{i=1}^N 
\left(\partial_x \phi_{n_i, 1, L}\right)^2
\\
&+i v_n\psi_{1,L}\partial_x \psi_{1,L}.
\end{align}
The chirality of the gapless neutral modes at the edge is opposite to that of the neutral modes in the original state. Hence, the topological orders with
the Chern numbers $\nu_C$ and $-\nu_C$ can be related by the above coupled-stripe construction. 
This relationship is illustrated by horizontal arrows in Fig.~\ref{fig:non_Abelian}.
The electron operators on the edge 

\begin{align}
\Psi_\psi\sim&\exp(2i\phi_{\rho, 1, L})\psi_{1,R}[\psi_{1,R}\psi_{1,L}] \nonumber\\
&\sim\exp(2i\phi_{\rho, 1, L})\psi_{1,L};\\
\Psi_n\sim&\exp(2i\phi_{\rho, 1, L})\exp(i\phi_{n_i, 1, R})
\nonumber\\
&\times[\exp(-i\phi_{n_i, 1, R})\exp(i\phi_{n_i, 1, L})] \nonumber\\
&\sim\exp(2i\phi_{\rho, 1, L})\exp(i\phi_{n_i, 1, L}).
\end{align}

\subsubsection{Effective coupled-stripe construction for particle-hole conjugation}
\label{sec:PH_conjugation_CW}

\begin{figure}[htb]
\includegraphics[width=2.8 in]{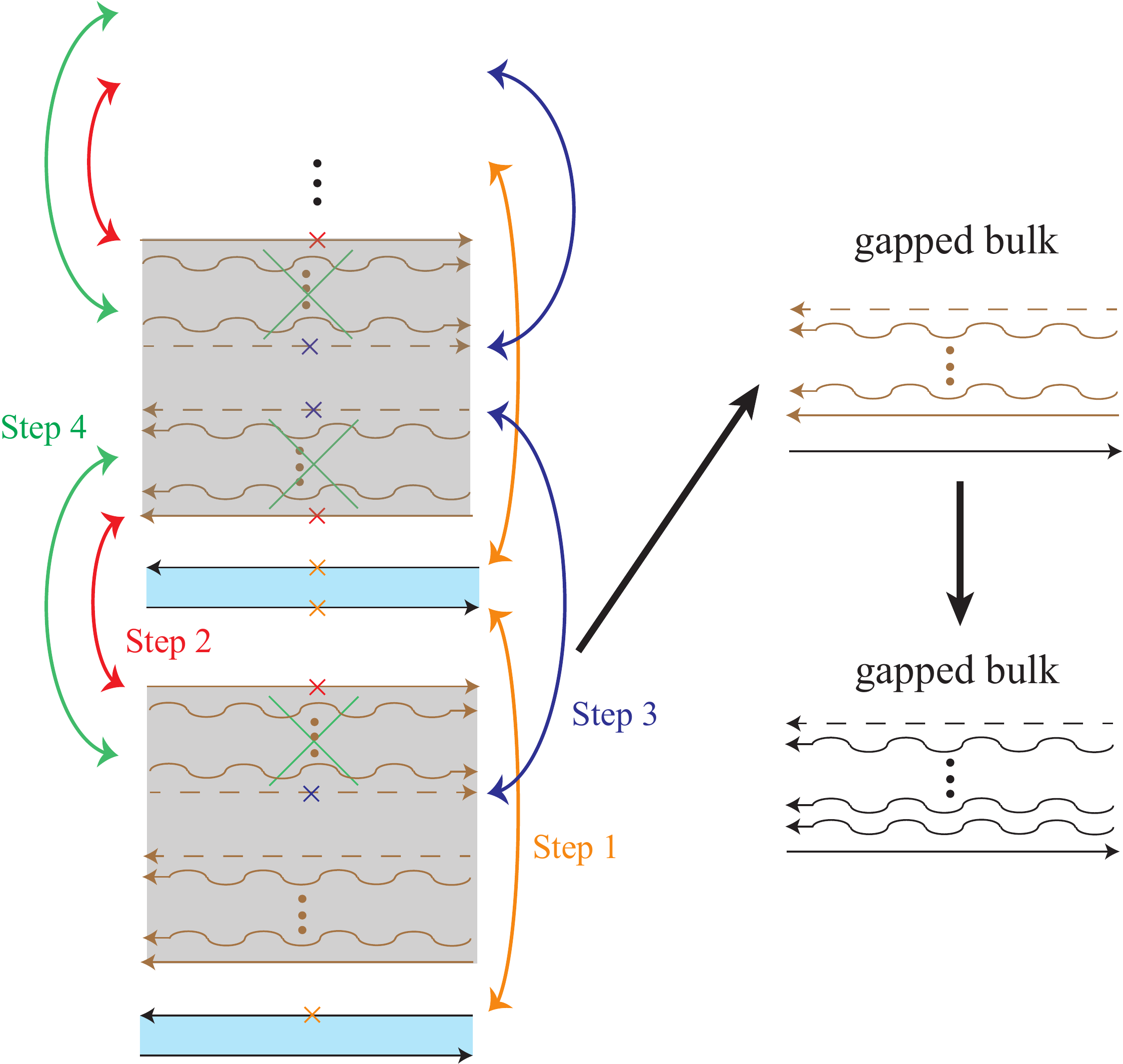}
\caption{(Color online) Effective coupled-stripe construction for particle-hole conjugation of a topological order with filling factor $\nu=1/2$. In the left panel, the system consists of $\nu=1$ integer quantum Hall stripes (narrow) and $\nu=1/2$ fractional quantum Hall stripes (wide), arranged in an alternating pattern. The upper right panel shows the resulting gapless edge structure after turning on tunneling between the stripes. The lower right panel is the edge structure in the presence of a density-density interaction on the edge.}
\label{fig:PH_conjugation}
\end{figure}

Another connection among the orders in the sixteenfold way is particle-hole conjugation. For example, the Pfaffian and anti-Pfaffian orders are particle-hole conjugates.
As shown in Fig.~\ref{fig:PH_conjugation}, we consider a collection of alternating stripes in the $\nu=1$ IQH state and in the $\nu=1/2$ FQH state to formulate a coupled-stripe construction for particle-hole conjugation. 

We begin by gapping out the modes from the $\nu=1$ IQH stripes with the following electron tunneling process:
\begin{align}
\nonumber
\mathcal{H}_1
=&\frac{\Gamma_1}{2} e^{i\phi^{\nu=1}_{\rho,j,L}}e^{-i\phi^{\nu=1}_{\rho,j+1,R}}
+\text{H.c.}
\\
=&\Gamma_1\cos{\left(\phi^{\nu=1}_{\rho,j,L}-\phi^{\nu=1}_{\rho,j+1,R}\right)}.
\end{align}
Here, $\phi_{\rho,j,L/R}^{\nu=1}$ denotes the charged mode in the $j$-th $\nu=1$ stripe. As illustrated in Fig.~\ref{fig:PH_conjugation}, the coupling gaps out the modes in the bulk of our system but leaves a single gapless charged mode $\phi^{\nu=1}_{\rho,1, R}$ at the edge of the first stripe. 

The modes in the $\nu=1/2$ FQH stripes can be gapped out by coupling the stripes in the same way as in the previous construction for neutral-mode flipping. 
We introduce three  tunneling terms $\mathcal{H}_2$, $\mathcal{H}_3$, and $\mathcal{H}_4$.
\begin{align} \label{eq:half_gap_charge}
\nonumber
\mathcal{H}_2
&=\Gamma_2
\sum_{j=1}^{\infty}
\left(\psi_{j,R} e^{2i\phi^{\nu=1/2}_{\rho, j, R}}\right)^2
\left(\psi_{j+1,L}e^{-2i\phi^{\nu=1/2}_{\rho, j+1, L}}\right)^2
+\text{H.c.}
\\
&=~2\Gamma_2
\sum_{j=1}^{\infty}
\cos{\left(4\phi^{\nu=1/2}_{\rho, j, R}-4\phi^{\nu=1/2}_{\rho, j+1, L}\right)}
\end{align}
gaps out the charged modes.
\begin{align} \label{eq:half_gap_Majorana}
\nonumber
\mathcal{H}_3
&=\Gamma_3
\sum_{j=1}^{\infty}
\left(\psi_{j,R} e^{2i\phi^{\nu=1/2}_{\rho, j, R}}\right)
\left(\psi_{j+1,L} e^{-2i\phi^{\nu=1/2}_{\rho, j+1, L}}\right)
+\text{H.c.}
\\
&=\tilde{\Gamma}_3
\sum_{j=1}^{\infty}
\psi_{j,R}\psi_{j+1,L}
+\text{H.c.}
\end{align}
gaps out the Majorana modes.
\begin{align} \label{eq:half_gap_neutral}
\nonumber
\mathcal{H}_4
&=\Gamma_4
\sum_{j=1}^{\infty}
\sum_{i=1}^N
\left(e^{-i\phi^{\nu=1/2}_{n_i, j, R}} 
e^{2i\phi^{\nu=1/2}_{\rho, j, R}}\right)
\\ \nonumber
&\quad\quad\quad\quad\quad\quad
\left(e^{i\phi^{\nu=1/2}_{n_i, j+1, L}} 
e^{-2i\phi^{\nu=1/2}_{\rho, j+1, L}}\right)
+\text{H.c.}
\\
&=2\tilde{\Gamma}_4
\sum_{j=1}^{\infty}
\sum_{i=1}^N
\cos{\left(\phi^{\nu=1/2}_{n_i, j+1, L}-\phi^{\nu=1/2}_{n_i, j, R}\right)}
\end{align}
gaps out the bosonic neutral modes. After the introduction of the couplings $\mathcal{H}_{1,2,3,4}$, only the integer charged mode $\phi^{\nu=1}_{\rho,1, R}$, and the fractional modes $\phi^{\nu=1/2}_{\rho,1,L}$, $\phi^{\nu=1/2}_{n_i,1,L}$, and $\psi_{1,L}$ remain gapless at the edge. This edge structure is shown in the upper right panel in Fig.~\ref{fig:PH_conjugation}.

To complete our procedure, we add a density-density interaction of the two charged modes  $\phi^{\nu=1}_{\rho,1, R}$ and $\phi^{\nu=1/2}_{\rho,1,L}$. Its energy density is
\begin{equation}
\label{eq:dima-w}
\mathcal{H}_w=\frac{2w}{4\pi}\partial_x\phi^{\nu=1}_{\rho,1, R}\partial_x\phi^{\nu=1/2}_{\rho,1,L}.
\end{equation}
The two charged modes decouple from the rest of the modes. The Lagrangian density of the charged modes becomes
\begin{align}
\label{eq:dima-c-n-nA}
\mathcal{L}
=&-\frac{1}{4\pi}[\partial_t\phi^{\nu=1}_{\rho,1, R}\partial_x\phi^{\nu=1}_{\rho,1, R}+v_1(\partial_x \phi^{\nu=1}_{\rho,1, R})^2] 
\nonumber \\
&+\frac{2}{4\pi}[\partial_t\phi^{\nu=1/2}_{\rho,1,L}\partial_x\phi^{\nu=1/2}_{\rho,1,L}
-v_{1/2}(\phi^{\nu=1/2}_{\rho,1,L})^2] 
\nonumber\\
&-\frac{2w}{4\pi}\partial_x\phi^{\nu=1}_{\rho,1, R}\partial_x\phi^{\nu=1/2}_{\rho,1,L}.
\end{align}
We introduce a new charged mode $\phi_\rho=\phi^{\nu=1}_{\rho,1, R}-\phi^{\nu=1/2}_{\rho,1,L}$ and a new neutral mode $\phi_{N+1}=\phi^{\nu=1}_{\rho,1, R}-2\phi^{\nu=1/2}_{\rho,1,L}$. We also choose $w=-2(v_1+v_{1/2})/3$ and $v_1=2v_{1/2}-3v_n$. The action then becomes
\begin{align}
\label{eq:dima-c-n-nA-2}
\mathcal{L}
=&-\frac{2}{4\pi}[\partial_t\phi_\rho\partial_x\phi_\rho+v_\rho(\partial_x\phi_\rho)^2] \nonumber\\
&+\frac{1}{4\pi}[\partial_t\phi_{N+1}\partial_x\phi_{N+1}-v_n(\partial_x\phi_{N+1})^2],
\end{align}
where $v_n$ is the same velocity as the speed of the rest of the neutral modes, and $v_\rho=v_{1/2}-2v_n$. To make sure that the Hamiltonian is positive definite, we assume that
$v_{\rho},v_{1/2},v_1\gg v_n$. The action (\ref{eq:dima-c-n-nA-2}) shows two decoupled modes. Adding the rest of the neutral modes, we arrive to the edge structure, depicted in the lower right panel of Fig. \ref{fig:PH_conjugation}.  This corresponds to the contribution of any of the edges of the stripes to Eq. (\ref{eq:dima-decoupled}). This was precisely our goal.
The structure of the allowed electron operators on the edge remains the same as before the tunneling between the stripes was turned on since the gapless edge structure is simply inherited from the lowest edges of the lowest wide and narrow stripes in Fig. \ref{fig:PH_conjugation}.

%A disorder interaction between the two charged modes $\phi^{\nu=1}_{\rho,1, R}$ and 
%$\phi^{\nu=1/2}_{\rho,1,L}$ can further equilibrate them and lead to a new charged mode
%$\tilde{\phi}^{\nu=1/2}_{\rho,1,R}$ and a new additional bosonic neutral mode
%$\tilde{\phi}^{\nu=1/2}_{n,1,L}$~\cite{Kane-disorder-dominated}. As illustrated in the lower right panel in Fig.~\ref{fig:PH_conjugation}, the final edge structure contains one charged %mode (downstream), $N+1$ bosonic neutral modes (upstream) and one Majorana mode (upstream). By comparing to the edge structure of the original $\nu=1/2$ FQH stripe, one can conclude %that the final edge structure after the coupled-stripe construction is equivalent to a particle-hole conjuguation. In other words, from a topological order with Chern number $\nu_C$ to 
%$-(\nu_C+2)$. Combining with the previous discussion on effective construction for neutral-mode flipping, we conclude that effective Hamiltonians for different non-Abelian topological% %orders in the 16-fold way can be constructed iteratively according to their mother-daughter relation as shown in Fig.~\ref{fig:non_Abelian}.

\subsection{Second coupled-stripe construction CW2 for non-Abelian topological orders}

\begin{figure}[htb]
\includegraphics[width=3.3 in]{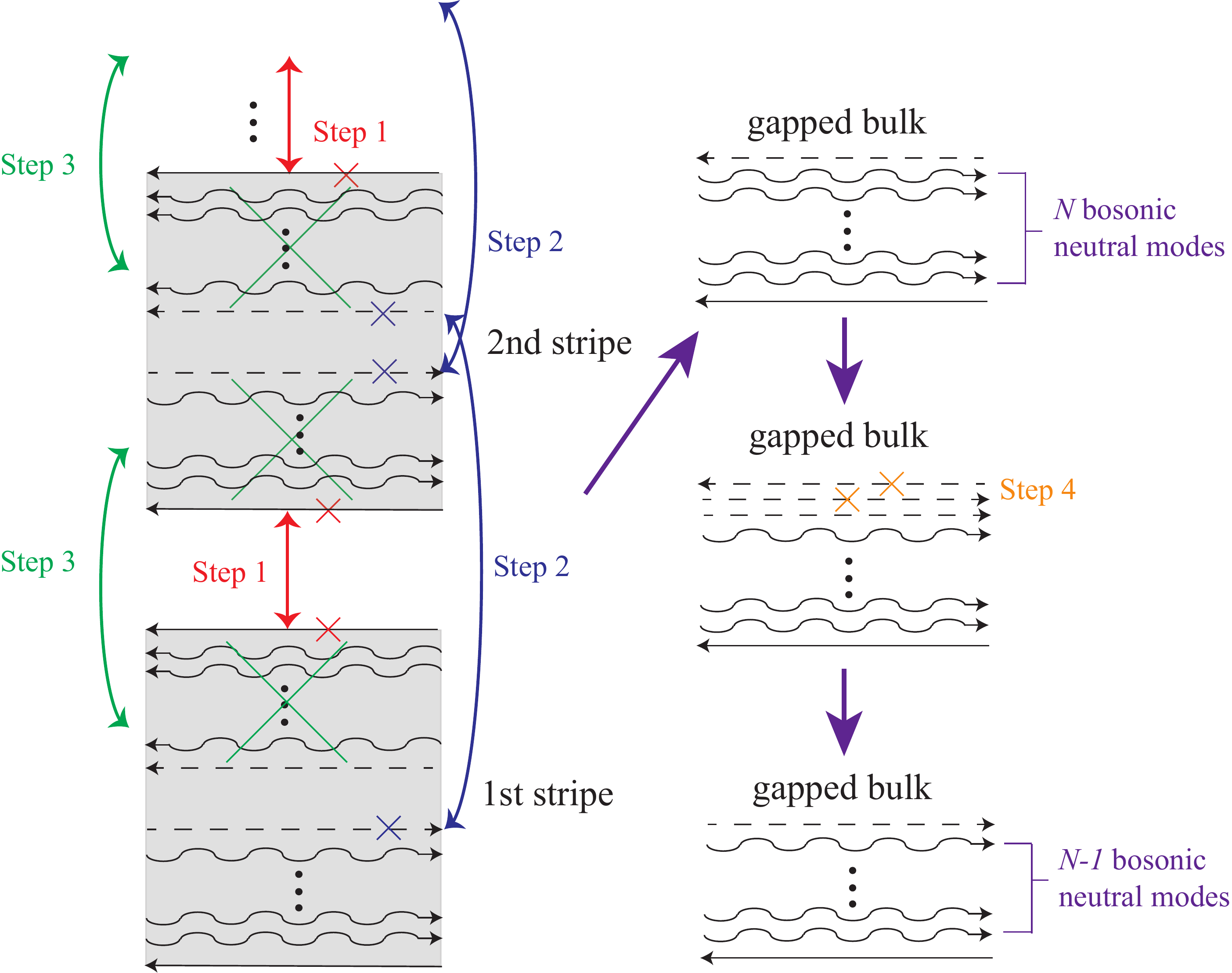}
\caption{(Color online) Second coupled-stripe construction (CW2) for non-Abelian topological orders of the sixteenfold way. On Step 4, we fermionize a neutral bosonic mode into two Majorana fermions, then gap out one of them by coupling it to the counter-propagating Majorana mode. Thus, the number of the bosonic neutral modes decreases by one.}
\label{fig:CW2_non_Abelian}
\end{figure}

Here, we provide a short discussion of another iterative coupled-stripe construction to relate different non-Abelian orders of the sixteenfold way. This  construction is called CW2 in Fig.~\ref{fig:non_Abelian}. It differs from CW1 in two ways. First,  neutral bosonic modes are gapped out in a different way on step 3 (cf. step 3 in Fig.~\ref{fig:CW1_non_Abelian} and Fig.~\ref{fig:CW2_non_Abelian}). Second, an additional step 4 is introduced.

After the coupling of the Hall stripes with three tunneling processes as shown in Fig.~\ref{fig:CW2_non_Abelian}, a gapless Majorana mode is left at the edge. 
Its propagation direction is opposite to the direction of the remaining gapless neutral Bose modes.
This ``wrongly-moving'' mode can be gapped out by coupling it to a Majorana mode obtained by fermionizing one of the neutral bosonic modes at the edge. Indeed, a Bose mode can be seen as two co-propagating Majoranas. As a result, this construction reduces the number of the bosonic neutral modes by one. Thus, it provides a way to relate the effective Hamiltonians of the orders from the sixteenfold way with the Chern numbers $\nu_C$ and $\nu_C-2$, as shown in Fig.~\ref{fig:non_Abelian}.

\subsection{Coupled-stripe construction for Abelian topological orders}   

A coupled-stripe construction can also be employed to construct effective Hamiltonians for the Abelian orders from the sixteenfold way. We  first construct an effective Hamiltonian for the 
331 order from the Pfaffian order. After this is done, the same tricks as in the previous subsection produce effective Hamiltonians for all the other Abelian orders. 

\subsubsection{Pfaffian order and $331$ order}
\label{sec:Pf_331}

\begin{figure}[htb]
\includegraphics[width=3.2 in]{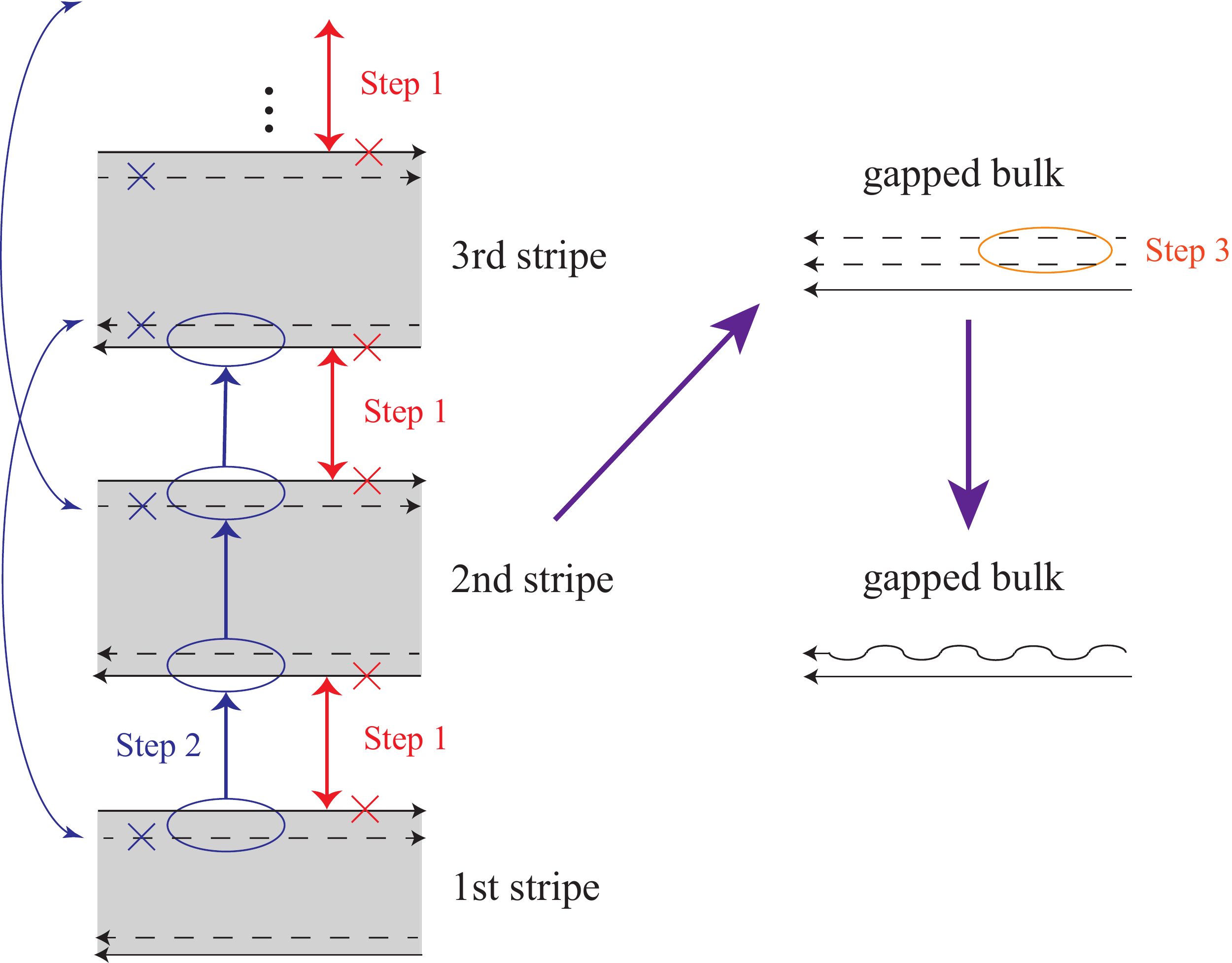}
\caption{(Color online) Coupled-stripe construction for the Abelian $331$ order from a collection of quantum Hall stripes in the Pfaffian state. On Step 3, two co-propagating Majorana modes form one neutral bosonic mode.}
\label{fig:Pfaffian_to_331}
\end{figure}

In Fig.~\ref{fig:Pfaffian_to_331}, we illustrate the coupled-stripe system and the corresponding couplings for constructing the $331$ order from the Pfaffian state. On Step 1, charged modes are gapped out by the tunneling operator from Eq.~\eqref{eq:gap_charge}. Since the edge of the $331$ liquid has one downstream neutral bosonic mode which is equivalent to two downstream Majorana modes, the Majorana modes in the stripes should be gapped by coupling the $j$th stripe and the 
$(j+2)$th stripe on Step 2 as shown in the figure. 
%This part of the construction consists of two steps, which are labeled as Step 2 and Step 3 in the figure.
More precisely, we introduce transfer of a pair of electrons among three stripes $j$, $j+1$, $j+2$. The Hamiltonian density of the tunneling term is

\begin{align}
\nonumber
\mathcal{H}_2
=&~\Gamma_2
\sum_{j=1}^{\infty}
\left[\left(\psi_{j,R} e^{2i\phi_{\rho, j, R}}\right)
\left(\psi_{j+1,L} e^{-2i\phi_{\rho, j+1, L}}\right)\right]
\\ \nonumber
&\quad\quad\quad
\times\left[\psi_{j+1,L}\psi_{j+1,R}\right]
\\ \nonumber
&\quad\quad\quad
\times\left[\left(\psi_{j+1,R} e^{2i\phi_{\rho, j+1, R}}\right)
\left(\psi_{j+2,L} e^{-2i\phi_{\rho, j+2, L}}\right)\right]
\\ \nonumber
&+\text{H.c.}
\\ 
=&~\tilde{\Gamma}_2\sum_{j=1}^\infty\psi_{j,R}\psi_{j+2,L}+\text{H.c.}
\end{align}
This operator is allowed since it conserves the total electric charge and the topological charge. Indeed, all four expressions in the parentheses are topologically trivial electron operators. The middle square brackets transfer a Majorana fermion between the edges of the same stripe and hence is allowed.

%The net effect of the transfer process is gapping out the combination:
%$4\phi_{\rho, j, R}-4\phi_{\rho, j+2, L}$. When $\Gamma_2$ is sufficiently strong, one obtains:
%\begin{eqnarray}
%\mathcal{H}_2
%=2\Gamma_2\cos{\left[4\left(\phi_{\rho, j, R}-\phi_{\rho, j+2, L}\right)\right]}.
%\end{eqnarray}
%Then, we proceed to gap out the Majorana modes in the bulk by the following coupling term:
%\begin{align}
%\nonumber
%\mathcal{H}_3
%&=\Gamma_3
%\sum_{j=1}^{\infty}
%\left(\psi_{j,R} e^{2i\phi_{\rho, j, R}}\right)
%\left(\psi_{j+2,L} e^{-2i\phi_{\rho, j+2, L}}\right)
%+\text{h.c.}
%\\
%&=\tilde{\Gamma}_3
%\sum_{j=1}^{\infty}
%\psi_{j,R}\psi_{j+2,L}
%+\text{h.c.}
%\end{align}

Steps 1 and 2 gap out all modes, except for $\phi_{\rho,1,L}$, $\psi_{1,L}$, and $\psi_{2,L}$. Notice that the two gapless Majorana modes have the the same chirality. Hence, they can be combined to form a single Dirac fermion. In the bosonization language, it is equivalent to a bosonic neutral mode
$\phi_{n,1,L}$. Finally, the effective Hamiltonian density for the edge modes is given by
\begin{eqnarray}
\mathcal{H}_{\text{edge}}
=\frac{2v_{\rho}}{4\pi}
\left(\partial_x \phi_{\rho, 1, L}\right)^2
+\frac{  v_n}{4\pi}
\left(\partial_x \phi_{n, 1, L}\right)^2.
\end{eqnarray}
This is  the edge structure of the $331$ order. The electron operators can be chosen in the form

\begin{align}
\Psi_e\sim\exp(2i\phi_{\rho,1,L})\psi_{1,L}
%\left
\{1\pm i[\psi_{1,L}\psi_{1,R}]\nonumber\\
\times[\psi_{1,R}\cos(2[\phi_{\rho,1,R}-\phi_{\rho,2,L}])\psi_{2,L}]%
%\right
\},
\end{align}
or, equivalently,

\begin{align}
\Psi_e\sim \exp(2i\phi_{\rho,1,L})[\psi_{1,L}\pm i\psi_{2,L}].
\end{align}

\begin{figure}[htb]
\includegraphics[width=3.3 in]{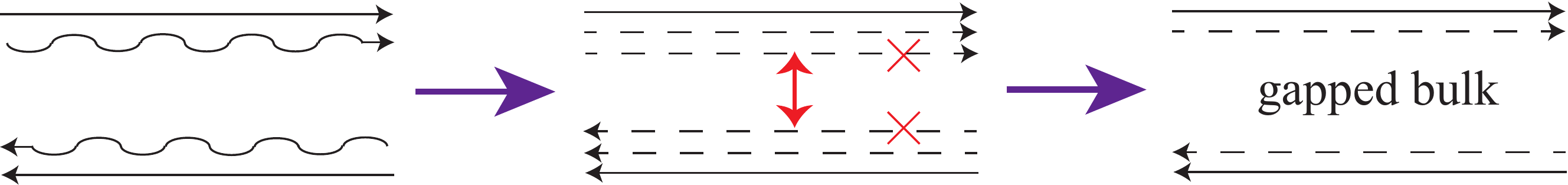}
\caption{(Color online) Obtaining the edge structure of the Pfaffian state from the $331$ state. For simplicity, we consider a single stripe. The bosonic neutral mode (wavy line) at the edge is fermionized into two Majorana modes (dashed lines). One of the Majorana modes couples with a Majorana mode on the opposite edge. The coupled Majorana modes are gapped out.}
\label{fig:331_to_Pfaffian}
\end{figure}

It is also possible to construct the Pfaffian state from the $331$ state. We illustrate this by an example of a single stripe, as shown in Fig.~\ref{fig:331_to_Pfaffian}.
Our example only shows how to get the Pfaffian edge structure from the 331 edge structure. Multiple stripes are needed to produce the bulk Pfaffian order.
 Recall that a bosonic neutral mode at the edge of the $331$ state can be fermionized into two copropagating Majorana modes. Since quasiparticles can tunnel between the two opposite edges of the same quantum Hall liquid (but not across two different quantum Hall liquids), two counterpropagating Majorana modes can be directly coupled and gapped out. The resulting edge structure consists of a charged mode and one downstream Majorana mode on each edge.  This is the edge structure of the Pfaffian state.

\subsubsection{Iterative construction for other Abelian orders}

In this sub-subsection, we consider in detail the construction of the $113$ order from the $331$ order. The construction is very similar to the one we used in the non-Abelian case. We then briefly address a generalization to an arbitrary Abelian topological order.

\begin{figure}[htb]
\includegraphics[width=3.0 in]{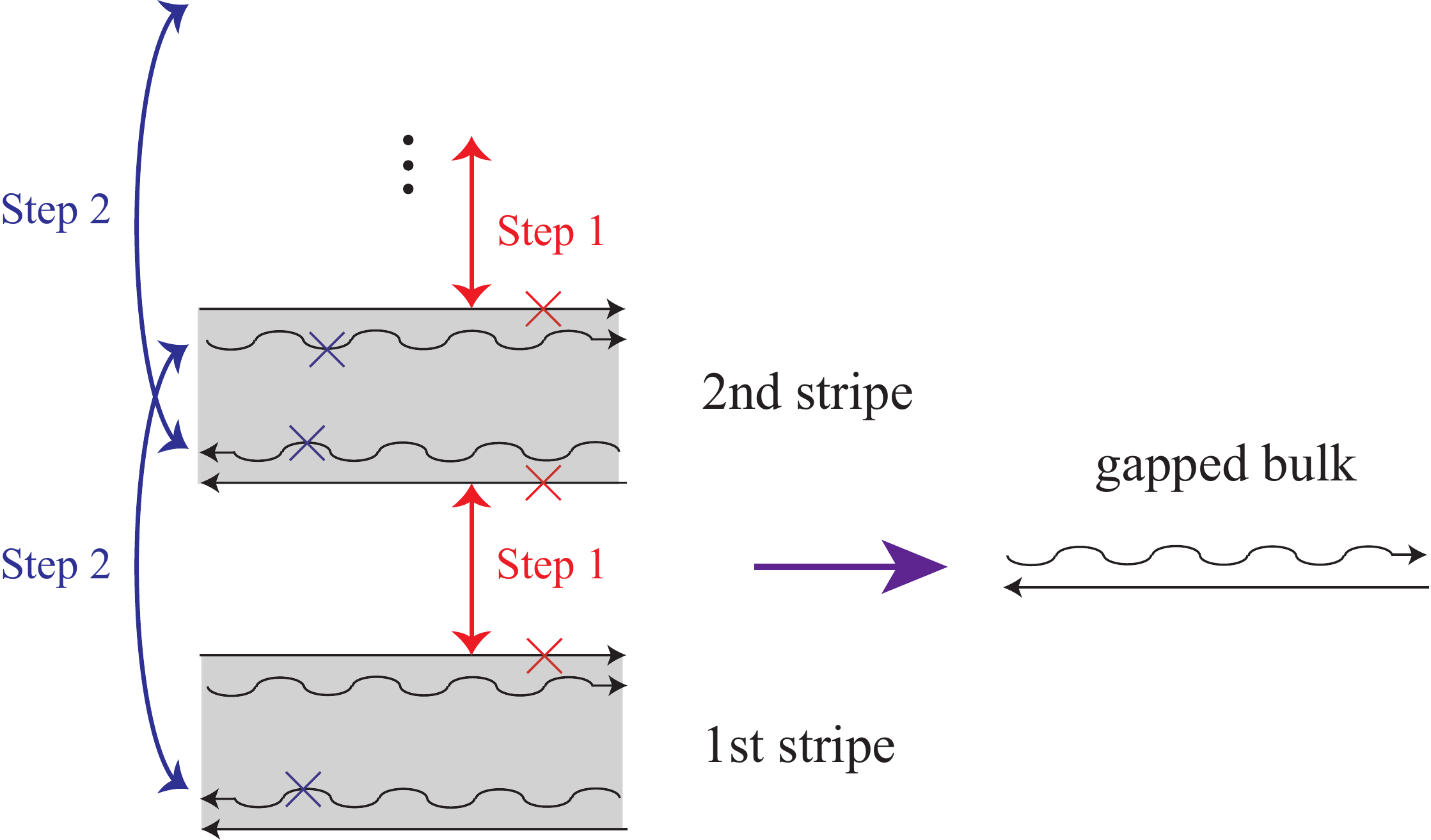}
\caption{(Color online) Coupled-stripe construction for the $113$ order from a collection of quantum Hall stripes in the $331$ state. Here, the solid lines (wavy lines) represent the charged modes (neutral modes).}
\label{fig:331_113}
\end{figure}

The construction of the $113$ order from the $331$ order is parallel to the construction of PH-Pfaffian order from the Pfaffian order. We illustrate the construction in Fig.~\ref{fig:331_113}.

Since a Majorana mode is absent on the edges of Abelian stripes, the electron operators are $\hat{\Psi}_\pm=e^{\pm i\phi_n}e^{2i\phi_\rho}$.
% and 
%$\hat{\Psi}_2=e^{i\phi_n}e^{2i\phi_\rho}$.
Here, $\phi_n$ can be any one of the neutral modes on the edge. Thus, it is possible to gap out the charged modes in the $331$ quantum Hall stripes with the following interaction when $\Gamma_1$ is sufficiently strong: 
\begin{align} 
\nonumber
\mathcal{H}_1
=&~\Gamma_1
\sum_{j=1}^{\infty}
\left[\left(e^{-i\phi_{n ,j,R}} e^{2i\phi_{\rho, j, R}}\right)
\left(e^{i\phi_{n,j,R}} e^{2i\phi_{\rho, j, R}}\right)\right]
\\ \nonumber
&\quad\quad\quad
\times\left[
\left(e^{i\phi_{n,j+1,L}} e^{-2i\phi_{\rho, j, R}}\right)
\left(e^{-i\phi_{n,j,L}} e^{-2i\phi_{\rho, j, R}}\right)\right]
\\ \nonumber
&+\text{H.c.}
\\
=&~2\Gamma_1
\sum_{j=1}^{\infty}
\cos{\left(4\phi_{\rho, j, R}-4\phi_{\rho, j+1, L}\right)}.
\label{eq:gap_charge_Abelian} 
\end{align}

Next, we proceed to gap out the neutral modes in the bulk as shown in the figure (Step 2). The corresponding interaction term is given by:
\begin{align}
\nonumber
\mathcal{H}_2
=&~\Gamma_2
\sum_{j=1}^{\infty}
\left(e^{-i\phi_{n,j,L}} e^{2i\phi_{\rho, j, R}}\right)
\left(e^{i\phi_{n,j+1,R}} e^{-2i\phi_{\rho, j+1, L}}\right)
\\ \nonumber
&+\text{H.c.}
\\
=&~2\tilde{\Gamma}_2
\sum_{j=1}^{\infty}
\cos{\left(\phi_{n, j+1, R}-\phi_{n, j, L}\right)}.
\end{align}
%We emphasize again that the left- and right-moving neutral modes carry the same. 
By gapping out the modes in the stripes with  $\mathcal{H}_1$ and $\mathcal{H}_2$, one recovers the edge structure of the 113 order. Following the non-Abelian case, one can also easily verify that the correct structure of the electron operators on the edge is reproduced by this procedure.

\begin{figure}[htb]
\includegraphics[width=3.3 in]{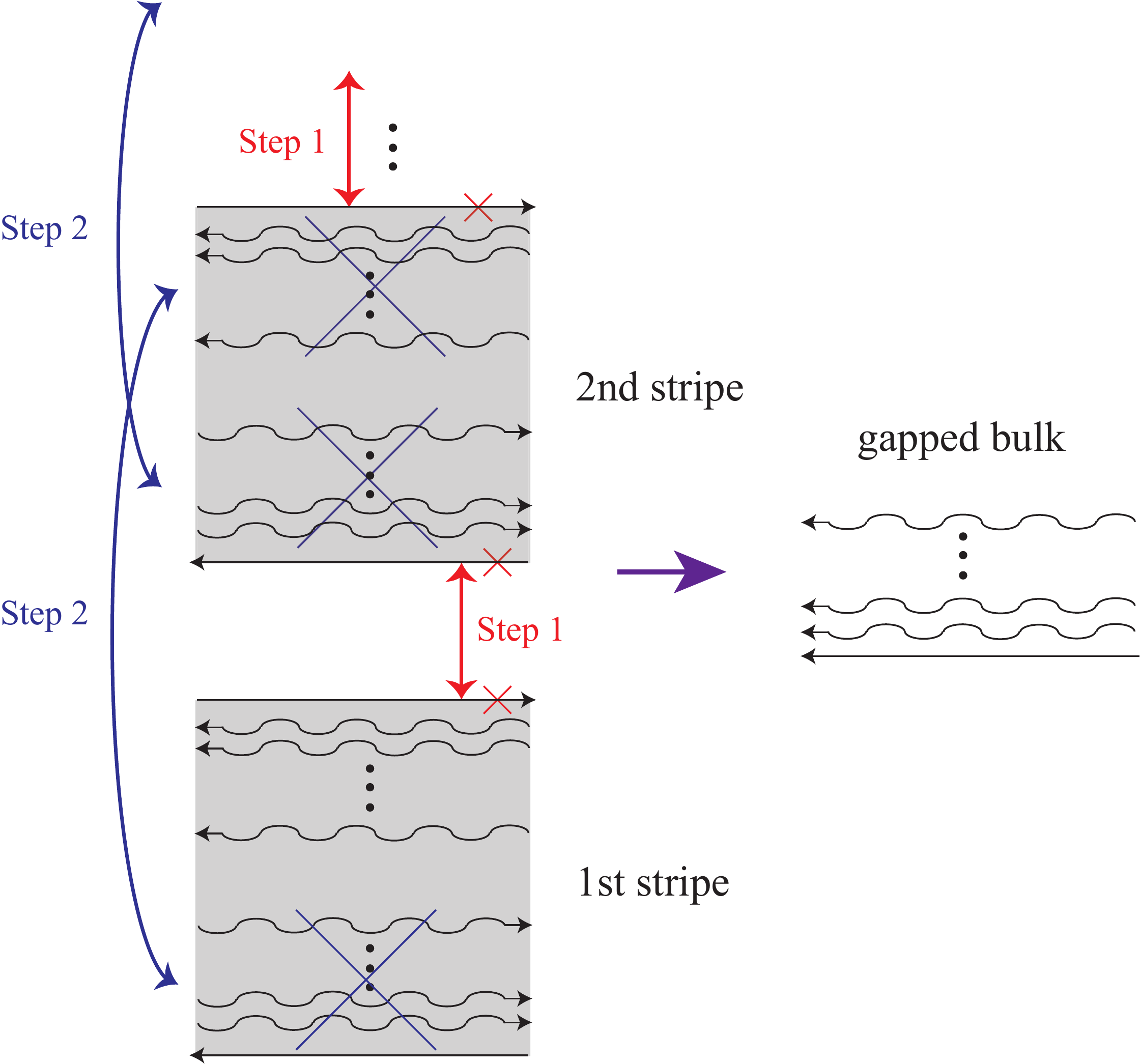}
\caption{(Color online) Neutral-mode flipping construction for Abelian topological orders.}
\label{fig:CW1_Abelian}
\end{figure}

For Abelian orders with more neutral modes on the edge, the same procedure can be applied to construct effective Hamiltonians for the topological order with the Chern number $\nu_C$ from a collection of stripes in the state with the Chern number $-\nu_C$ as shown in Fig.~\ref{fig:CW1_Abelian}. The charged modes can be gapped out by the interaction from 
Eq.~\eqref{eq:gap_charge_Abelian}. Equation~\eqref{eq:gap_neutral_modes} shows a way to gap out the bulk neutral modes. 

Aside from neutral-mode flipping (Fig. \ref{fig:CW1_Abelian}), we also need to perform particle-hole conjugation. The procedure is essentially identical to the non-Abelian case. The only important difference is that the interaction (\ref{eq:gap_charge_Abelian}) is used to gap out fractional charged modes.

% ------------------------------ Conclusion -----------------------------------------

\section{Conclusions} \label{sec:conclusion}

Composite fermions give an intuitive and powerful approach to FQHE. At odd-denominator filling factors, a difficult FQHE problem reduces to the much simpler integer QHE of composite fermions. In the latter problem, the single-particle spectrum is gapped. As a result, the basic properties of the QHE liquid are robust. In particular, similar physics is expected for a great variety of microscpic Hamiltonians. As long as the filling factor is the same, one can realistically expect the same topological order in a complicated experimental system and in a system with a highly simplified Hamiltonian, suitable for numerical simulations. 

Such picture cannot be generalized to half-integer filling factors, where the simplest application of the composite-fermion idea predicts a gapless liquid. This simplest behavior is compatible with the experiment at some filling factors but not at the others. This is not surprising, since gapless states are not as robust as gapped ones. Indeed, a gapless liquid can be unstable to various weak interactions. Kitaev's classification reveals 16 instabilities which lead to 16 possible topological orders. All 16 orders are close relatives since they all emerge from Cooper pairing of the same type of composite fermions. Importantly, the existing numerical evidence does support a state of the sixteenfold way and hence the composite fermion description of half-integer quantum Hall plateaus. Given a close relation of the 16 orders, it is much harder to narrow down the list of possibilities to a single state. This subtle problem goes beyond the sort of questions one has to tackle at simpler filling factors like $1/3$. The current debate about the Pfaffian, PH-Pfaffian, and anti-Pfaffian orders in GaAs at $\nu=5/2$  illustrates this point \cite{comment2018}.

One cannot help wondering whether all 16 topological orders of the sixteenfold way may be present in some physical systems. Only experiment can shed light on this question. This motivates a review of possible experimental signatures in this paper.

The PH-Pfaffian order gives rise to a curious situation. That topological order is compatible with the particle-hole (PH) symmetry \cite{Son2015}, yet, it appears that PH symmetric Hamiltonians break the PH symmetry in their ground states \cite{x1-Morf, new-num, Zaletel2015, Rezayi-PRL}. 
On the other hand, disorder and Landau level mixing break the PH symmetry of a Hamiltonian. The symmetry-from-no-symmetry principle \cite{Zucker2016} suggests that those symmetry-breaking effects stabilize the PH-symmetric order. Indeed, mechanisms \cite{Milovanovic_LLM, Mross2018, Wang2018, Lian2018} have been proposed for the stabilization  of the PH-Pfaffian topological order by LLM and disorder. Moreover, existing coupled-wire constructions for the PH-Pfaffian order also break the particle-hole symmetry (see Ref. \onlinecite{Kane-Stern-Halperin}).
In fact, our coupled-stripe construction for getting the PH-Pfaffian order from Pfaffian stripes is rather similar to the stabilization of the PH-Pfaffian order by disorder in the mechanisms \cite{Mross2018, Wang2018, Lian2018}. The coupled-stripe construction involves no disorder, but it breaks the PH symmetry in a way similar to how it is broken by disorder in those mechanisms \cite{Mross2018, Wang2018, Lian2018}. This is another manifestation of the symmetry-from-no-symmetry principle  \cite{Zucker2016} for the PH-Pfaffian liquids.

In conclusion, we give a uniform description of different proposed topological orders for the half-integer fractional quantum Hall states. The candidate orders can be seen as 
arising from Cooper pairing between composite fermions in different pairing channels. We introduce a mother-daughter relation between the topological orders, which relates them iteratively via particle-hole conjugation and neutral-mode flipping. The same mother-daughter relation allows us to iteratively construct wave functions and effective Hamiltonians for all orders. We also verify explicitly that all resulting topological orders belong to Kitaev's sixteenfold way~\cite{Kitaev}. This  is used to predict experimental signatures of all 16 orders in multiple types of experiments, as summarized in Table~\ref{tab:summary}. 
%Finally, we point out that the tunneling currents and the Fano factors from the Mach-Zehnder interferometry for all orders are consistent with the Byers-Yang Theorem~\cite{Byers-Yang}. 

% ------------------------------ Acknowledgements -----------------------------------------

\begin{acknowledgments}
We acknowledge useful discussions with P. T. Zucker. This research was supported in part by the National Science Foundation under Grant No. DMR-1607451.
 
\end{acknowledgments}

%%%%%%%%%%%%%%%

%\newpage
\appendix

%----------------------- Fabry-Perot Interferometry ------------------------------

%--------------------- Finite T MZ interferometer ------------

\section{Finite-temperature Mach-Zehnder interferometry}
\label{app:Finite_T_MZ}

In Sec.~\ref{sec:MZ_experiment}, we discussed experimental signatures of topological orders in Mach-Zehnder interferometry at zero temperature. In this appendix, the discussion is generalized to finite-temperature systems on the basis of the kinetic equation approach~\cite{KT_noise, KT2008}.

\subsection{Review of kinetic equations}

We introduce the symbol $P_{s,i}(t)$ for the probability that the charge $sq$ was transferred from source S1 to drain D2 during the time $t$. Here $q$ is the charge of the quasiparticle which dominates tunneling, and $i$ labels the topological charge of drain D2 at the time $t$. The topological charge is not affected by the transfer of an integer number of electrons to D2 ($s\rightarrow s+ne/q$).
% The resulting particle is in the superselection sector $i$, with $1\leq i \leq \mathcal{N}$ and $\mathcal{N}$ is the number of possible superselection sectors for D2. Here, $q$ stands %for . We define the superselection sector for D2 as $i=1$ at $t=0$ (for example, $i=1$ for the superselection sector $(-e/4,\sigma)$ in Fig.~\ref{fig:MZ_e/4}). In order for D2 to be in %the $i$-th superselection sector, $k$ quasiparticles are required to be transferred from S1 to D2. 
The probability satisfies the following kinetic equation:
\begin{align} \label{eq:kinetic_eqn_P}
\nonumber
\frac{d}{dt} P_{l,i} (t)
=~&\sum_{j=1}^{\mathcal{N}} \left[P_{l-1,j}(t)w^{+}_{j\rightarrow i}
+P_{l+1,j}(t)w^{-}_{j\rightarrow i}\right]
\\
&-\sum_{j=1}^{\mathcal{N}} P_{l,j} (t)
\left(w^{+}_{i\rightarrow j}+w^{-}_{i\rightarrow j}\right).
\end{align}
In the above equation, $\mathcal{N}$ labels the number of possible topological charges.
% which depends on the topological order and the type of quasiparticle participating in the tunneling process. 
The symbol $w_{i\rightarrow j}$ labels the transition rate from sector $i$ to sector $j$. The superscript ``$+$'' corresponds to tunneling from the edge with the higher electrochemical potential to the edge with the lower electrochemical potential (edge 1 to edge 2 in Fig.~\ref{fig:MZ}).  We call this type of tunneling ``forward tunneling''. At a non-zero temperature, thermal fluctuations allow backward tunneling from edge 2 to edge 1. The corresponding transition rates carry the superscript ``$-$''.

The calculation of $w_{i\rightarrow j}$ consists of two steps. First, we assume that the tunneling anyon and the initial topological charge $i$ of the drain are in the 
fusion channel $j$. We compute the tunneling rate $p^{+}_{i\rightarrow j}$ under this assumption. On the second step, we multiply the outcome $p^{+}_{i\rightarrow j}$ of the first step
by the probability of the fusion channel $j$.  The bare tunneling rate $p^{-}_{j\rightarrow i}$ is defined in a similar way. It is related to the rate of the forward process by the detailed balance principle:
\begin{eqnarray}
p^{-}_{j\rightarrow i}
=\exp{\left(-\frac{qV}{k_B T}\right)} p^{+}_{i\rightarrow j}.
\end{eqnarray}
Again, the above result must be multiplied by the probability of the fusion outcome $i$.
%Let $\alpha$ be the topological charge of the drain in the superselection sector $i$. Let the topological charge of the tunneling particle be $x$. 
%We assume that $\alpha$ and $x$ must fuse to $\beta$ in order to reach the final superselection sector $j$. 
Let $x$ be the topological charge of the tunneling particle. The fusion probability of $i$ and $x$ into $j$ is known from the algebraic theory of anyons~\cite{Kitaev, KT_noise, KT2008, Feldman2006}:
\begin{eqnarray}
p^{j}_{i x}
=N^j_{ix}\frac{d_j}{d_i d_x},
\end{eqnarray}
where $N^j_{ix}$ is the fusion multiplicity and $d_c$ labels the quantum dimension of anyon $c$. 
This fusion probability is independent of the temperature. As an example of its calculation, consider tunneling between the states $(-e/4,\sigma)$ and $(0,\psi)$. The fusion probability of forward tunneling $(-e/4\rightarrow 0)$ equals $1/2$ since $\sigma\times\sigma=\psi+I$, $N^\psi_{\sigma\sigma}=1$, $d_\psi=1$, and $d_\sigma=\sqrt{2}$. However, the fusion probability from $(0,\psi)$ to $(-e/4,\sigma)$ for   the backward tunneling is $1$ since $\psi\times\sigma=\sigma$. The total transition rates are given by
\begin{eqnarray}
w^{+}_{i\rightarrow j}=p^{\beta}_{\alpha x}p^{+}_{i\rightarrow j},~
w^{-}_{j\rightarrow i}=p^{\alpha}_{\beta \bar{x}}p^{-}_{j\rightarrow i},
\end{eqnarray}
where $\bar{x}$ is the antiparticle of $x$.
%Therefore, it is incorrect to simply reverse all the arrows in Fig.~\ref{fig:MZ_e/4} and directly use the total transition rates there when one needs to include backward tunneling for %systems at finite temperature.

We introduce a generating function
\begin{eqnarray}
f_i(z,t)
=\sum_{n=-\infty}^{\infty} P_{k+ne/q,i}(t)z^{k+ne/q}.
\end{eqnarray}
Here $k$ is uniquely determined by the topological sector $i$.
In terms of $f_i$, the average charge transmitted during the time interval $t$ and its variance are given by
\begin{eqnarray}
\langle Q(t)\rangle
=q\left.\left(\frac{d}{dz}\sum_{i=1}^{\mathcal{N}} f_i\right)\right|_{z=1}
\end{eqnarray}
and
\begin{eqnarray}
\langle \delta Q^2(t)\rangle
=q^2\left.\left(\frac{d}{dz}z\frac{d}{dz}\sum_{i=1}^{\mathcal{N}} f_i\right)\right|_{z=1}
-\langle Q(t)\rangle^2
\end{eqnarray}
From equation~\eqref{eq:kinetic_eqn_P}, we obtain a kinetic equation for $f_i(z,t)$ as
\begin{align} \label{eq:kinetic_eqn_f}
\nonumber
\frac{d}{dt} f_i (z,t)
=~&\sum_{j=1}^{\mathcal{N} }
\left[zf_j(z,t)w^{+}_{j\rightarrow i}
+\frac{1}{z} f_j(z,t)w^{-}_{j\rightarrow i}\right]
\\
&-\sum_{j=1}^{\mathcal{N}} f_i (z,t)
\left(w^{+}_{i\rightarrow j}+w^{-}_{i\rightarrow j}\right)
\end{align}
The above equation can be written in the matrix form: 
$\dot{\mathbf{f}} (z,t)=\mathbf{A}\cdot\mathbf{f} (z,t)$. At $z=1$, the kinetic matrix 
$\mathbf{A}$ satisfies the Rohbrach theorem~\cite{Rorbach_theorem}. Therefore, all its eigenvalues are non-positive at $z=1$. Besides, one of the eigenvalues must be zero and non-degenerate. We denote it as $\lambda(z)$. This eigenvalue dominates the long-term behavior of the solution of equation~\eqref{eq:kinetic_eqn_f}. With this idea, the tunneling current and the Fano factor can be evaluated as
\begin{eqnarray} \label{eq:finite_T_current}
I
=\lim_{t\rightarrow\infty} \frac{\langle Q(t)\rangle}{t}
=q\left.\lambda'(z)\right|_{z=1}
\end{eqnarray}
and
\begin{eqnarray} \label{eq:finite_T_Fano}
e^*
=\lim_{t\rightarrow\infty} 
\frac{\langle \delta Q^2(t)\rangle}{\langle Q(t)\rangle}
=q\left[1+\frac{\left.\lambda''(z)\right|_{z=1}}
{\left.\lambda'(z)\right|_{z=1}}\right].
\end{eqnarray}

In practice, it is not straightforward to obtain $\lambda(z)$. Nevertheless, $\lambda'(1)$ and 
$\lambda''(1)$ can be determined from the characteristic equation: 
$\text{det}[\mathbf{A}(z)-\lambda(z)\mathbf{I}]=0$~\cite{KT2008}. Suppose the characteristic equation takes the form $C_0(z)+C_1(z)\lambda(z)+C_2(z)\lambda(z)^2+\cdots=0$. Using the condition that $\lambda(1)=0$ and the product rule, we have
\begin{align}
\label{eq:first_D_lambda}
\lambda'(1)
&=-\frac{C_0'(1)}{C_1(1)},
\\
\label{eq:second_D_lambda}
\lambda''(1)
&=-\frac{C_0''(1)+2C_1'(1)\lambda'(1)+2C_2(1)\lambda'(1)^2}{C_1(1)}.
\end{align}
From the above results, the tunneling current and the Fano factor at finite temperatures can be evaluated systematically.

\subsection{$e/4$ quasiparticle tunneling}

Now, we evaluate the tunneling current and the Fano factor at a finite temperature 
when the tunneling process is dominated by charge-$e/4$ quasiparticles. We focus on non-Abelian orders. In this case, there are $\mathcal{N}=6$ superselection sectors as depicted in 
Fig.~\ref{fig:MZ_e/4}. For simplicity, we separate the kinetic matrix into three pieces: 
$\mathbf{A}=\mathbf{A}_F+\mathbf{A}_B+\mathbf{A}_L$. The first matrix corresponds to forward tunneling from state $j$ to state $i$. By ordering the superselection sectors as
$\left(-e/4,\sigma\right)$, $\left(0,\psi\right)$, $\left(0,I\right)$, $\left(e/4,\sigma\right)$, 
$\left(e/2,\psi\right)$, and $\left(e/2,I\right)$, we have
\begin{eqnarray}
\nonumber
\mathbf{A}_F
=z
\begin{pmatrix}
0 & 0 & 0 & 0 & p(-\frac{\pi}{2}) & p(\frac{\pi}{2}) 
\\
p_{\mathbf{1}}/2 & 0 & 0 & 0 & 0 & 0
\\
p_{\mathbf{3}}/2 & 0 & 0 & 0 & 0 & 0
\\
0 & p(\pi) & p(0) & 0 & 0 & 0
\\
0 & 0 & 0 & p_{\mathbf{2}}/2 & 0 & 0
\\
0 & 0 & 0 & p_{\mathbf{4}}/2 & 0 & 0
\end{pmatrix}.
\\
\end{eqnarray} 
The symbols $p_{\mathbf{k}}$ and $p(\phi_s)$ are defined in Sec.~\ref{sec:MZ_experiment}. The second matrix represents backward tunneling from state $j$ to state $i$:
\begin{eqnarray}
\nonumber
\mathbf{A}_B
=\frac{1}{z} \mu
\begin{pmatrix}
0 & p_{\mathbf{1}} & p_{\mathbf{3}} & 0 & 0 & 0 
\\
0 & 0 & 0 & p(\pi)/2 & 0 & 0
\\
0 & 0 & 0 & p(0)/2 & 0 & 0
\\
0 & 0 & 0 & 0 & p_{\mathbf{2}} & p_{\mathbf{4}}
\\
p(-\frac{\pi}{2})/2 & 0 & 0 & 0 & 0 & 0
\\
p(\frac{\pi}{2})/2 & 0 & 0 & 0 & 0 & 0
\end{pmatrix},
\\
\end{eqnarray} 
where $\mu=e^{-eV/(4k_B T)}$. Lastly, $\mathbf{A}_L$ is the diagonal piece of the kinetic matrix with the following matrix elements:
\begin{align}
\nonumber
(\mathbf{A}_L)_{11}
&=-\left[\frac{p_{\mathbf{1}}}{2}+\frac{p_{\mathbf{3}}}{2}\right]
-\frac{\mu}{2}\left[ p(-\frac{\pi}{2})+ p(\frac{\pi}{2})\right],
\\
\nonumber
(\mathbf{A}_L)_{22}
&=-p(\pi)-\mu p_{\mathbf{1}},
\\
\nonumber
(\mathbf{A}_L)_{33}
&=-p(0)-\mu p_{\mathbf{3}},
\\
\nonumber
(\mathbf{A}_L)_{44}
&=-\left[\frac{p_{\mathbf{2}}}{2}+\frac{p_{\mathbf{4}}}{2}\right]
-\frac{\mu}{2}\left[ p(\pi)+ p(0)\right],
\\
\nonumber
(\mathbf{A}_L)_{55}
&=-p(-\frac{\pi}{2})-\mu p_{\mathbf{2}},
\\
(\mathbf{A}_L)_{66}
&=-p(\frac{\pi}{2})-\mu p_{\mathbf{4}}.
\end{align}

Using Eqs.~\eqref{eq:finite_T_current} and~\eqref{eq:first_D_lambda}, we obtain the generalization of Eqs.~\eqref{eq:case1_I_e/4}-\eqref{eq:case3_I_e/4} to 
a finite temperature. From the top to the bottom, $\nu_C=1,3,5,7~({\rm mod}~8)$:

\begin{widetext}
\begin{align}
\nonumber
&I_{e/4}
=\frac{er}{4}(\left|\Gamma_1\right|^2+\left|\Gamma_2\right|^2)\left(1-\mu\right)
\left[
\frac{(1+\mu)^2 (1-s^2)+\frac{s^4}{8}(\mu^2+6\mu+1)
-\frac{s^4}{8}(\mu-1)^2\cos{4\gamma}}
{(1+\mu)^2(1-\frac{3s^2}{4})
+\frac{s^4}{16}(\mu^2+6\mu+1)
-\frac{s^4}{16}\left[(\mu-1)^2\cos{4\gamma}-(\mu^2-1)\sin{4\gamma}\right]}
\right],
\\ \\
&I_{e/4}
=\frac{er}{4}(\left|\Gamma_1\right|^2+\left|\Gamma_2\right|^2)\left(1-\mu\right)
\left[
\frac{1-s^2+\frac{s^4}{4}\sin^2{2\gamma}}
{1-\frac{s^2}{2}}
\right]
=I_{e/4}(0)\left[1-e^{-eV/(4k_B T)}\right],
\\  \nonumber
&I_{e/4}
=\frac{er}{4}(\left|\Gamma_1\right|^2+\left|\Gamma_2\right|^2)\left(1-\mu\right)
\left[
\frac{(1+\mu)^2 (1-s^2)+\frac{s^4}{8}(\mu^2+6\mu+1)
-\frac{s^4}{8}(\mu-1)^2\cos{4\gamma}}
{(1+\mu)^2(1-\frac{3s^2}{4})
+\frac{s^4}{16}(\mu^2+6\mu+1)
-\frac{s^4}{16}\left[(\mu-1)^2\cos{4\gamma}+(\mu^2-1)\sin{4\gamma}\right]}
\right],
\\ \\
&I_{e/4}
=\frac{er}{4}(\left|\Gamma_1\right|^2+\left|\Gamma_2\right|^2)\left(1-\mu\right)
=I_{e/4}(0)\left[1-e^{-eV/(4k_B T)}\right].
\end{align}
\end{widetext}
%From the top to the bottom, the results are valid for the case when $\nu_C~(\text{mod }8)$ equals to $1$, $3$, $5$ and $7$, accordingly. 
Notice that the zero-temperature results can be recovered in all cases by setting $\mu=0$. 
Note also that the coefficients $r$ and $s$ can contain an additional dependence on the voltage and temperature.
 For Abelian topological orders, the calculation is essentially the same. 
However, the results are too lengthy to be displayed here. 

\subsubsection{Fano factor for the PH-Pfaffian order}

In principle, the Fano factor can be calculated from Eqs.~\eqref{eq:finite_T_Fano}, \eqref{eq:first_D_lambda}, and~\eqref{eq:second_D_lambda}. However, a simple analytic expression only exists when $\nu_C\equiv 7~(\text{mod }8)$. This covers the PH-Pfaffian case. One can show ~\cite{Zucker-thesis} that
\begin{eqnarray}
e^*(T)
=\frac{e}{2}\text{csch}{\left(\frac{eV}{4k_B T}\right)}
+e^*(0)\tanh{\left(\frac{eV}{8k_B T}\right)},
\end{eqnarray}
where $e^*(0)$ is given by Eq. (\ref{eq:dima-PH-F}), and $s$ in Eq. (\ref{eq:dima-PH-F}) may depend on $T$ and $V$.

\subsection{$e/2$-quasiparticle tunneling}

If the tunneling process is dominated by $e/2$ quasiparticles, the calculation  simplifies dramatically. There are only two superselection sectors as shown 
in Fig.~\ref{fig:MZ_e/2}. We limit our discussion to the case, represented in the left panel of Fig.~\ref{fig:MZ_e/2}. From Eq.~\eqref{eq:kinetic_eqn_f} with $\mathcal{N}=2$, one can derive the following kinetic matrix:
\renewcommand\arraystretch{1.5}
\begin{eqnarray}
\mathbf{A}
=
\begin{pmatrix}
-p(-\frac{\pi}{2})-\mu' p(\frac{\pi}{2}) & zp(\frac{\pi}{2})+\frac{1}{z}\mu' p(-\frac{\pi}{2})
\\
zp(-\frac{\pi}{2})+\frac{1}{z}\mu' p(\frac{\pi}{2}) & -p(\frac{\pi}{2})-\mu' p(-\frac{\pi}{2})
\end{pmatrix}
\end{eqnarray}
Here, $\mu'=e^{-eV/(2k_B T)}$. Following the previous procedure, we determine the tunneling current at a finite temperature as
\begin{eqnarray}
I_{e/2}(T)
=I_{e/2}(0)
\left[1-e^{-eV/(2k_B T)}\right],
\end{eqnarray}
where $I_{e/2}(0)$ is given in Eq.~\eqref{eq:tunneling_e/2: case 1}. Furthermore, the Fano factor at a finite temperature is evaluated as
\begin{eqnarray}
\nonumber
e^*_{e/2}(T)
=e~\text{csch}{\left(\frac{eV}{2k_B T}\right)}
+e^*_{e/2}(0)\tanh{\left(\frac{eV}{4k_B T}\right)}.
\\
\end{eqnarray}
As always, $r$ and $s$ in the expressions for $I_{e/2}(0)$ and $e^*_{e/2}(0)$ may depend on $T$ and $V$.

%--------------- References ------------------------------


\begin{thebibliography}{99}
\bibitem{foot-counterclockwise}
``Counterclockwise'' means counterclockwise when looking in the direction of the magnetic field. We will generally assume that negatively charged electrons move in a magnetic field, pointing down.
\bibitem{footnote-dima-1} 
Indeed, the wave function before a projection has the correct angular momentum. This implies the right electron density and shift after the projection, even if the projection is only performed once at the last step. We expect the topological order to also be correct.
\bibitem{foot-disorder}
Note that the order and disorder operators $\sigma$ and $\mu$ in the Ising chiral CFT have the same $h=1/16$.
\bibitem{footnote-03-20}
Adding $(e, 0)$ to the drain shifts all interference phases by $\pi$. This shift can be ignored if one identifies the states $(ne/4,a)$ and $([n+4]e/4,a+\psi)$ of the drain.
\end{thebibliography}
\end{document}